\documentclass[twocolumn, appendixfloats, numberedappendix, iop]{openjournal}

\makeatletter
\long\def\@makecaption#1#2{%
 \noindent\begin{minipage}{0.9999\linewidth}
   \if\csname ftype@\@captype\endcsname 2 
   \vskip 2ex\noindent \centering\@table@type@size{\@eapj@cap@font  #1: }%
    \if@chicago\@eapj@cap@font\fi #2\par\medskip
   \else
   \vspace*{\abovecaptionskip}\noindent\footnotesize #1 #2\par\vskip \belowcaptionskip
   \fi
 \end{minipage}\par
 }
\makeatother

\makeatletter
\def\LT@makecaption#1#2#3{%
  \LT@mcol\LT@cols c{\hbox to\z@{\hss\parbox[t]\LTcapwidth{%
      \centering
      #2: #3%
    \endgraf\vskip\baselineskip}%
  \hss}}}
\let\LT@makecaption@rtx\LT@makecaption
\makeatother

\makeatletter
\newcommand{\forcehref}[1]{\Hy@MakeCurrentHref{#1}}
\makeatother

\usepackage[pass]{geometry}
\usepackage{natbib}
\usepackage{longtable}
\usepackage{graphicx} 
\usepackage{threeparttable}
\usepackage[colorlinks=true,
            linkcolor=red,
            citecolor=blue,
            urlcolor=blue]{hyperref}
\let\oldhref\href
\renewcommand{\href}[2]{\oldhref{#1}{\textbf{#2}}}
\renewcommand{\ref}[1]{\textbf{\hyperref[#1]{\ref*{#1}}}}

\usepackage{xcolor}
\usepackage{orcidlink}
\usepackage{float}

\newcommand\kms{km~s$^{-1}$}
\newcommand\npassed{1,899,553}
\newcommand\nsample{1,901,834}
\newcommand\nhighew{167,466}
\newcommand\nlowew{1,734,368}
\newcommand\BPTpass{94,183}
\newcommand\ndupes{197,521}
\newcommand\nlineshigh{80}
\newcommand\nlineslow{27}

\shorttitle{The eBOSS-DAP}
\shortauthors{O. Matthews Acu\~{n}a et al.}

\begin{document}

\title{\vspace{-0.25cm}Precision Spectroscopy for 1.7 Million Galaxies from SDSS-IV:  Improved Spectral Measurements and Catalogs for eBOSS\vspace{-1.5cm}}
\author{Owen S. Matthews Acu\~{n}a*\orcidlink{0000-0001-9225-972X}$^{1}$,
Christy A. Tremonti\orcidlink{0000-0003-3097-5178}$^{1}$,
Kyle B. Westfall\orcidlink{0000-0003-1809-6920}$^{2}$,
Shea DeFour-Remy\orcidlink{0009-0003-7427-9614}$^{1,3}$,
Aleksandar M. Diamond-Stanic$^{4}$,
Zach J. Lewis\orcidlink{0000-0003-0695-4414}$^{1}$,
Britt Lundgren\orcidlink{0000-0002-6463-2483}$^{5}$,
Drake Miller III\orcidlink{0009-0006-2178-1178}$^{1,6}$,
Lizhou Sha\orcidlink{0000-0001-5401-8079}$^{1}$\vspace{0.5em} 
}
\affiliation{
  $^{1}$Department of Astronomy, University of Wisconsin-–Madison, Madison, WI 53706, USA\\
  $^{2}$University of California Observatories, University of California, Santa Cruz, 1156 High Street, Santa Cruz, CA 95064, USA\\
  $^{3}$Steward Observatory, University of Arizona, Tucson, AZ 85719, USA\\
  $^{4}$Department of Physics and Astronomy, Bates College, Lewiston, ME 04240, USA\\
  $^{5}$Department of Physics and Astronomy, University of North Carolina Asheville, Asheville, NC 28804, USA\\
  $^{6}$Department of Astrophysical and Planetary Sciences, University of Colorado Boulder, Boulder, CO 80309, USA\\
  }
\thanks{Corresponding author: Owen S. Matthews Acu\~{n}a:\\ \href{matthewsacun@wisc.edu}{matthewsacun@wisc.edu}}
\keywords{galaxies: Emission Line Catalog, Absorption Index Catalog}

\begin{abstract}
The Sloan Digital Sky Survey IV DR17 Extended Baryon Oscillation Spectroscopic Survey (eBOSS) consists of 2,233,939 high‑quality optical galaxy spectra obtained through 2" fibers, providing a rich spectroscopic resource for studying galaxy evolution across a broad redshift range. 
eBOSS was designed primarily for large‑scale structure and BAO measurements and, as such, focused on galaxy redshifts, leaving much of the information contained in the spectra unexplored. 
In addition to the trove of spectra, the large number of repeat observations (\ndupes~ duplicate spectra) enables evaluation of the survey's spectrophotometric quality.
To unlock this potential, we introduce the eBOSS Data Analysis Pipeline (eBOSS‑DAP), adapted from the MaNGA‑DAP, which delivers measurements of emission‑line fluxes and equivalent widths, stellar and gas kinematics, continuum spectral indices, and stellar population fits.
Using the eBOSS-DAP, we successfully analyze \npassed\ high-quality galaxy spectra below a redshift of $z < 1.12$ to produce an extensive spectroscopic catalog for the eBOSS galaxy sample. 
We characterize the calibration performance, quantify the reliability of the derived measurements, and release a suite of data products that fully exploit the power of the eBOSS dataset. 
These catalogs open the door to a new generation of studies in galaxy evolution and cosmology.
\end{abstract}

\section{Introduction} \label{sec:intro} 
The era of publicly available big data, such as the first Sloan Digital Sky Survey data release \citep[SDSS-I;][]{York:2000}, has demonstrated that innovative applications of surveys beyond their intended scope can lead to significant advancements in astronomy.
Initially designed to study large-scale structures, SDSS-I was quickly used to further the field of galaxy evolution studies following the release of the MPA-JHU catalog of galaxy emission lines, star formation rates, and stellar masses \citep{Brinchmann2004}.

 The MPA-JHU catalog enabled one of the first analyses of the star formation main sequence at $z\sim0.1$ \citep{Brinchmann2004}. These data were also used to study the stellar mass--metallicity relation \citep{mass-metal_1, mass-metal_2, mass-metal_3}, the fundamental metallicity relation \citep{FMR1,Salim2014}, the properties of Type II Active Galactic Nuclei 
\citep[AGN;][]{Kauffmann2004, TypeIIAGN_2, TypeIIAGN_3}, and the impact of galactic environment on galaxy evolution \citep{Weinmann2006,env2, Peng2010}.

The SDSS-IV Extended Baryon Oscillation Spectroscopic Survey \citep[eBOSS;][]{Dawson:2016}, a component of Sloan Digital Sky Survey IV \citep[SDSS IV;][]{SDSS_IV}, is the successor of SDSS I-III. 
Intending to map the Universe's structure by studying galaxies, quasi-stellar objects (QSOs), and the Lyman-alpha forest, eBOSS sports a much broader and higher redshift range than SDSS-I. 
eBOSS produced about 4 million spectra, providing a rich resource for studying galaxy and QSO properties at various redshifts. 
However, unlike the SDSS-I catalog, eBOSS has only been used for a few galaxy evolution studies to date \citep[c.f.,][]{Zhu2015, Lan2018, Huang2019, Anand2021}.
One reason is the lack of a high-quality and well-documented emission line catalog. 
The current eBOSS emission line catalog ({\tt spZline} files) provides only 18 emission lines and has relatively minimal documentation of the critical facets of the stellar continuum and emission line fitting routines \citep{Bolton:2012}. Additionally, this catalog used empirical rather than theoretical stellar population templates (See \S\ref{subsubsec:C3K}) and Principal Component Analysis to fit the stellar continuum rather than stellar population modeling.

There are three primary goals in this work: 1) to provide an overview of the properties of the eBOSS galaxy sample; 2) to characterize the spectroscopic calibration of the survey; and 3) to measure and produce a catalog of emission lines, spectral indices, and stellar kinematics.  eBOSS is a combination of at least seven different individual surveys, thus giving the eBOSS galaxy sample unique and complex characteristics beyond those of other comparable surveys  \citep{Dawson:2013}. To analyze the survey calibration, we make use of a unique facet of the eBOSS sample: duplicate spectra. eBOSS has a remarkable quantity of repeat observations of single objects, allowing us to assess the precision of the spectrophotometic calibration.
Our final goal, to provide a catalog of emission lines, spectral indices, and stellar kinematics, is accomplished through the creation of the eBOSS - Data Analysis Pipeline (eBOSS-DAP).  The eBOSS-DAP was built from the well-vetted algorithms developed for the 
Mapping Nearby Galaxies at Apache Point Observatory - Data Analysis Pipeline  \citep[MaNGA-DAP;][]{MaNGA-DAP, Belfiore_MaNGA}.

An outline of the paper is as follows: 
In \hyperref[sec:Data]{Section 2}, we discuss the properties of the eBOSS galaxy sample and we quantify the survey's spectrophotometric quality. 
Then, in \hyperref[sec:methods]{Section 3}, we describe how we have adapted the MaNGA-DAP to work with eBOSS data and we outline eBOSS-DAP workflow. 
Next, in \hyperref[sec:results]{Section 4}, we discuss the eBOSS-DAP's results and provide some figures that demonstrate the scientific potential of the data. 
In \hyperref[sec:DUC]{Section 5}, we include a summary table containing recommendations regarding best practices when using the data from the eBOSS-DAP catalog.
Finally, we provide a brief summary in \hyperref[sec:summary]{Section 6}.

The data products detailed in the paper are publicly available through the NOIRLab \href{https://datalab.noirlab.edu/}{Astro Data Lab} \citep{noirlab1, noirlab2}.
The source code, spectral template models, and customized emission line lists are available on GitHub.
Details of the data/code access are described in \hyperref[sec:data_availability]{Data Availability}.

\section{Data} \label{sec:Data} 

\subsection{SDSS-IV Data} \label{subsec:SDSS-IV_Data} 
The galaxy spectra analyzed in this work are drawn from the Sloan Digital Sky Survey \citep{York:2000} Data Release 17 \citep[SDSS DR17;][]{Abdurrouf2022}. 
SDSS-I obtained imaging of $\sim1/4$ of the sky in the $ugriz$ filters \citep{Fukugita:1996} with a drift scan imaging camera \citep{Gunn1998} mounted in the SDSS 2.5-m telescope \citep{Gunn:2006} at Apache Point Observatory. 
The $ugriz$ photometry \citep{Lupton:2001} was used to target a variety of stars, galaxies, and quasars for follow-up spectroscopy \citep{Stoughton2002}.

In this work, we focus on the roughly 2 million spectra obtained as part of the SDSS-III Baryon Oscillations Spectroscopic Survey \citep[BOSS;][]{Eisenstein2011} and the SDSS-IV Extended Baryon Oscillation Spectroscopic Survey \citep[eBOSS;][]{Dawson:2016}. 
The goal of BOSS and eBOSS \footnote{Hereafter, we will reference the sum of both of these surveys as eBOSS.} was to map the 3D structure of the universe to measure baryon acoustic oscillations at different redshifts \citep[c.f.,][]{Eisenstein2005}. 
This was accomplished by targeting three different types of objects for spectroscopy based on their photometric colors: Luminous Red Galaxies (LRGs) in several different redshift slices \citep{Eisenstein2001, Reid2016, Prakash2016}; 
Emission Line Galaxies (ELGs) at $z\sim0.6 - 1.1$ \citep{Raichoor2017, Raichoor2021}, and Quasars at $0.9 < z < 2.2$ \citep{Myers2015}. SDSS-IV also conducted two smaller programs, SPIDERS \citep{Clerc2016, Dwelly2017} and TDSS \citep{Morganson2015, MacLeod2018} which target X-ray sources and variable objects respectively. 

\begin{figure}[!ht]
\plotone{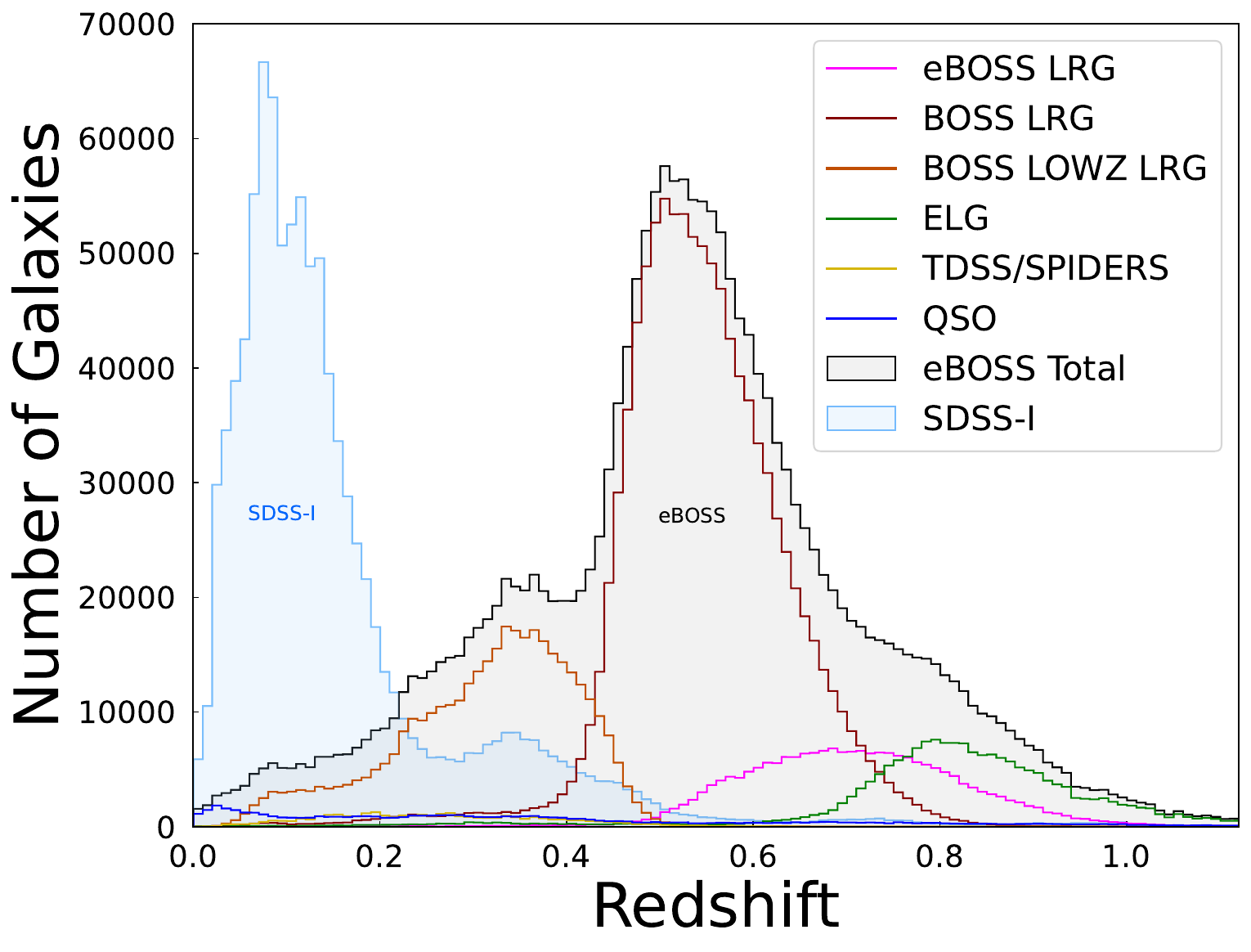}
\caption{
Redshift distribution of the SDSS-I (light blue) and eBOSS (gray) galaxy samples. The eBOSS galaxies are further divided based on how they were targeted for spectroscopy. LRGs are Luminous Red Galaxies; these galaxies are further broken down by the survey from which they originate. ELGs are Emission Line Galaxies and take up a large portion of our sample at the highest redshifts. TDSS/SPIDERS are targeted based on variable photometry and X-ray data. Finally, the histogram labeled QSO represents objects that were initially targeted as quasars but, upon further analysis, were revealed to be compact blue galaxies. In total, the eBOSS galaxy sample peaks at \textit{z} = 0.55 and contains \nsample~ spectra. For comparison, the SDSS-I galaxy sample contains 987,729 spectra with two peaks at \textit{z} = 0.1 and \textit{z} = 0.35.
\label{fig:redshift_hist}}
\end{figure}

\begin{figure}[!ht]
\plotone{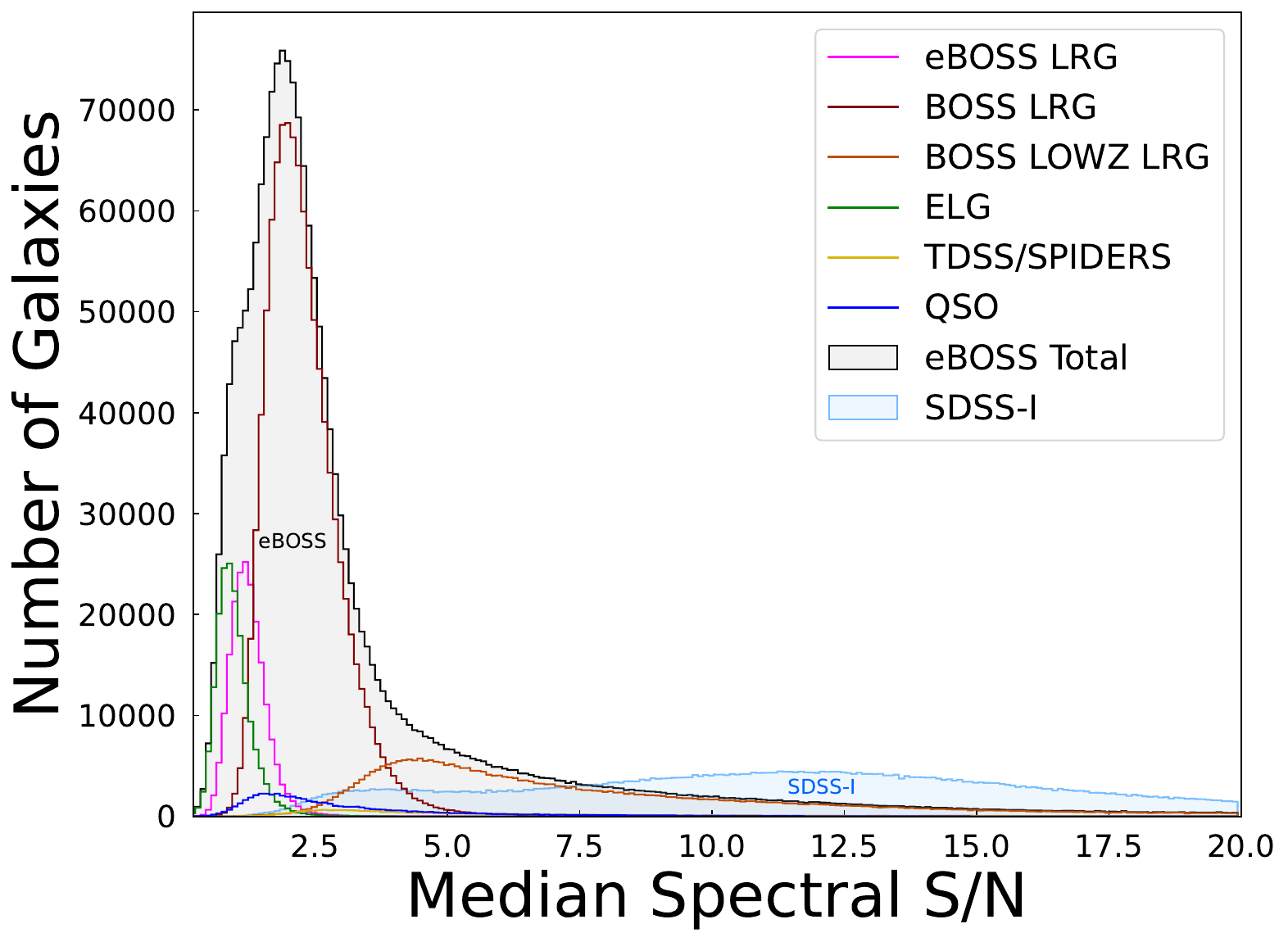}
\caption{
Median signal-to-noise (S/N) per pixel distribution of the SDSS-I (light blue) and eBOSS (gray) galaxy spectra. The eBOSS galaxies are further divided based on how they were targeted for spectroscopy. In total, the eBOSS galaxy sample peaks at S/N = 1.88. For comparison, the SDSS-I galaxy sample has two peaks at S/N = 3.9 and S/N = 12.1.
\label{fig:snr_hist}}
\end{figure}

Targets were observed through 2\arcsec\ diameter optical fibers positioned at the SDSS telescope's focal plane using custom-drilled aluminum `plates' that span 3 degrees on the sky. 
The plates enable one thousand targets (900 objects, 80 sky, 20 standard stars) to be observed simultaneously. 
 The hole positions for fibers placed on galaxy and standard star targets were centered to account for predicted atmospheric differential refraction \citep{Filippenko1982}. 
 For the majority of the survey, fibers were centered to maximize throughput for 5400 \AA\ light. However, fibers on plates targeting ELGs were centered on the predicted position for 7500~\AA\ light to maximize throughput for redshifted [OII]~$\lambda3727$. In addition, some quasar targets were centered to maximize throughput at 4000\AA\ for Lyman-alpha forest studies.

 The light from the fibers was sent to the upgraded SDSS-BOSS spectrographs \citep{Smee:2013, Dawson:2013}, which have wavelength coverage from 3610 -- 10140~\AA\ at a spectral resolution of $R = 1560 - 2650$. Spectra were reduced by the {\tt idlspec2D} pipeline version v5\_13\_2\footnote{\url{https://svn.sdss.org/public/repo/eboss/idlspec2d/tags/v5_13_2/}} \citep{Dawson:2016, Morrison2026}.
The {\tt spectro1d} pipeline classified each eBOSS spectrum as a star, galaxy, or quasar and measured its redshift \citep{Bolton:2012}.

\subsection{Sample Selection} \label{subsec:SampleSelect}

We selected our sample using the SDSS DR17 summary file {\tt spAll-v5\_13\_2}. 
We chose spectra that were classified as galaxies ({\tt CLASS = GALAXY}), resulting in 2,233,939 eBOSS galaxy spectra. The {\tt GALAXY} class includes Type~II AGN, but not Type-I (broadline) AGN \citep{Bolton:2012}. 
We then selected only the galaxy spectra which met the SDSS survey quality standards ({\tt PLATEQUALITY = GOOD}) and applied a redshift cut of $0.0005 < z < 1.12$ and required that no redshift warning flags were set ({\tt ZWARNING=0}). 
The lower redshift limit ($\sim150$ km~s$^{-1}$) eliminates some Milky Way stars that are mistakenly classified as galaxies. 
Our upper redshift limit was adopted because our fitting code requires either H$\alpha$ or H$\beta$ to fit the other emission lines (see \S\ref{subsubsec:linelist}) and H$\beta$ redshifts out of the spectra at $z\gtrsim1.12$. 
With these quality and redshift cuts, we trim the 2,233,939 eBOSS galaxy spectra to a sample of \nsample. Of these spectra, \ndupes~ are duplicates (e.g., only 1,704,313 unique galaxies were observed) with the max number of duplicates being 64 for one object .

In Figure~\ref{fig:redshift_hist}, we show the redshift distribution of our galaxy spectra with the sample divided and color-coded by target class (e.g., e/BOSS LRG, ELG, QSO, etc.).  
We defined the target classes to encompass several broad categories by which galaxies were photometrically selected for follow-up spectroscopy (see Appendix Table~\ref{table:target_classes}).
Details of the target selection for each object are preserved in the SDSS {\tt spAll} file and described in \citet{Dawson:2013, Dawson:2016} and references therein. 
The objects labeled `QSO' were originally targeted as quasars but were spectroscopically classified as galaxies.

 In Figure~\ref{fig:snr_hist}, we show the median signal-to-noise ratio (S/N) distribution of the eBOSS galaxy spectra in our sample divided and color-coded by target class. Both figures also show the distribution for the widely used SDSS-I sample. 
 By comparing these two, we see that eBOSS is roughly two times larger with a much higher and wider redshift range while also being, on average, lower S/N than SDSS-I.

\begin{figure}[!ht]
\plotone{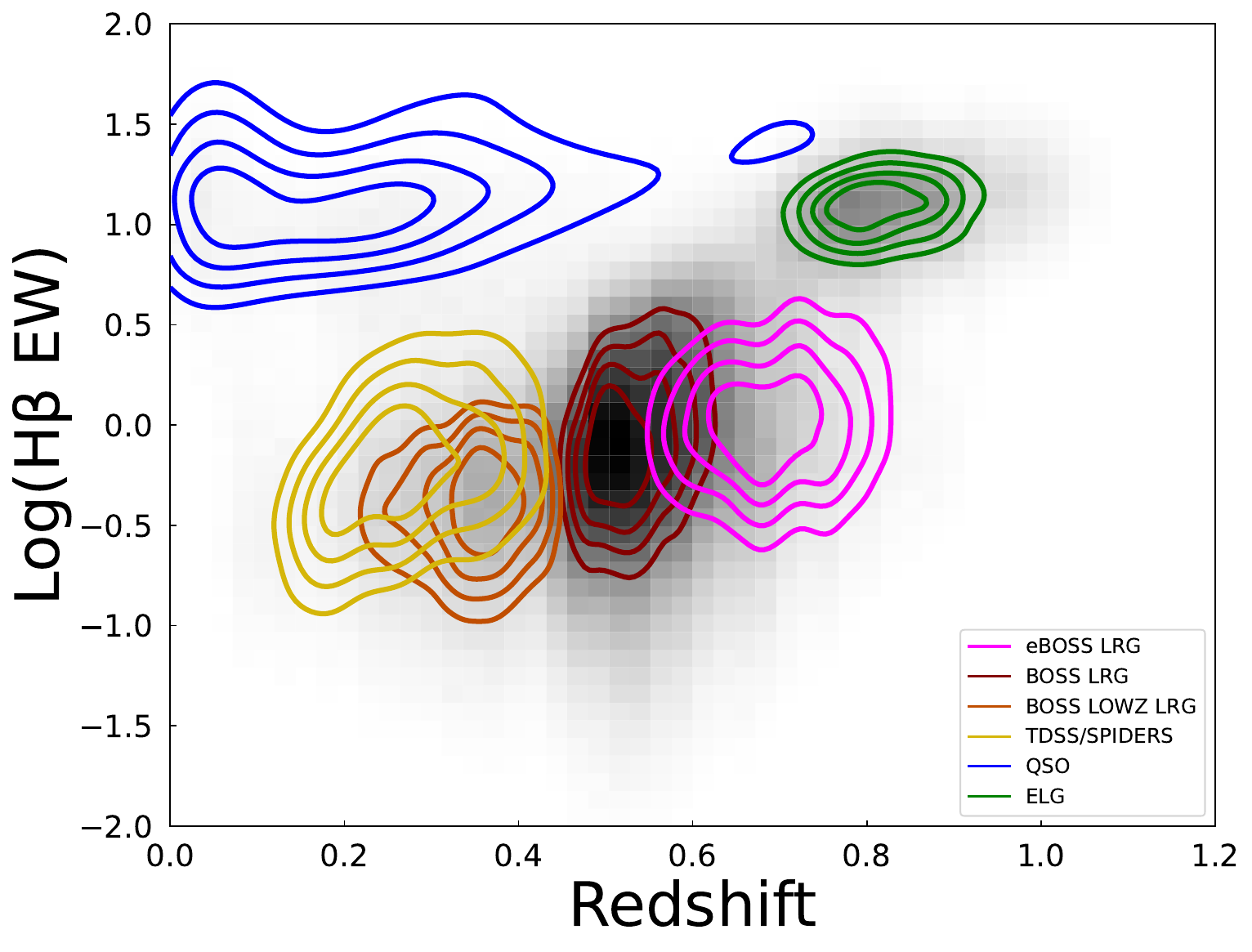}
\caption{ H$\beta$ emission line Equivalent Width vs.\ redshift for all six target categories and a 2D histogram of the full sample. The target contours begin at 50\% of the data.
\label{fig:HB_EW}}
\end{figure}

To illustrate some of the differences between the various eBOSS target classes, we plot H$\beta$ emission line Equivalent Width (EW) vs.\ redshift in Figure \ref{fig:HB_EW}. ELGS and QSOs have much higher EWs than the LRG portion of the sample, as expected based on their color-selection.
Additionally, there is a small upward trend with redshift in the LRG sample where the higher $z$ galaxies show marginally higher H$\beta$ EWs.

\subsection{Aperture Bias}\label{subsec:appature_bias}
eBOSS observed galaxies through 2\arcsec\ diameter fibers and calibrated the spectra such that their fluxes would be correct for targets that were point sources. However, the majority of eBOSS galaxies are spatially extended, leading to some aperture bias and loss of light. 

\begin{figure}[ht]
\plotone{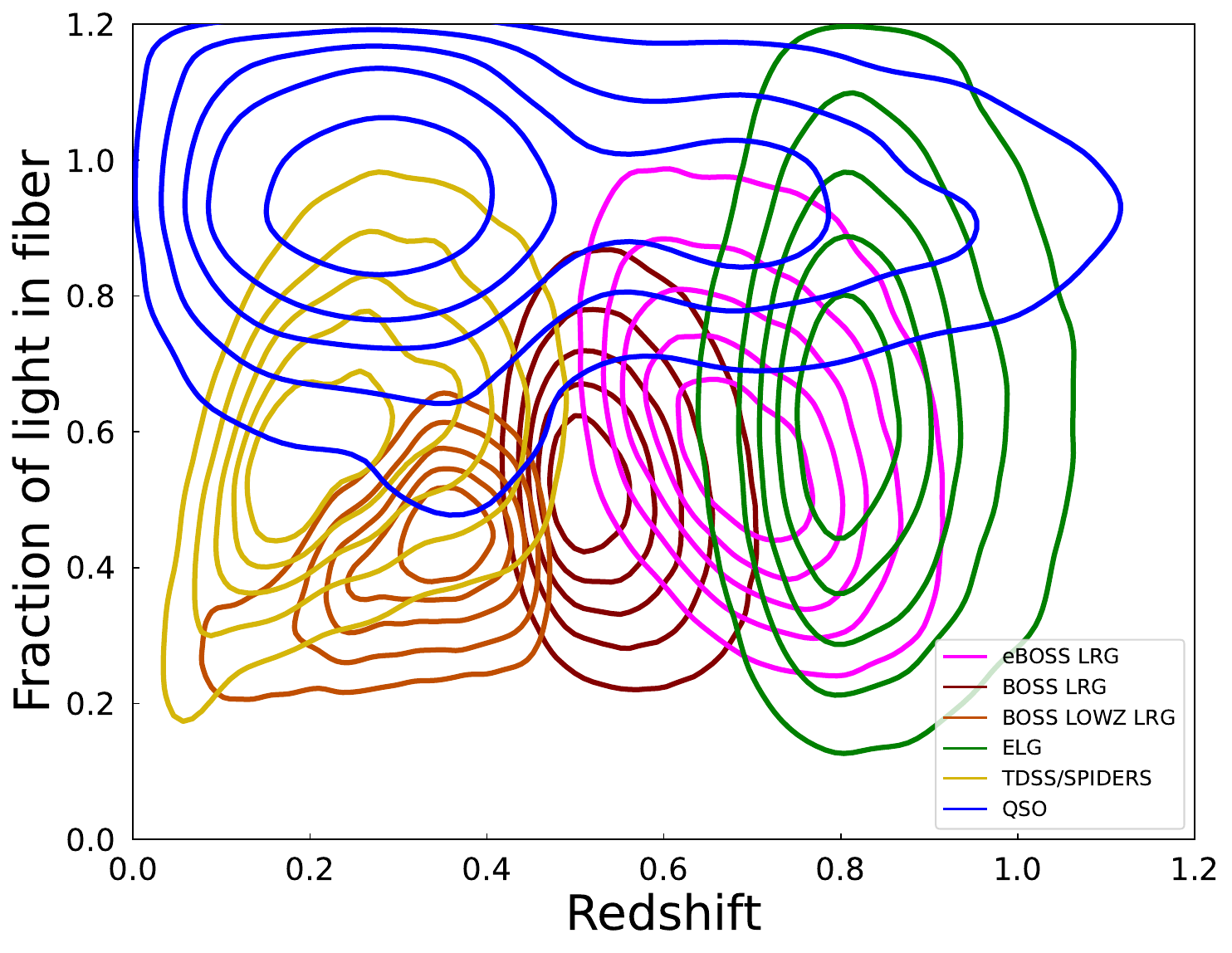}
\caption{A two-dimensional kernel density estimate plot of the fraction of galaxy light encompassed by the SDSS fiber for the different target categories as a function of redshift. The fraction of light in the fiber is computed from the ratio of the SDSS $i$-band flux measured from the spectrum over that measured from photometry. This shows that different targeting categories are more likely to be fully captured by the fiber than others. The outer contours enclose 90\% of the sample.
\label{fig:infiber}}
\end{figure}

To quantify this effect, we computed the fraction of the total $i$-band galaxy light captured by the fiber by comparing synthetic photometry computed from the spectra (found in the spAll field {\tt spectroflux}) to the observed photometry ({\tt modelflux}). The $i$-band ($\lambda_{eff} = 7480$~\AA) was selected because it provides the highest S/N measurement given the red colors of most of the eBOSS galaxies.  We explore trends between the fraction of light in the fiber, redshift, and galaxy target class in Figure \ref{fig:infiber}. The fraction of light in the fiber ranges from $\sim0.2 - 1.1$ with a median of 0.51. This is larger than the typical fraction of light in the fiber for SDSS-I galaxies ($\sim0.3)$ since SDSS-I primarily targeted galaxies at lower redshifts ($z\lesssim 0.25$; see Fig. \ref{fig:redshift_hist}.) 

Galaxies that were targeted as QSOs (blue contours in Fig.~\ref{fig:infiber}) have large fractions of light in the fiber since many of the quasar target selection algorithms required them to be point sources. Notably, for most other target classes, the fraction of light in the fiber increases as a function of redshift due to the smaller apparent diameter of the galaxies. Roughly 5.3\% of objects have a light fraction greater than 1, implying that the spectrum is brighter than expected from the imaging data. This comes about for three reasons: i) flux calibration errors in the spectra (see \S\ref{subsec:Spectra_qual}); ii) errors in the photometry, which is near the magnitude limit of the survey; and iii) errors in photometric deblending which cause large galaxies to be shredded into smaller pieces, causing the fluxes from the imaging data to be underestimated (i.e., an H~II region in a nearby galaxy might be categorized as a compact galaxy; see Fig.\ref{fig:cutout}). 

In summary, for a typical eBOSS galaxy, the spectrum captures 40 - 75\% of the total galaxy light. This implies that the bias due to the finite fiber aperture is not extreme; the spectra are closer to global galaxy spectra than to nuclear spectra. However, it will be essential to correct for the missing light when computing absolute quantities such as H$\alpha$-based star formation rates. 

\subsection{Spectrophotometric Quality} \label{subsec:Spectra_qual}

A major source of spectrophotometric calibration error for eBOSS --- and most ground-based spectroscopic surveys --- is atmospheric dispersion. Atmospheric dispersion occurs when light from celestial objects passes through the Earth's atmosphere at an angle, causing the light to refract and spread out into its component colors. The resulting image is dispersed in a direction perpendicular to the horizon by an amount that depends on the elevation of the object. At an airmass of 1.5 (elevation of 42$^{\circ}$), light at 4000~\AA\ and 1 $\mu$m are separated by about 1.7" \citep{Filippenko1982}. This can affect the accuracy of spectroscopic measurements made through small apertures that can miss a fraction of the highly dispersed blue or red light. 

The eBOSS data are spectrophotometrically calibrated by the SDSS {\tt idlspec2D} pipeline as described in \citet{ssds_photo_calib} and \citet{Dawson:2016}. 
In brief, on any given plate, 20 color-selected `standard stars' are observed. These A-to-G-type stars are not classical standards like those of \citet{Oke1990} because their calibrated spectrum is unknown. Flux calibration is achieved by matching each star to a grid of stellar atmosphere models that have been rescaled to match the observed $r$-band PSF magnitude of the standard star. The use of multiple standards per plate (which are averaged with rejection) mitigates the effect of the occasional problematic star. 
The standard and object spectra experience approximately the same atmospheric dispersion, so, blue and red light losses are --- in principle --- automatically corrected by the flux calibration routine. 

\begin{figure}[!ht]
\plotone{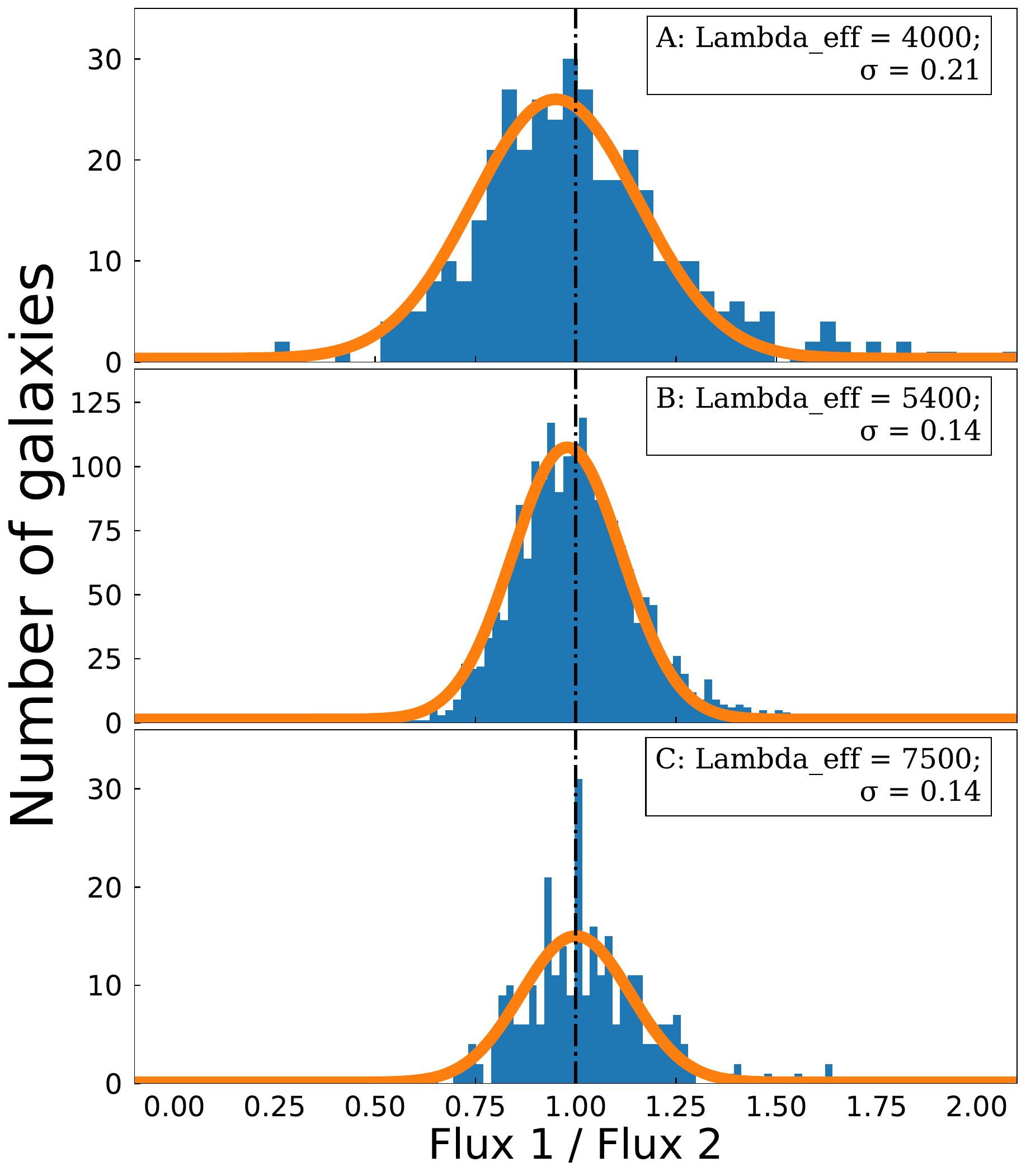}
\caption{
An estimate of the relative flux calibration error of the eBOSS spectra based on comparisons of repeat spectra of the same galaxies observed at two or more different epochs. Here we compare the sum of the flux in a 200~\AA~ window centered at the wavelength, {\tt LAMBDA\_EFF}, that the spectrophotometric calibration is optimized for. For the majority of galaxies, this is 5400 \AA~ (\textbf{Panel B}), however most ELG targets have {\tt LAMBDA\_EFF}=7500~\AA~ (\textbf{Panel C}) and most QSO targets classified as galaxies have {\tt LAMBDA\_EFF} = 4000~\AA~ (\textbf{Panel A}). The orange lines show Gaussian fits; the 1-$\sigma$ spread of the data is listed in each plot. This represents the minimum uncertainty of a line flux measurement due to spectrophotometric calibration errors. These errors can be larger at wavelengths that are far from {\tt LAMBDA\_EFF} as shown in Figure~\ref{fig:Bowtie}.
\label{fig:flux_cal_hist}}
\end{figure}

\begin{figure*}[!ht]
\centering
\includegraphics[width=0.8\textwidth]{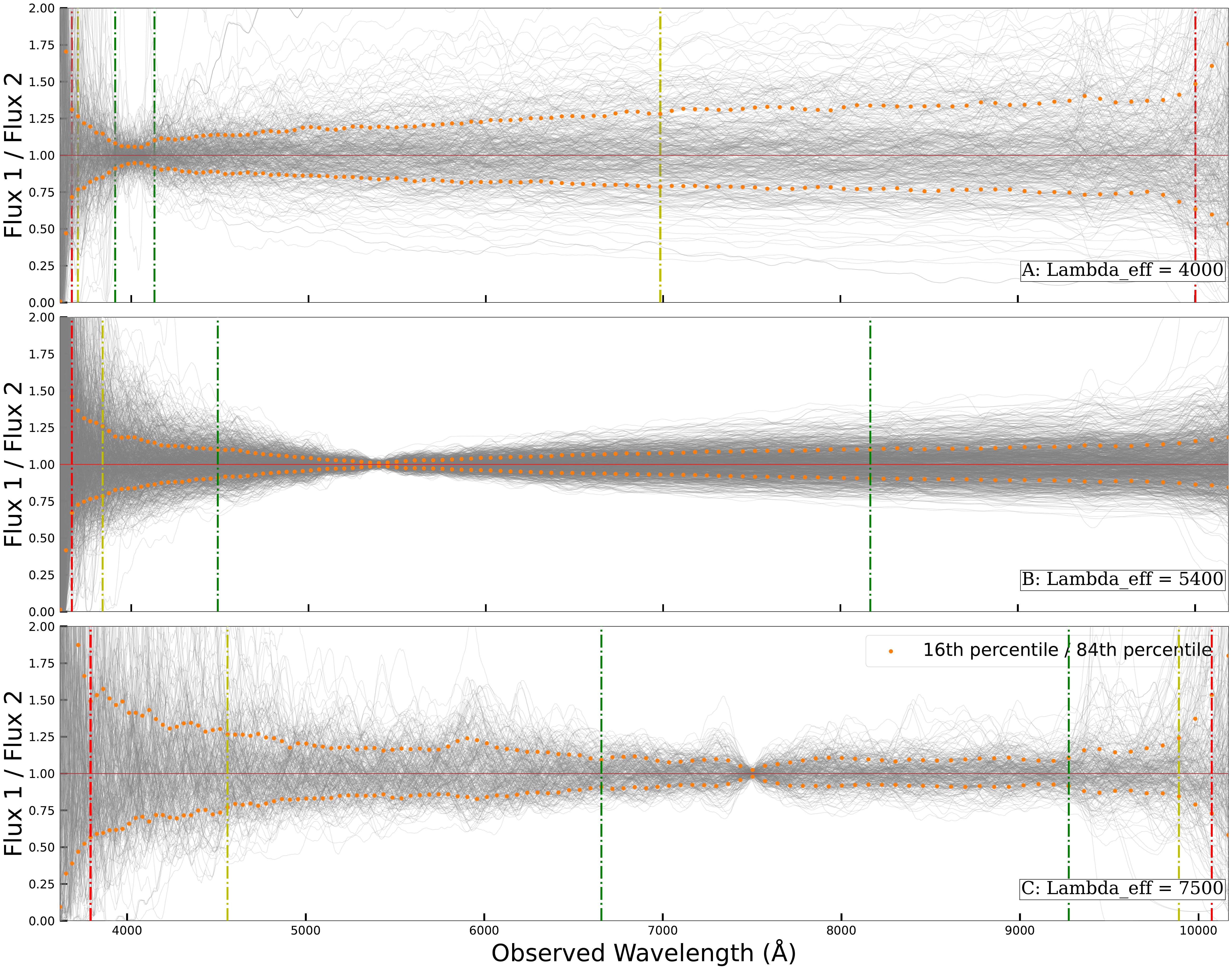}
\caption{
An analysis of the typical error in the spectral shape made by dividing pairs of repeat spectra after normalizing them at {\tt LAMBDA\_EFF}, the wavelength the flux calibration has been optimized for. We show galaxies with {\tt LAMBDA\_EFF} = 4000~\AA, 5400~\AA, and 7500~\AA\ in panels A, B, and C, respectively. Each gray line represents a pair of spectra that were taken of the same galaxy in different epochs. Spectra in each pair were normalized at their respective {\tt LAMBDA\_EFF} and smoothed individually before computing their ratio. Small-scale wiggles in the gray lines are primarily due to the low S/N per pixel of the spectra. The $1\sigma$ spread of the data in 40-pixel-sized bins is shown as the orange dotted lines. Green, yellow, and red vertical lines have been added, marking where the percent error is 10\%, 25\%, and 50\%, respectively. For most galaxies, spectrophotometric calibration errors will lead to line flux ratio errors $<25$\%. However, caution is advised when using emission lines observed near the blue and red ends of the spectrum. For access to summary tables for this figure see \hyperref[sec:data_availability]{Data Availability}
\label{fig:Bowtie}}
\end{figure*}

\begin{figure}[!ht]
\plotone{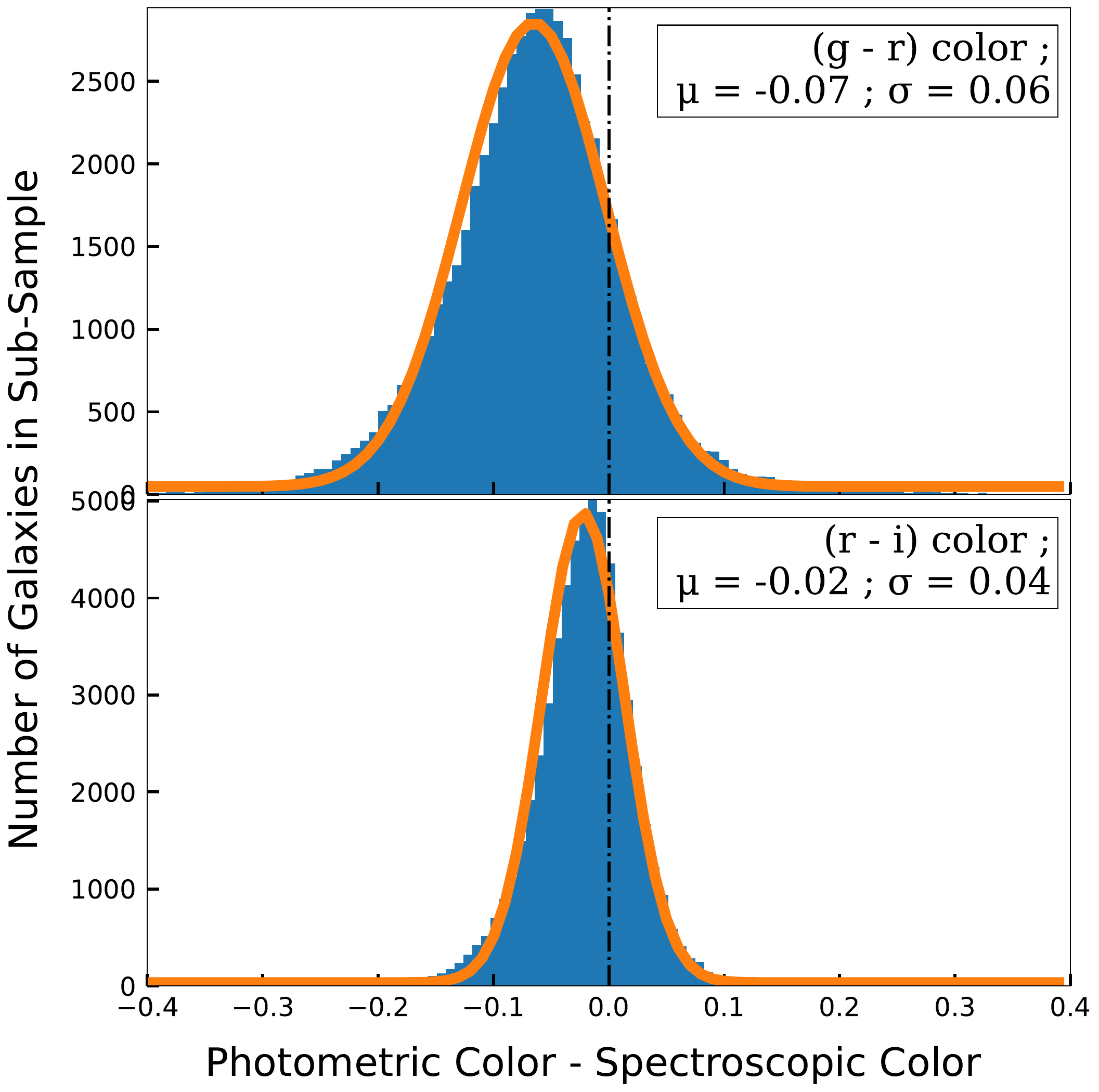}
\caption{Comparisons of galaxy colors synthesized from the spectra with photometric colors for galaxies with $r< 17.7$~mag and spectra with {\tt LAMBDA\_EFF} = 5400~\AA. This comparison measures the absolute calibration quality in eBOSS spectroscopy. Relatively bright galaxies are chosen for this comparison to minimize random error.  In the \textbf{Top Panel}, ($g - r$) colors are compared, whereas in the \textbf{Bottom Panel}, ($r - i$) colors are compared. Lower $\sigma$ values and $\mu$ values closer to zero indicate better absolute calibration.
\label{fig:abs_cal_hist}}
\end{figure}

In practice, this calibration approach works well: in the SDSS-I survey a flux calibration accuracy of $\sim5$\% was achieved at 4000~\AA\footnote{\url{ https://classic.sdss.org/dr7/products/spectra/spectrophotometry.php}}. The main failure mode results from the miscentering of objects in the fibers, which can occur due to the deformation of the aluminum plug plates, which are warped to match the curved focal plane of the telescope \citep{York:2000}. For example, when the hole for an object fiber is centered too high (and the majority of the standards are centered correctly in their fibers), more blue light and less red light is captured by the object fiber than for the standards, leading to an error in the relative flux calibration. 
The SDSS-I and II surveys used 3" diameter fibers, which mitigated the fiber centering problems. For SDSS-III/IV (BOSS and eBOSS), the fibers were reduced to 2\arcsec\ in diameter to improve sky subtraction for the distant galaxies and quasars targeted. To evaluate the quality of the flux calibration in SDSS-IV/eBOSS, we take advantage of the large number of repeat galaxy spectra obtained. 

In eBOSS, fibers were positioned on the plates to account for the expected atmospheric differential refraction, with different spectroscopic targets centered at different wavelengths in an attempt to mitigate light losses in particular parts of the spectrum. Most objects were centered on the position of 5400~\AA\ light (1,716,144 spectra). 
However, objects targeted as QSOs were centered around 4000~\AA~(23,870 spectra), while objects targeted as ELGs were centered at 7500~\AA~(161,820 spectra). 
This is reflected in the parameter {\tt LAMBDA\_EFF} found in the {\tt spAll-v5\_13\_2.fits} file. To analyze the relative and absolute calibration for each unique {\tt LAMBDA\_EFF}, we selected pairs of galaxy spectra after applying a cut on median S/N per pixel. 
For {\tt LAMBDA\_EFF} = 4000 we used a S/N $>=$ 3 cut, resulting in 386 unique pairs.
For {\tt LAMBDA\_EFF} = 5400 we used a S/N $>=$ 6 cut, resulting in 1094 unique pairs.
Finally, for {\tt LAMBDA\_EFF} = 7500 we used a S/N $>=$ 1.5 cut, resulting in 279 unique pairs.
The S/N cuts were implemented to reduce problematic negative pixels; the S/N thresholds are different for each {\tt LAMBDA\_EFF} to enable a reasonable number of pairs in each bin. 
In principle, these S/N cuts do not affect our analysis since the spectrophotometric quality is independent of S/N.

In Figure~\ref{fig:flux_cal_hist}, we show the ratios of a 200~\AA~flux window centered at the {\tt LAMBDA\_EFF} in pairs of galaxy spectra. 
The pairs represent spectra of the same galaxy obtained at different epochs. 
In some cases, the same plate was observed on multiple dates; in others, the galaxies were targeted on two different plates with different sets of standards. 
The ratio of the galaxy fluxes at the {\tt LAMBDA\_EFF} provides a valuable assessment of the repeatability of the absolute flux calibration at this wavelength. A Gaussian fit to the distribution has a width of $\sigma = 0.21, 0.14, 0.14$ for {\tt LAMBDA\_EFF} = 4000, 5400, and 7500, respectively.
Thus, for most eBOSS galaxies, the flux calibration is accurate to about 15-20\% at wavelengths near {\tt LAMBDA\_EFF}.

In Figure~\ref{fig:Bowtie}, we explore the relative flux calibration of the spectra as a function of observed-frame wavelength for each {\tt LAMBDA\_EFF}. This is the error on the spectral \emph{shape} that would impact ratios of emission lines widely separated in wavelength. Galaxy spectra are normalized at their {\tt LAMBDA\_EFF}, median smoothed by 100 pixels, followed by a 40-pixel average smooth, and then divided. The orange dotted line shows the 1-$\sigma$ width of the distribution of all pairs in bins of 40 pixels. 

According to Fig.~\ref{fig:Bowtie}, For {\tt LAMBDA\_EFF} = 4000 the relative flux calibration error is less than 10\% between 3909 -- 4208~\AA, and less than 25\% between 3733 -- 7449~\AA. The error is over 50\% at wavelengths less than 3665 or greater than 10002 \AA. 
For {\tt LAMBDA\_EFF} = 5400, the relative flux calibration error is less than 10\% between 4488 -- 8397~\AA. The error is less than 25\% blueward 3838~\AA~ and over 50\% at wavelengths less than 3665~\AA. 
Finally, for {\tt LAMBDA\_EFF} = 7500, the relative flux calibration error is less than 10\% between 6595 -- 9273~\AA, and less than 25\% between 4562 -- 9890~\AA. The error is over 50\% at wavelengths less than 3830 or greater than 10074 \AA. 
These statistics suggest caution in using emission lines observed near the edge of the spectrum. 

In Figure~\ref{fig:abs_cal_hist}, we show the difference in color between the photometry and spectroscopy for galaxies with {\tt LAMBDA\_EFF} = 5400~\AA\ and an $r$-band AB magnitude of 17.7 or smaller. The colors of the spectra have been synthesized using the SDSS filter curves \citep{Fukugita:1996}.
We compare synthetic photometry computed from the spectra (found in the spAll field {\tt spectroflux}) to the observed photometry ({\tt modelflux}). For this sample, $\sim$27.3\% of the galaxy's light falls in the 2'' fiber aperture, and thus we do not expect galaxy color gradients to impact our comparison significantly. At $r<17.7$, the photometric errors on the $g-r$ and $r-i$ colors should be minor, and, as such, any deviations between the colors of the photometry and the spectroscopy likely indicate errors in the spectrophotometric calibration. 
Whereas Figure~\ref{fig:Bowtie} demonstrates the consistency of the calibration, i.e., the precision, Figure~\ref{fig:abs_cal_hist} shows us the absolute calibration accuracy. 
In Figure~\ref{fig:abs_cal_hist}, the further left the data lies, the more blue it is relative to the photometry.
The ($g-r$) color has a larger offset ($\mu = -0.07$) and standard deviation ($\sigma = 0.06$) than the ($r-i$) color ($\mu = -0.02$; $\sigma = 0.04$).The difference in $\sigma$ between the two colors is consistent with the larger spectrophotometric errors seen at blue wavelengths in the repeat spectra comparison (Fig.~\ref{fig:Bowtie}). The slight offset in ($g-r$) color ($\mu=-0.07$~mag) implies that there are some minor systematic issues in the calibration, as previously found for SDSS-I ($\mu_{g-r} = 0.02$ mag)\footnote{\url{https://classic.sdss.org/dr6/products/spectra/spectrophotometry.php}}. 
A similar analysis could not be performed on {\tt LAMBDA\_EFF} = 4000 and 7500 spectra, as there are an insufficient number of bright galaxies available for these {\tt LAMBDA\_EFF}.

\subsection{Known issues with the SDSS-IV data}\label{subsec:known_errs}

\begin{figure}[ht]
\plotone{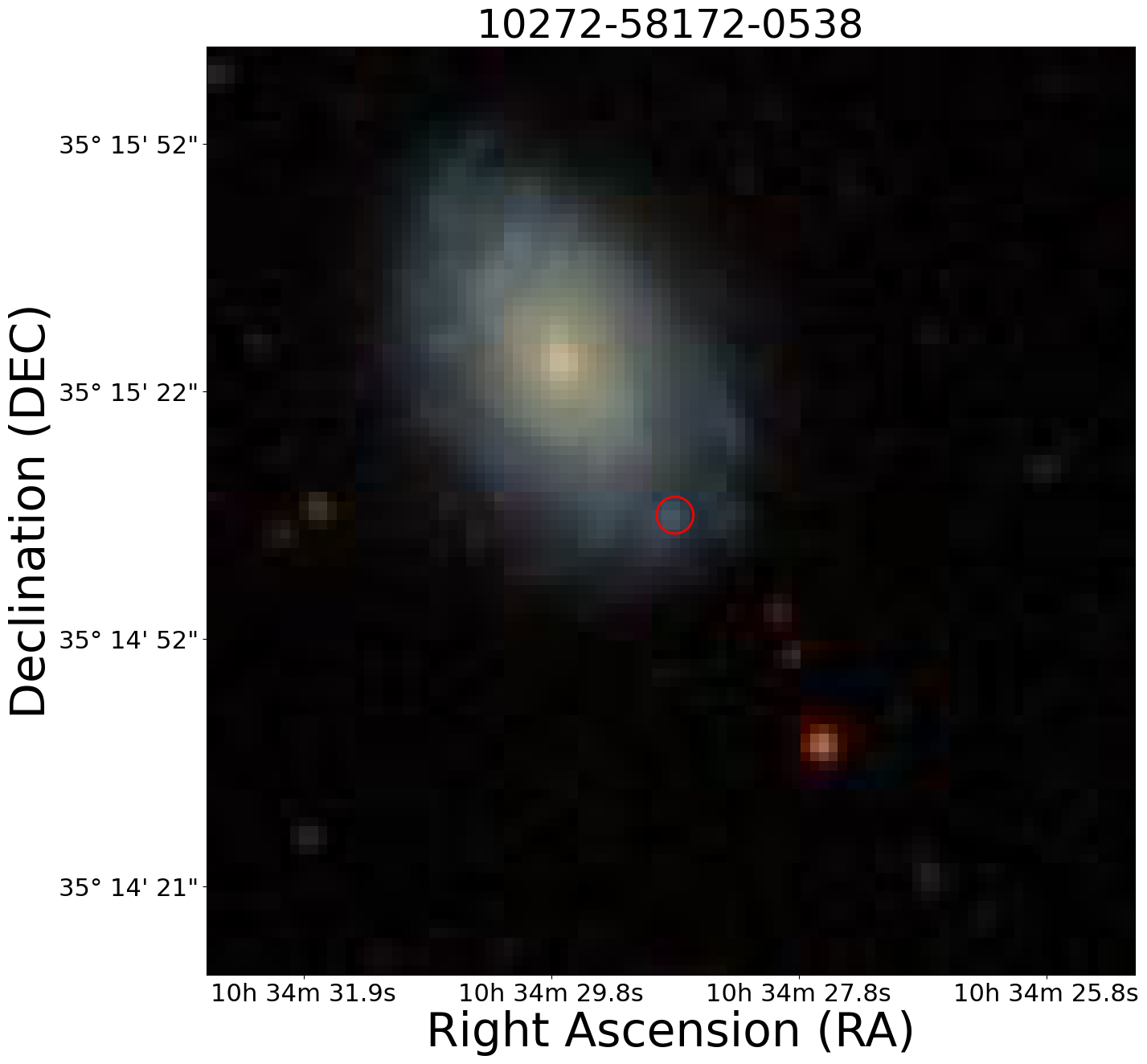}
\caption{ 
An example of a galaxy incorrectly deblended by the SDSS photometric pipeline. The H~II region circled in red was considered a separate photometric object and targeted for spectroscopy as a quasar. (The circle represents the size of the eBOSS fiber.) The spectrum (Plate: 10272; MJD: 58172; Fiber: 0538) is classified as a galaxy and otherwise free of problems. However, quantities derived from the photometry will pertain only to the galaxy sub-region. For example, this object has a photometrically-derived stellar mass of log(M$_{*}$) = 5.2 when it is clearly part of a much larger galaxy.
\label{fig:cutout}}
\end{figure}

There exist two caveats to our data resulting from known issues in SDSS-IV that we did not rectify. The first is that some large nearby galaxies are incorrectly de-blended by the SDSS photometric pipeline, leading to incorrect photometric measurements. An example is shown in Figure \ref{fig:cutout}, where an H~II region has been incorrectly classified as a separate photometric object and selected for spectroscopy as a quasar target. The spectrum was subsequently classified as belonging to a galaxy; however, the galaxy's stellar mass and other properties calculated from the photometry are incorrect. These deblending problems occur frequently enough at $z<0.1$ to produce a substantial group of outliers in the eBOSS stellar mass--metallicity relation (i.e., galaxies with high metallicity and unrealistically low stellar masses; see Fig.~\ref{fig:MZR}). 
The debending problem impacts both SDSS \citep{Lupton:2001} and Legacy Survey \citep{Dey:2019} photometry. (The latter is used for our stellar masses.) Visual inspection of the Legacy Survey data shows that 5\% of star-forming galaxies below $\log(M_*/M_{\odot}) = 7.3$ are incorrectly deblended \citep{Scholte:2024}. 
Unfortunately, objects with deblending issues are not straightforward to identify without visual inspection of the photometric images. A potential approach to finding such sources would be to determine whether an object lies within the Petrosian radius of another photometric object.

\begin{figure*}[!t]
\plotone{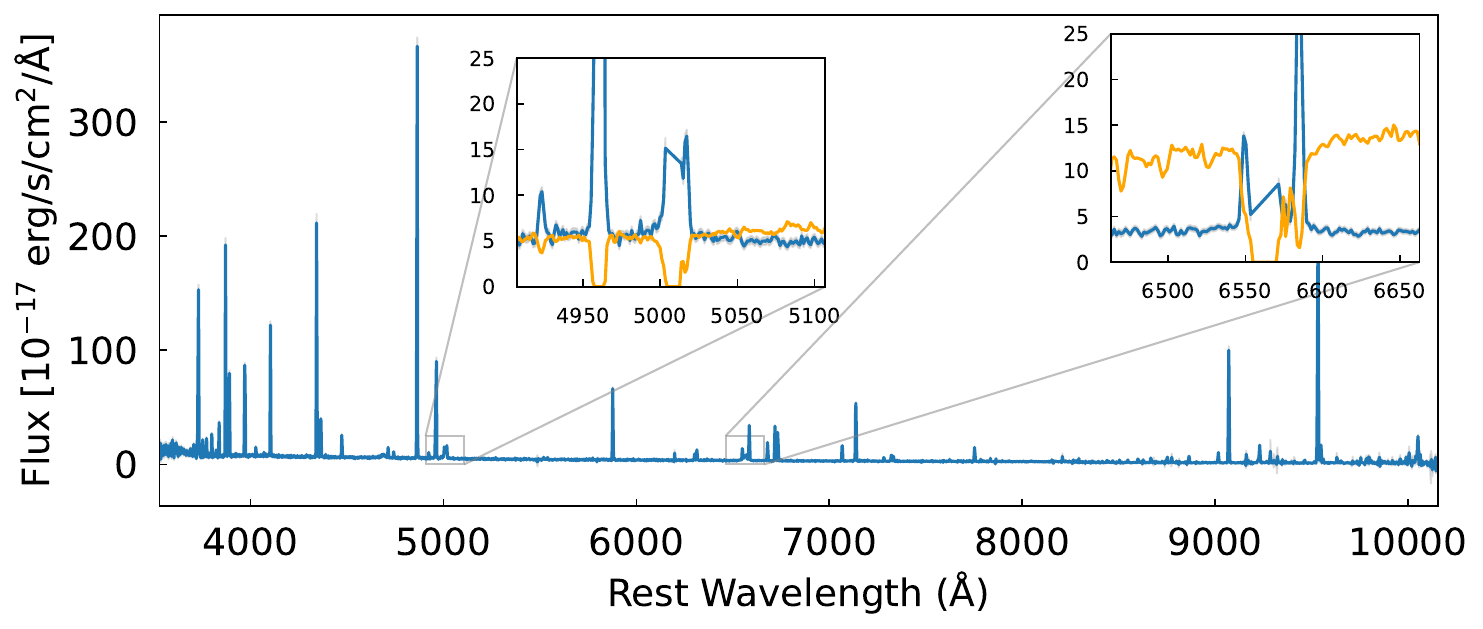}
\caption{ The spectrum (Plate: 11294; MJD: 58451; Fiber: 0164) for a galaxy with clipped emission lines. The spectrum is shown in blue, the error in light gray, and the inverse variance in orange. Two zoomed-in windows at [OIII]~$\lambda$5008 (left) and H$\alpha$ (right) have been added to demonstrate that the flux for these two lines has been masked out, as shown in orange, where the clipped lines coincide with an inverse variance of 0, which is a marker of bad data in SDSS. This effect is present in some galaxies with high EW emission lines and low gas velocity dispersions. See \S\ref{subsec:known_errs}.
\label{fig:cutoff_spec}}
\end{figure*}

Our second caveat applies to galaxies with very high emission line EWs and low gas velocity dispersions (i.e., star-bursting dwarf galaxies). In the eBOSS spectra of these galaxies, the strongest emission lines such as [O III]~5008 and H$\alpha$ are sometimes clipped off. An example is shown in Figure \ref{fig:cutoff_spec}. The missing portions of the emission lines typically have their inverse variance set to zero (i.e., they are flagged as bad data.) However, the lines are sometimes fit by the eBOSS-DAP if only the core of the line profile is flagged as bad; the accuracy of such fits is unclear. 

The problem of clipped emission lines was well known in the SDSS-I; some code changes in DR7 improved, but did not eliminate, the issue \citep[see \S6.3 in ][]{Abazajian2009}. The clipped lines arise from a combination of data reduction problems. One issue can be the saturation of the detector at the position of the brightest lines. Another issue can arise when the {\tt spectro2D} data reduction pipeline combines the 3 - 12 individual spectroscopic exposures that make up the final spectrum. Slight mismatches between the exposures (due to small pointing differences, for example) cause the pixels associated with bright lines to be rejected because they differ by more than the errors allow between exposures. 

The fully calibrated individual spectroscopic exposures are stored in the final FITS spectrum output by SDSS (the ``full", not the ``lite" version). Custom software for combining these exposures could likely recover many of the clipped lines. However, this is beyond the scope of the current work. In some cases, clipped emission lines can be identified by looking for discrepancies between line EWs estimated by Gaussian fits vs. straight integration (see \S\ref{subsubsec:ew_comp}) or by flagging spectra with unusual line ratios (i.e., H$\alpha <$~H$\beta$). However, this method will not identify all cases of clipped lines, and therefore caution is advised in working with very low-mass galaxies with narrow  ($\sigma_{line} < 100$ \kms) high-EW (EW$ > 100$~\AA) emission lines.

\section{Methods} \label{sec:methods} 
We aimed to produce a catalog of emission line fits, spectral indices, and stellar population template weights from the eBOSS galaxy spectra. 
To this end, we developed the eBOSS-DAP, an analysis pipeline built from the Mapping Nearby Galaxies at Apache Point Observatory - Data Analysis Pipeline (MaNGA-DAP) code package \citep{MaNGA-DAP, Belfiore_MaNGA}.
We begin by briefly describing the MaNGA-DAP in \S\ref{subsec:MaNGA-DAP}. In \S\ref{subsec:eBOSS-DAP} and its subsections, we describe the modifications that we made to create the eBOSS-DAP. In \S\ref{subsec:workflow} we provide an overview of the eBOSS-DAP workflow.

\subsection{The MaNGA-DAP} \label{subsec:MaNGA-DAP}
The MaNGA-DAP is a multi-stage spectral fitting software tool designed to process and analyze the data collected by the MaNGA survey \citep{MaNGA_OG_paper}.
It handles the complex data cubes produced by the survey, which contain spatially resolved spectroscopy of nearby galaxies. 
The pipeline performs various tasks, including full-spectral fitting of the galaxy spectra to measure stellar kinematics and absorption-line indices from the stellar continuum and gas kinematics, fluxes, and equivalent widths (EWs) from Gaussian modeling of the emission lines. 
Within the MaNGA-DAP, stellar continuum fitting is done with the Python implementation of the Penalized PiXel-Fitting method \citep[{\tt pPXF};][]{ppxf} using a mix of template libraries and a multiplicative polynomial.
The code follows a multistage process. 
First, the stellar kinematics are fit with the emission lines masked. Second, the stellar continuum and emission lines are fit simultaneously, with the stellar kinematics fixed to the result of the first step. Third, the best-fit emission-line model is subtracted from the galaxy spectra before measuring the absorption-line indices. 
Detailed algorithms for all modules in the MaNGA-DAP are described by \citet[][Sections 6-10]{MaNGA-DAP} and \citet[][Section 2]{Belfiore_MaNGA}.
We use Version 4.5 of the MaNGA-DAP available on Github.\footnote{ \url{https://github.com/sdss/mangadap}}
The {\tt examples/fit\_one\_spec.py} file from the MaNGA-DAP was adapted and used as the core component of the eBOSS-DAP pipeline. 

We chose to base our code on the MaNGA-DAP for several reasons. 
The eBOSS and MaNGA datasets use the same spectrograph and have similar, but not identical, data structures (plates vs. cubes), allowing many methods to be used with minimal modification. 
We also chose the MaNGA-DAP for its modular capabilities. 
We were able to fully customize our template libraries, line lists, and spectral indices, as well as the data that is saved. 
Another boon is the ability of the MaNGA-DAP pipeline to run on many spectra simultaneously, significantly reducing computational time for our 1.9 million spectra. 
Another key feature is MaNGA-DAP's 
ability to tie various emission line properties (line widths, velocity offsets, flux ratios) across multiple lines \citep{Belfiore_MaNGA}, using the implementation provided by pPXF.
This allows us to use stronger lines to ensure robust fits for weaker lines by limiting some of their free parameters.

\subsection{The eBOSS-DAP} \label{subsec:eBOSS-DAP}
To adapt the MaNGA-DAP for the goals of the eBOSS-DAP, we made several key changes to the code and supporting files, which fall into two classes: infrastructure changes and scientifically motivated changes. 

Our first major infrastructure change was to enable the code to accept folders of one-dimensional spectra obtained on the same plate instead of single spectra or cubes. 
The details surrounding this change are found in \S\ref{subsec:workflow}.
Additionally, we changed the preprocessing for the foreground Milky Way dust extinction correction. We use the {\tt dustmaps} package \citep{dustmaps} to retrieve the Milky Way E(B-V) value at the position of each target from the Corrected Schlegel, Finkbeiner \& Davis (1998) dust maps \citep[CSFD;][]{CSFD}. We remove the foreground reddening from each spectrum using the {\tt dust\_extinction} code \citep{dust_extinction} with the \citet{dust_extinction_law} Milky Way dust attenuation law instead of the \citet{manga_dust} curve used for MaNGA. 

An additional change was modifying the masking routine for the data, as eBOSS has slightly different mask bits than MaNGA. 
Our final set of infrastructure changes was fixing some minor issues we uncovered through our testing. For example, we found that if an emission line had only a few unmasked pixels, poor Gaussian fits could result. 
We, therefore, only report the results of the Gaussian fit for emission lines with less than 50\% of their pixels masked out. We discuss the masking procedure further in \S\ref{subsubsec:masking}.
In addition, we only report spectral indices with less than 10\% of their pixels masked out, as spectral indices are more sensitive to masking.

The first scientifically motivated change was to modify the spectral template libraries that the code relies upon.
The MaNGA-DAP is configured to be able to use several different stellar population template libraries \citep[see discussion in ][]{Abdurrouf2022}. These templates can be composed of individual stellar spectra or simple stellar population (SSP) models. We added a new SSP library called C3K that is discussed in detail in \S\ref{subsubsec:C3K}. 
We also made changes to the list of emission lines and line indices as detailed in \S\ref{subsubsec:linelist} and \S\ref{subsubsec:spind}.

\subsubsection{Masking} \label{subsubsec:masking}
The eBOSS data reduction pipeline creates a mask array that encodes data quality information differently from MaNGA.\footnote{for further reference, see \url{https://www.sdss4.org/dr17/algorithms/bitmasks/}} Accordingly, we had to implement our own masking routine.
When fitting the data, we chose to use a combination of masks, primarily the AND\_MASK\footnote{The particulars of the AND\_MASK are found here: \url{https://www.sdss4.org/dr17/algorithms/bitmasks/\#SPPIXMASK}} and our own mask based on the inverse variance. 
The AND\_MASK is a bitmask that flags pixels that were masked in every exposure contributing to the final spectrum; the bits encode the reason for masking. 
This mask was then trimmed down by removing certain bits that we found via visual inspection to be masking apparently good data, namely, bit 23: BRIGHTSKY, as well as bits 0-12, which are so-called {\tt FIBERMASK\_BITS}\footnote{Definitions of the individual {\tt FIBERMASK\_BITS} (bits 0--12) are given here: \url{https://www.sdss4.org/dr17/algorithms/bitmasks/\#SPPIXMASK}} that mask the full spectrum. These bits are flagged when an error that would affect the full spectrum occurs, such as when the flat field is flagged as bad. 
There are 27,620 spectra or $\sim$ 1.5\% of our sample where a {\tt FIBERMASK\_BIT} is set. We chose to fit these spectra, but we add a boolean column, {\tt FIBERMASKED}, to the final output catalog indicating whether any {\tt FIBERMASK\_BIT} was set for that spectrum, rather than reporting the underlying hexadecimal mask value.
After masking the pixels identified in our trimmed mask, we added any pixels for which the inverse variance array was negative, infinite, or zero. 

In addition to masking pixels on \textit{input} to the eBOSS-DAP, the analysis of the data leads to additional masking. The two primary masks indicate where the model is invalid (e.g., where the rest wavelengths of the template and galaxy spectra do not overlap) and which pixels in the galaxy spectra were rejected as outliers during the full-spectral fitting.

Our final additional mask consists of pixels in the window 3662 -- 3706.3~\AA\ and it is set only for stellar continuum and emission line fitting. 
This mask is created to avoid contamination of the continuum fit by high-order Balmer emission lines, which, as they are unfit by our code, pull up the continuum fit artificially. This mask is removed during spectral index measurement.
The final mask is saved in the final output file for all spectra.
 
\subsubsection{C3K Spectral Libraries} \label{subsubsec:C3K}
\begin{figure*}[!ht]
\plotone{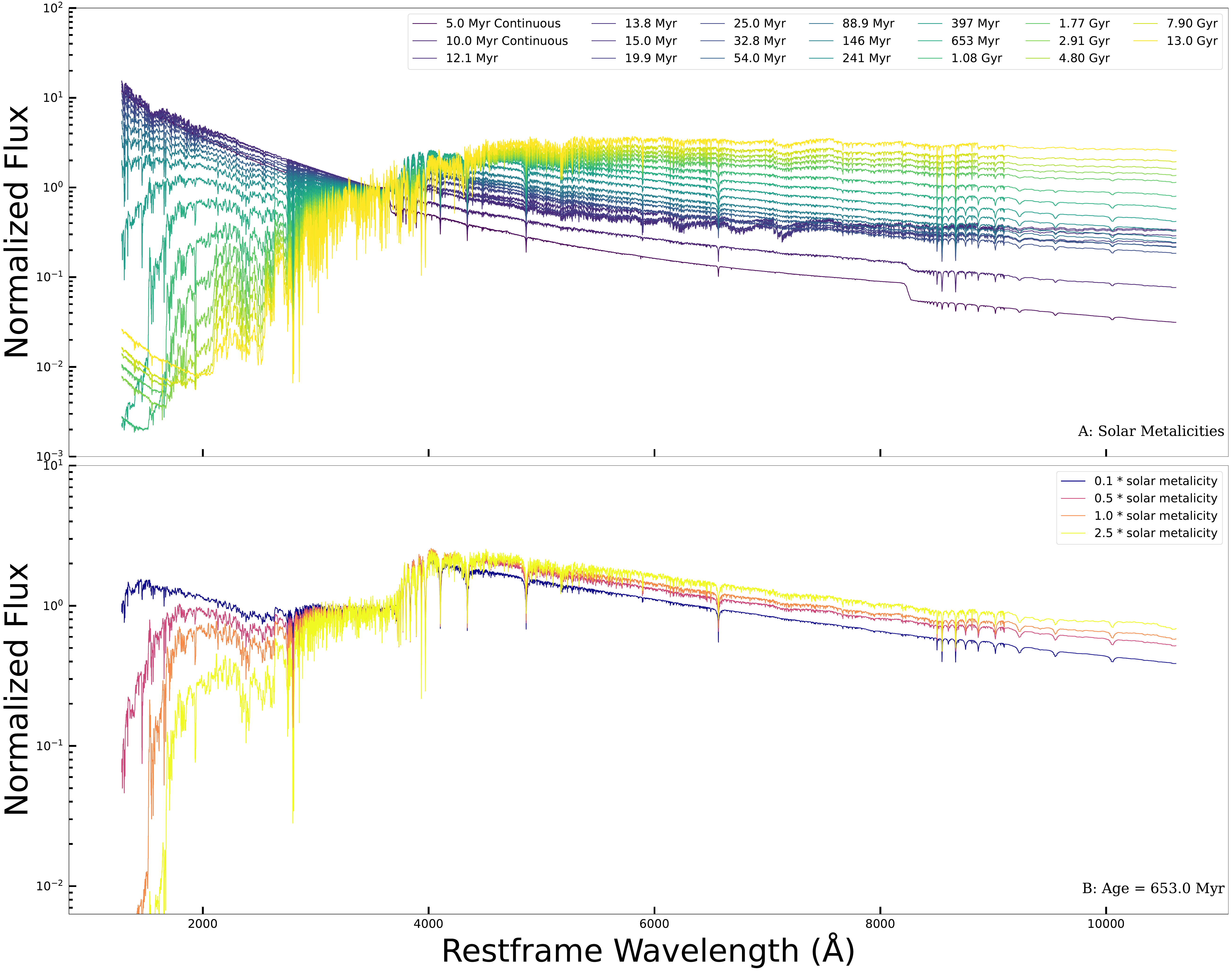}
\caption{Stellar population templates employed by the eBOSS-DAP.  The templates have been generated with FSPS \citep{fsps_1,fsps_2} using the C3K stellar library (Conroy et al., in prep.). {\bf Top:} Solar metallicity SSPs and two continuous models color-coded by age. {\bf Bottom:} Four SSPs that share the same age of 653 Myr but vary in metallicity from $Z=0.1 - 2.5~Z_{\sun}$. The full template library includes all ages shown at all four metallicities (80 templates, in total). All templates are normalized to their flux at 3500~\AA. 
\label{fig:c3k}}
\end{figure*}

For our template library, we chose to generate SSPs and continuous star formation models based on the C3K stellar library (Conroy et al., in prep.) and MIST isochrones \citep{MIST_0,MIST_1} found in the Flexible Stellar Population Synthesis (FSPS) package \citep{fsps_1,fsps_2}.
The C3K model library, described in \citet{ssps_analysis}, consists of fully theoretical stellar spectra created using the ATLAS12 \citep{atlas12_a, atlas12_b} and SYNTHE \citep{Synthe} codes.
These models have stellar effective temperatures of 2500 $\leq$ T$_{eff}$ $\leq$ 50,000, surface gravities of -1 $\leq \log g \leq$ 5.5 and metallicities 0.003 $\leq \frac{{\mathrm{Z}}}{{\mathrm{Z_{\odot}}}} \leq$ 3.16.

Approximately 80\% of the eBOSS galaxy sample is at $z = 0.4 - 1.1$, which means that the blue edge of the spectrum is at restframe 1700 -- 2600~\AA.
We chose to use the C3K SSPs primarily because most empirical models have limited UV coverage and/or low UV spectral resolution. 
Theoretical models also provide a denser exploration of parameter space ($T_{eff}$, $\log g$, $Z$), which can be important in achieving an optimal fit \citep{Coelho2020}. 
Additionally, the C3K models have higher spectral resolution than our data over most of the relevant wavelength range. 
Unlike observational data, they are unaffected by factors like flux calibration errors or sky subtraction residuals. 
Additionally, theoretical SSP models have been shown to better recover the ages and metallicities of star clusters \citep{synth_models}.
However, their precision is limited by the precision of the underlying atomic and molecular data, as well as the computational models used to generate them. \citep{Conroy_review, Coelho2020}.

For our SSP model grid, we chose four metallicity bins and 18 age bins as seen in Figure \ref{fig:c3k}.
These ages were selected such that periods of rapid evolution received more finely binned SSPs to provide good coverage of different spectral shapes.  Our youngest SSP model is 12.1 Myr old. To fit galaxies with ongoing star formation, we add two models that represent 5 and 10 Myr of continuous star formation. We adopted this approach to minimize the number of models needed. Because we are fitting galaxy spectra, they are unlikely to have changes in their star formation histories on timescales shorter than 5 Myr. 

\begin{figure*}[!ht]
\plotone{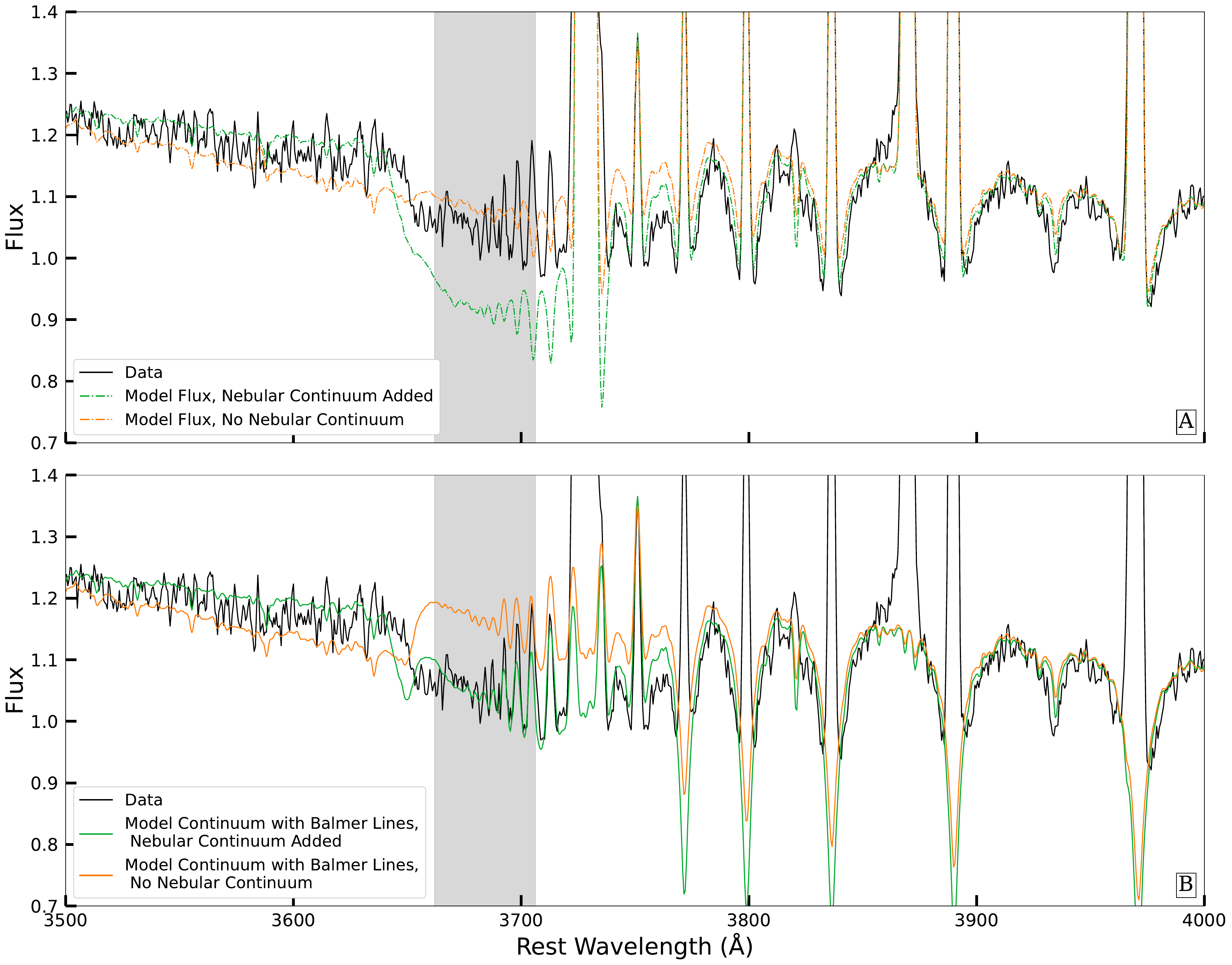}
\caption{A comparison of stellar population models with and without nebular continuum. In both panels, the black line shows a stacked spectrum made from galaxies with very high H$\beta$ EW, the orange line is the best-fit stellar population model when templates that do not include nebular continuum emission are used, and the green line is the best-fit model with nebular continuum included. The vertical gray band indicates the mask that is applied during fitting because the eBOSS-DAP does not model Balmer lines above H12 (see \ref{subsubsec:masking}).
\textbf{Panel A}: The full best-fit model fluxes for this stack, including all fitted emission lines. The models without nebular continuum (orange) underestimate the continuum blueward of 3662~\AA\ and overestimate it redward of 3706~\AA. The models with nebular continuum correctly estimate the continuum outside of the masked region.
\textbf{Panel B}: The stellar continuum from the same best-fit models, with flux from high-order Balmer lines added to the continuum fits by scaling to the measured H12 flux assuming Case B line ratios. When the high-order Balmer flux is included, the models with nebular continuum provide a superior fit across the full wavelength range shown.
\label{fig:neb_comp}}
\end{figure*}

We use FSPS to add nebular continuum to our continuous star formation models. 
(We did not add it to our SSPs because nebular continuum does not contribute significantly to stellar populations older than 10~Myr \citep{byler_neb_cont}.)  
We found that for galaxies with high levels of ongoing star formation, nebular continuum is needed to properly fit the spectrum at 3600-3700~\AA\ as seen in Figure \ref{fig:neb_comp}, panel \textbf{A}.
If nebular continuum flux is not present in the models, the continuum will be underestimated blue-ward of the Balmer jump ($\sim3645$~\AA) while being overestimated red-ward.
Additionally, this can cause a loss of measured flux in [O II]$\lambda\lambda~3727, 3729$, a critical emission line, illustrating how differences in the continuum treatment can propagate into emission-line measurements
A second concern is that, although the eBOSS-DAP does not measure Balmer lines bluer than H12, there is significant continuum infilling from closely packed high-order Balmer lines (H14$\rightarrow$H30) as shown in Figure \ref{fig:neb_comp} panel \textbf{B}.
If the nebular continuum is not included in the fit, all of this flux is inappropriately attributed to the stellar continuum.
If the high-order Balmer line fluxes are added by scaling the lines to H12, the models with nebular continuum included provide a superior fit in the 3500-3800~\AA\ region.

The wavelengths of the C3K models natively span from 100 -- $10^7~$\AA; however, for our models used in emission line fitting, we trimmed to 1281 -- 10616~\AA.
The resolution of the C3K SSPs created by FSPS is $R \sim 1180$ between 1281 -- 2750~\AA, $R \sim 7060$ between 2750 -- 9100~\AA, and $R \sim 1180$ between 9100 - 10616~\AA. 
For models used to fit the stellar continuum kinematics, we further restrict them to the $\lambda$ = 3000~\AA\ -- 9000~\AA\ range, with $R \sim 7060$. The spectral sampling of the C3K templates is also not uniform, meaning that we need to resample the spectra before the eBOSS-DAP can use them. To match the average sampling step across the wavelength range of our models, we resample the templates such that they are uniformly sampled in log space with a spectral spacing of ${\rm d}\log\lambda = 7.239 \times10^{-5}$ or 49.97 \kms. The templates are then resampled again by the DAP to match the wavelength array of each plate.

\subsubsection{Emission Line Fitting} \label{subsubsec:linelist}
The MaNGA-DAP natively provides users with a flexible and modular file that details the emission line list. 
What sets this fitting routine apart is the ability to tie parameters of the lines, namely the flux, velocity, and velocity dispersion. 
Line tying (as described in \citet{MaNGA-DAP} with features implemented in \citet{ppxf_code, ppxf}) refers to requiring the parameters of two or more separate lines to be related to each other, for example: [OIII]~$\lambda$4960 flux = 0.331~[OIII]~$\lambda$5008 flux. More specifically, when tying the flux of two lines, one free parameter effectively governs the flux of both lines in the model, and the fitting algorithm adjusts that parameter to optimize the fit to both lines simultaneously. Table \ref{table:line_list} stores the emission line information, including the ratios and the lines used for tying. 

A new feature in the eBOSS-DAP (which was possible but unused in previous MaNGA-DAP analysis) is the ability to specify that two line parameters be within some fractional range of each other.
This can be seen with the example of [O~III]~$\lambda$5008 in Table \ref{table:line_list}. 
Here, the line list specifies that the [O~III]~$\lambda$5008 line must have equal velocity to H$\alpha$ (in the low redshift bin), but the velocity dispersion is free to be within 75\% of the H$\alpha$ dispersion (i.e., $1/1.75~\sigma_{\mathrm{H}\alpha} < \sigma_{[\mathrm{O}~\mathrm{III]}} < 1.75~\sigma_{\mathrm{H}\alpha}$).
We chose to impose this velocity dispersion tie on all lines, as it helps weak lines to be fit accurately and does not have adverse consequences for the stronger lines. 

The allowed range in dispersion ratios ($\sigma_{\mathrm{line}}$ = 0.57 - 1.75~$\sigma_{\mathrm{H}\alpha}$) was selected after empirically testing the range of dispersion ratios found for strong emission lines when no velocity dispersion tie was used. It accommodates the wavelength-dependent instrumental resolution and allows for real physical differences between emission lines produced in different regions of a galaxy, for example, [O~III] lines in the narrow line region of Type~II AGN. Note that Type~I AGN, which can have extreme velocity dispersion differences between Balmer and forbidden lines, are generally not included in our sample, as they are classified as quasars rather than galaxies by the SDSS {\tt spectro1d} pipeline which measures redshifts and spectral classifications. 

The eBOSS-DAP makes several additions to the MaNGA-DAP emission line list provided by DR17 \citep{MaNGA-DAP, sigmacorr_paper_2}. The list is increased to \nlineshigh~ lines, of which only 35 were fit in MaNGA. 
The 43 new eBOSS-DAP lines are demarcated by an asterisk in Column 1 of Table~\ref{table:line_list}. 
These lines include all iron lines found in the line list and all lines of $\lambda < 3600$~\AA. Of particular note are [Ne~V] $\lambda\lambda$3347, 3427 and [Fe~X] $\lambda$6374, which are used as AGN diagnostics.
The bluest line we fit is [C~III] $\lambda\lambda$1906,1909, which redshifts into the observed spectral range at $z= 0.87$.

Unlike the MaNGA-DAP, which uses one line list for the entire sample, we use four distinct line lists that split our sample by H$\beta$ equivalent width (EW) and by redshift. 
This was done in an effort to resolve two major problems: data size and restframe wavelength coverage. 

Initially, we were fitting all lines in all spectra, but we noticed that our weaker lines returned measurements consistent with noise in the majority of spectra. 
Therefore, we elected to split the sample into two subsamples: one with an H$\beta$ EW $>$ 10 \AA~ resulting in \nhighew\ spectra, and another with H$\beta$ EW $<$ 10 \AA\ resulting in \nlowew~ spectra.
All lines were fit in the high EW subsample, but in the low EW sample, a \nlineslow-line subset, made up of typically stronger lines, is fit.
These lines are flagged in the last column of Table ~\ref{table:line_list}. 

Our second division is a result of the wide redshift range of our data. 
A practical limitation of the MaNGA-DAP is that it must designate a reference line when setting up the set of tied emission lines. 
The line chosen is effectively irrelevant to the optimization algorithm, but the tying setup fails if the reference line is outside of the observed spectral range. 
This means that, while we may like to tie all of our lines to H$\alpha$, if it is outside of our observable window, all lines tied to it are not fit, regardless of whether the tied lines are observable.
To remedy this, we made a redshift cut, and for all spectra above $z = 0.4963$, we tied lines to H$\beta$ instead. 
We chose this redshift cut as this is when H$\alpha$ enters a zone of increased measurement error near the edge of the spectrum, as seen in Figure~\ref{fig:Bowtie}.
Roughly 60\% of our sample lies above this redshift threshold, drastically increasing the quantity of data we can successfully process.

To evaluate the impact of these divisions on the consistency of our emission--line measurements, we selected a random subsample of 4171 spectra satisfying $9 < \mathrm{EW}_{\mathrm{H}\beta} < 11$~\AA\ and $0.1 < z < 0.4$, ensuring that all objects lie near the equivalent--width boundary while retaining all relevant lines within a regime of low spectrophotometric error.
We find no systematic offset introduced by either the EW-based line-set division or the redshift-based anchor-line change; in both cases, the mean percent difference between configurations is consistent with zero for all six lines considered, and the mean difference normalized by the propagated measurement error reaches at most $\sim$8\% of the measurement uncertainty.
Any differences between configurations are consistent with random measurement noise.
Full details of these comparisons, including per-line statistics, are provided in Appendix~\ref{sec:survey_analysis} (Figures~\ref{fig:trumpet_ew} and \ref{fig:trumpet_z}; Tables~\ref{table:survey_analysis_ew} and \ref{table:survey_analysis_z}).
For access to the underlying per-object flux measurements for the two comparisons discussed above, see \hyperref[sec:data_availability]{Data Availability}.

\subsubsection{Spectral Indices}\label{subsubsec:spind}
Following \citet{MaNGA-DAP}, we adopt the term ``Spectral Index'' to refer to measurements of stellar continuum features.
We have not changed any details of the spectral index measurements or outputs as specified in Table 4 of \citet{MaNGA-DAP} and Section 5.2.2 of \citet{Abdurrouf2022}.
Our singular contribution is the addition of a ``BalmerBreak'' index. We adopt the definition of \citet{balmerbreak}, which is measured identically to D4000 but with a blue bandpass of $\lambda$ = 3620.0 -- 3720.0~\AA\ and a red bandpass of $\lambda$ = 4000.0 -- 4100.0~\AA\, in vacuum wavelengths.
This additional index allows us to measure the Balmer break, which appears as a discontinuity or ``jump'' in the stellar continuum spectrum at around 3645 \AA. 
This break is caused by the ionization of hydrogen atoms from the second energy level within stellar atmospheres, leading to a drop in intensity at wavelengths shorter than this limit. 
The Balmer break is prominent in SSPs with ages of 0.3 - 1 Gyr, and it is therefore used in estimating the age of stellar populations and inferring the star formation history of galaxies.

\subsection{Workflow}
\label{subsec:workflow}

Running the eBOSS-DAP first requires a few setup steps. For each galaxy, we collect the foreground Milky Way E(B-V) value from the CSFD \citep{CSFD} maps in the {\tt dustmaps} \citep{dustmaps} package and the galaxy redshift from the {\tt spAll-v5\_13\_2} file released in DR17. The sample is then divided into ``low-$z$'' and ``high-$z$'' samples, distinguished by those galaxies below and above $z = 0.4963$, respectively. Galaxies in the low-$z$ bin tie all emission lines to H$\alpha$, whereas the high-$z$ samples use H$\beta$; recall, this is largely a practical matter dictated by the use of the DAP. We also divide the sample into ``low-EW'' and ``high-EW'' samples based on a preliminary measurement (using the DAP) of the H$\beta$ EW being below or above 10~\AA, respectively. All emission lines are fit for galaxy spectra in the high-EW sample, whereas only a subset of the strongest lines are fit to the low-EW sample; see Table \ref{table:line_list}. The division by $z$ and EW leads to four samples of spectra to be fit.

Spectra in each sample are fit in groups based on their plate number. This grouping is also primarily a practical issue: Each spectrum is treated independently, but this grouping allows us to perform common preparation steps (like building the spectral template library) once for a large number of spectra, instead of once per spectrum. The detailed steps performed for each spectral group \citep[which closely follow the steps for the MaNGA DAP described by][]{MaNGA-DAP} are as follows:

\smallskip

\noindent\textbf{(1)} The spectral templates are resampled to the pixel scale of the galaxy data divided by an integer, N. Here we choose N=4 such that the line-spread function of the C3K templates (see \S\ref{subsubsec:C3K}) is still well-sampled. Due to C3K's varying spectral resolution, two spectral libraries are created. The first has coverage of $\lambda$ = 1280~\AA\ -- 10600~\AA\ ($R\sim1180-7060$) and is used to measure the stellar continuum and emission lines. The second has coverage of $\lambda$ = 3000~\AA\ -- 9000~\AA\ ($R\sim7060$) and is used to measure the stellar kinematics. 

\smallskip

\noindent\textbf{(2)} The galaxy spectra are packaged into two-dimensional arrays with a common wavelength vector by adding modest zero-padding. The padded spectral regions are included in the masks (see Section \ref{subsubsec:masking}). The observed fluxes and uncertainties are corrected for Milky Way extinction using the E(B-V) values and the {\tt dust\_extinction} \citep{dust_extinction} package using the \citet{dust_extinction_law} Milky Way extinction law. Masks from \S\ref{subsubsec:masking} are applied.

\smallskip

\noindent\textbf{(3)} The stellar kinematics are measured for each spectrum independently using pPXF \citep{ppxf}. The emission lines are masked during this fit, and we use an eighth-order polynomial in (additive) combination with the spectral templates with the shorter spectral range to model the continuum. See \citet{MaNGA-DAP}, Section 7.\footnote{Measurements of the stellar velocity dispersion that exceed 450 $\mathrm{km~s^{-1}}$ are considered erroneous. The output database replaces such measurements by imposing this upper limit, setting $\sigma_\ast = 450\,\mathrm{km~s^{-1}}$.}

\smallskip

\noindent\textbf{(4)} The emission-line properties (flux, velocity, and velocity dispersion) are modeled simultaneously with the continuum templates, where the stellar kinematics are fixed to the results from the previous step. The continuum templates used include nebular continuum emission in the two continuous models as discussed in \S\ref{subsubsec:C3K}. For this module, the continuum templates used are those with the longer wavelength range, and we include an eighth-order \textit{multiplicative} polynomial to apply low-order continuum modifications to address attenuation. Additionally, this polynomial helps minimize issues related to flux-calibration errors and differences between the template library and the data. The best-fitting model Gaussian profiles for each emission line are then used to calculate their equivalent widths. The continuum for the EW measurements is defined by a line connecting the median flux in two passbands (red and blue, as defined in the input file) to the (unweighted) center of each passband. The continuum level used in the EW calculation is the value of this line at the fitted center of the emission line. 

\smallskip

\noindent\textbf{(5)} In addition to the emission-line properties measured from the best-fitting Gaussian function, we also measure emission-line fluxes and EWs by direct integration of the continuum-subtracted spectrum. The code utilizes two sidebands to normalize any deviations from the continuum by drawing a line that passes through the coordinates defined by the center of each passband and the median flux within the passband. The continuum value is taken as the value of that line at the (measured) emission-line center. See \citet{MaNGA-DAP}, Section 9, and \citet{Belfiore_MaNGA} for further details. We only report the EW measurements, which are primarily used as a fit-quality metric; i.e., the EWs measured using the best-fit Gaussian and direct integration should be statistically identical for high-quality fits (see \S\ref{subsubsec:ew_comp}).
\smallskip

\noindent\textbf{(6)} The last analysis step is to measure the set of spectral indices described in \S\ref{subsubsec:spind} for our eBOSS spectra, after first subtracting the best-fit emission-line-only model. See \citet{MaNGA-DAP}, Section 10, and \citet{Abdurrouf2022}, Section 5.2.2.

\smallskip

\noindent\textbf{(7)} Finally, the analysis results are saved to a FITS file, one file per spectrum. Additionally, all non-detections and failures have both the signal and error overwritten with -999.0, the eBOSS-DAP error code. Once the analysis of all spectra is complete, the individual output files are aggregated into our final output catalogs; see Section \ref{subsec:catalogs}.

\begin{figure*}[!ht]

\centering
\includegraphics[width=0.9\textwidth]{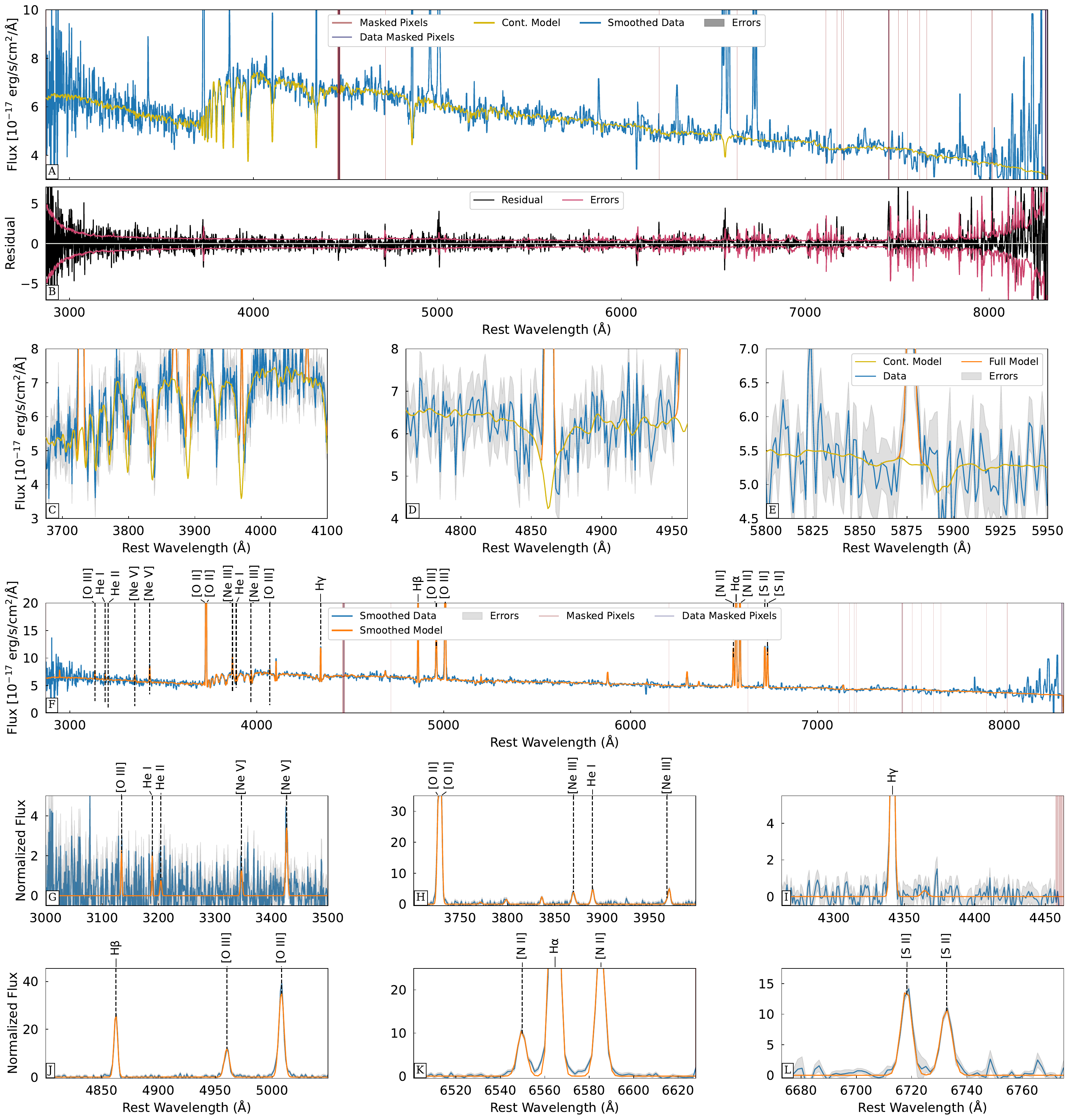}

\caption{Sample eBOSS-DAP fit for the spectrum spec-4037-55631-0920.fits, which was categorized as an AGN using the Baldwin, Phillips, and Terlevich (BPT) \citep{OG_BPT} diagnostic. In all panels, the processed data are shown in blue. The best-fit continuum model is shown in dark yellow. The full best-fit model, including emission lines, is shown in orange. The masked portions are shown as vertical red bars. Errors are shown in gray bands. {\bf Panel A:} A zoomed-in view on the smoothed continuum to demonstrate the quality of the fit. {\bf Panel B:} The residual of the full model fit (black). There is a white line at zero for legibility and a red line showing the errors on the residual. {\bf Panels C, D, and E:} Zoomed-in windows showing the continuum fit near some strong stellar absorption lines: the low-order Balmer lines (C), H$\beta$ (D), and Na~I (E). {\bf Panel F:} The full fit (stellar continuum plus emission lines) compared to the data. Emission lines are labeled. {\bf Panels G-L:} Various zoomed-in locations of important emission lines, which have been labeled. These panels are continuum-subtracted, so the continuum flux is zero at all points.
\label{fig:AGN_fit}}
\end{figure*}
\section{Results} \label{sec:results}
We ran the eBOSS-DAP on a sample of \nsample~ galaxy spectra with \npassed~ successfully passing the full pipeline, giving us a 99.88\% success rate. 
 Figure \ref{fig:AGN_fit} shows an example of the eBOSS-DAP's continuum and emission line fits. Two additional spectra are shown in the Appendix (Fig.~\ref{fig:SFG_fit}, Fig.~\ref{fig:RD_fit}). \footnote{These plots are not generated as part of the eBOSS-DAP; they are generated by a separate program found here: \url{https://github.com/owenmatthewsa/ebossdap/}}
Panels A -- E showcase the quality of our continuum fits, zooming in on some important spectral features. Similarly, panels F -- L zoom in on several important emission lines and show that our line fitting is robust. 

In \S\ref{subsec:catalogs} we discuss the catalogs generated by the eBOSS-DAP. In \S\ref{subsec:Validation}, we assess the quality of the catalog data by comparing measurements made on duplicate spectra and comparing to some existing catalogs in the literature. Finally, in \S\ref{subsec:characterization}, we present a series of plots that further characterize the eBOSS sample and illustrate its scientific potential. 

\subsection{Catalogs}\label{subsec:catalogs}

We use the eBOSS-DAP outputs to create two emission line catalogs:  
one for the full sample ({\tt eBOSS\_DAP\_emlines\_full\_v\#.fit}; \npassed~ spectra, which utilizes the reduced emission line list (\nlineslow~ lines), and one for the high equivalent width (H$\beta$~EW $>$ 10~\AA) sample ({\tt eBOSS\_DAP\_emlines\_high\_ew\_v\#.fit}; \nhighew\ spectra), which makes use of the full line list (\nlineshigh~ lines). 
The column descriptions of these catalogs are found in Table \ref{table:eml_table}. 
All emission lines are corrected for foreground Milky Way reddening as described in Step \textbf{(2)} in \S\ref{subsec:workflow}.
Notably, we include, along with the emission line velocity dispersions, the instrumental dispersion corrections made to the dispersions. In cases where the instrumental resolution exceeds the measured line width, the dispersion has been flagged as a poor fit.

The final two catalogs are the spectral indices and the stellar population template weights, along with auxiliary data. 
The {\tt eBOSS\_DAP\_spind} table contains the spectral index catalog and the column descriptions given in Table \ref{table:spind_table}.
To find the units of a particular spectral index, readers should refer to the table of spectral indices (Table 4) in \citet{MaNGA-DAP}. 
These spectral indices are not corrected to a fixed velocity dispersion, as they were in the MaNGA-DAP; however, the corrections are included in the catalog if needed.
The {\tt eBOSS\_DAP\_auxdata} table contains the template grid weights as well as the multiplicative polynomial values and other auxiliary data required to produce the model as described in Table \ref{table:auxdata_table}. 

\begin{figure*}[!ht]
\centering
\includegraphics[width=1.0\textwidth]{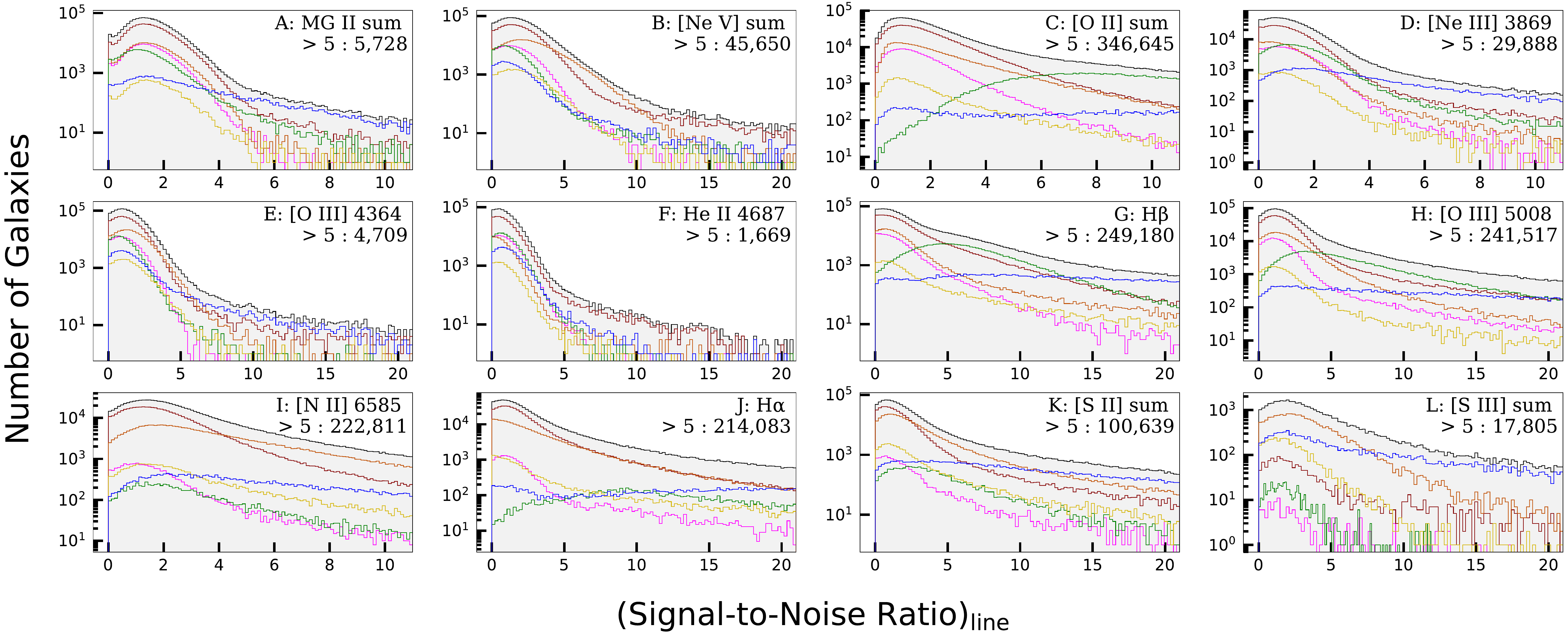}
\caption{The S/N histograms of 12 lines in the eBOSS-DAP full sample, split by target classification. The target classes are color-coded identically to Fig. \ref{fig:redshift_hist}. Additionally, the number of galaxies that have S/N $>$ 5 in the line flux has been included underneath the line name.
\label{fig:EML_HIST}}
\end{figure*}

\begin{figure*}[!ht]
\centering
\includegraphics[width=1.0\textwidth]{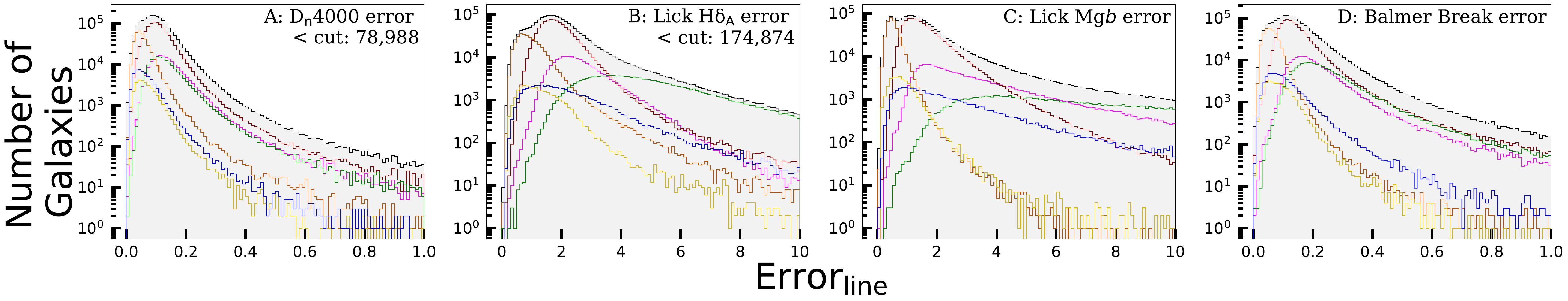}
\caption{Histograms of the errors in four spectral indices in our full sample, split by target classification (see Fig \ref{fig:redshift_hist} for color coding). Additionally, the number of galaxies that pass the quality cuts suggested by \citet{spind_cut} are included underneath the index name for D$_{\mathrm{n}}$4000 and H$\mathrm{\delta}_{\mathrm{A}}$ (see \S\ref{subsubsec:hdela-vs-dn4000} for descriptions of these cuts). 
\label{fig:SPIND_HIST}}
\end{figure*}

To help visualize the eBOSS-DAP catalog, we have produced the following two figures. 
In Figure \ref{fig:EML_HIST} we show a set of S/N histograms for our Gaussian fits to 12 commonly used emission lines in our catalog, annotated with the number of detections of S/N$>5$ in each line. 
Of note is the large quantity of galaxies with [Ne~V] S/N $>$ 5. We will explore these objects further in future work (Matthews Acu\~{n}a et al., in prep.).

In Figure \ref{fig:SPIND_HIST} we show a histogram of the errors in four popular spectral indices.
Due to the way spectral indices are measured, values may be positive or negative. S/N measurements are misleading when the index values approach zero, so we elected to simply show the errors in these indices. 
We additionally include the number of spectra in D$_{\mathrm{n}}$4000 and H$\mathrm{\delta}_{\mathrm{A}}$ which pass the quality cuts suggested in \citet{spind_cut} of D$_{\mathrm{n}}$4000 error $<$ 0.03 and H$\mathrm{\delta}_{\mathrm{A}}$ error $<$ 0.8.

\subsection{Data Quality Assessment} \label{subsec:Validation}
We will assess data quality in three ways: Comparing Gaussian and direct integrated EWs (\S\ref{subsubsec:ew_comp}), comparing eBOSS-DAP data with SDSS pipeline emission line measurements (\S\ref{subsubsec:sdss-comp}), and comparing measurements made on repeat spectra (\S\ref{subsubsec:trumpets}).

\subsubsection{Gaussian vs Integrated EW Comparisons}\label{subsubsec:ew_comp}

We test the integrity of our Gaussian emission line fits by comparing the emission line equivalent widths (EWs) computed from the Gaussian fits to those found by integrating the continuum-subtracted spectrum between the line boundaries specified in Table~\ref{table:eml_table}. 
These integrated EWs should be within a standard deviation of the Gaussian EWs found by the model fitting and, as such, offer a check on the quality of any particular fit. Following \citet{Belfiore_MaNGA}, we compare the Gaussian and integrated EWs by computing their percent difference as a function of S/N.

\begin{figure}[!ht]
\plotone{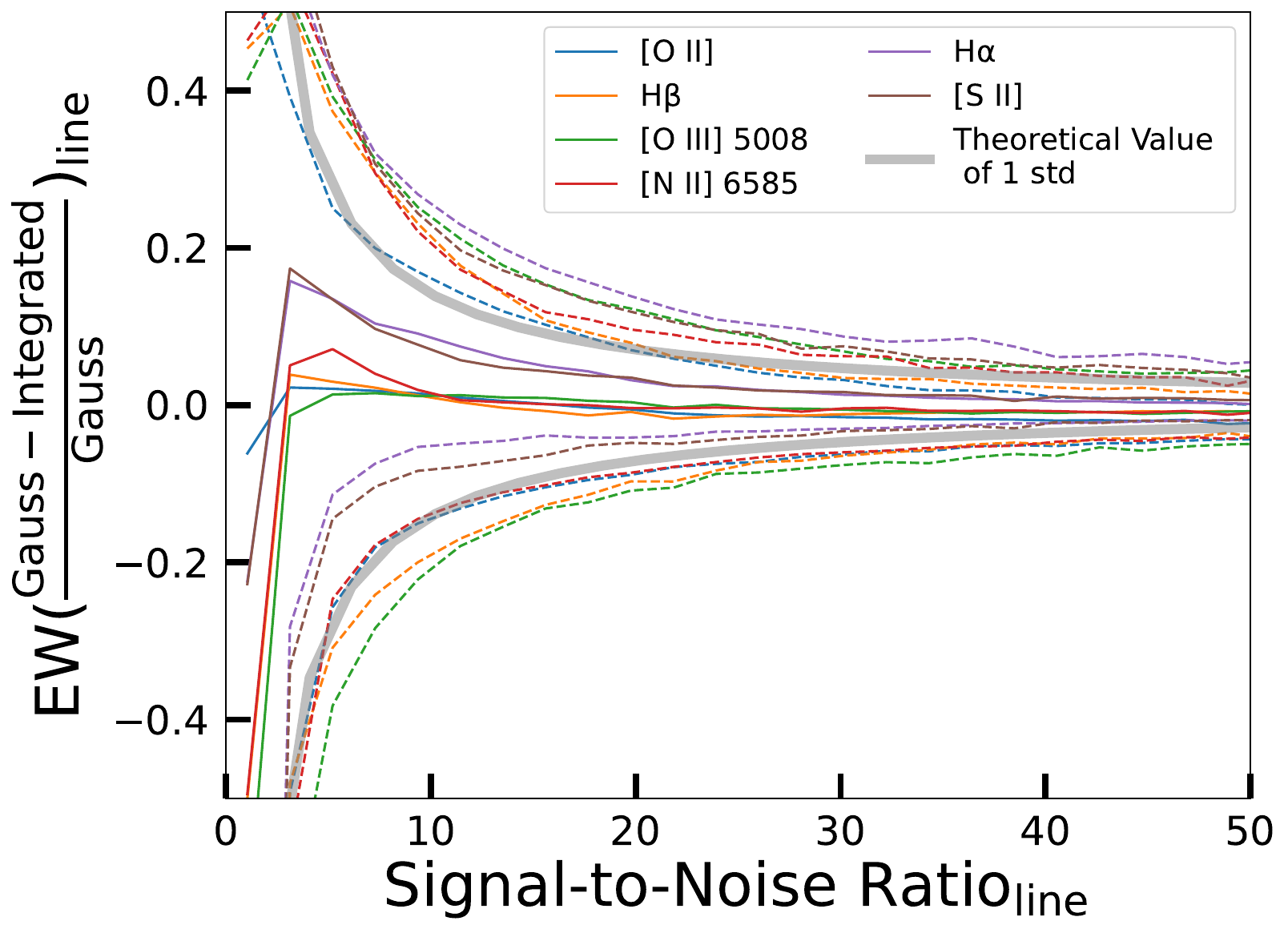}
\caption{A comparison of Gaussian and integrated equivalent widths (EWs) relative to S/N for six commonly used strong emission lines. 
We take the Gaussian EW, subtract the integrated EW, and divide the result by the Gaussian EW to find the percent difference. 
Solid lines represent the median, while the dashed lines represent the standard deviation.
A gray line has been included to show the expected value of the standard deviation given the S/N.
\label{fig:EW_comp}}
\end{figure}

In Figure \ref{fig:EW_comp} we see that as S/N increases, the integrated and Gaussian EWs grow closer, as expected. 
Of note is that both H$\alpha$ and [S II] have, on average, lower integrated EWs.
This is caused by the narrow integration boundaries used for these lines due to the presence of nearby lines (see Table ~\ref{table:eml_table}). In most of the lines, the Gaussian and integrated EWs agree within the expected margin of error, with slightly higher deviations in H$\alpha$ and [S~II], likely caused by broader components outside of the integration bounds of those lines. Additionally, the [O~III] and H$\beta$ lines show higher-than-expected deviations, likely due to substructure that is not well fit by a Gaussian -- for example, a low-level broad component.
Both Gaussian and integrated EWs are included in the eBOSS-DAP catalogs.

\subsubsection{SDSS measurement Comparisons}\label{subsubsec:sdss-comp}
SDSS provides an emission line catalog for the eBOSS sample stored in the {\tt spZline} files.\footnote{\url{https://data.sdss.org/datamodel/files/BOSS_SPECTRO_REDUX/RUN2D/PLATE4/RUN1D/spZline.html}}
However, there are several issues with the SDSS {\tt spZline} catalog that motivated the development of the eBOSS-DAP: the SDSS line list comprises only 27 lines compared to our \nlineshigh; the emission lines are not corrected for Milky Way dust attenuation; and stellar absorption indices are not measured. Moreover, there is only minimal documentation available on the methods used to create the {\tt spZline} catalog \citep{Bolton:2012}.
Nonetheless, it provides a valuable point of comparison for the eBOSS-DAP catalog.

\begin{figure*}[!ht]
\centering
\includegraphics[width=0.8\textwidth]{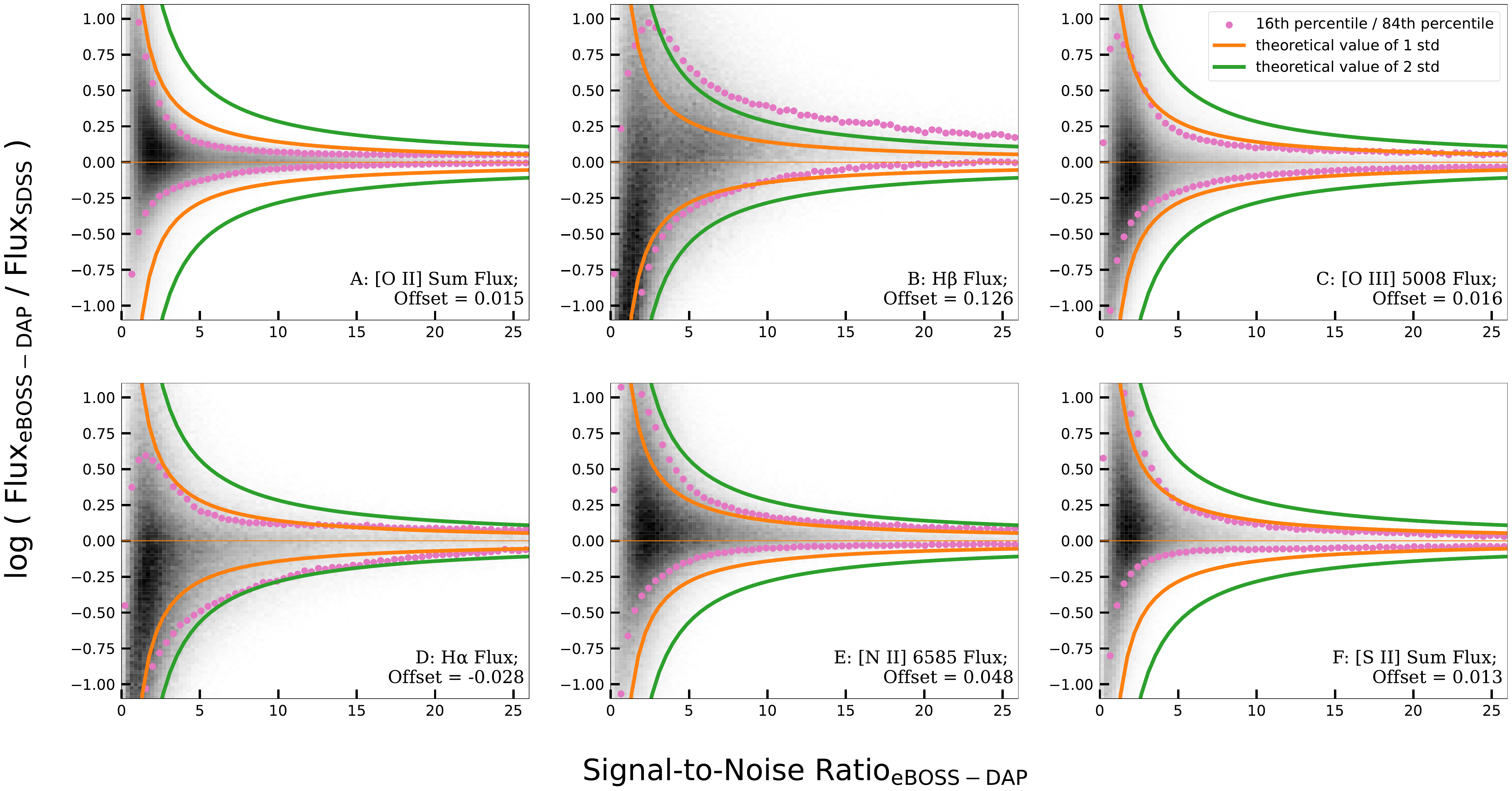}
\caption{
A comparison of line flux measurements in the eBOSS-DAP and the SDSS {\tt spZline} catalog for six relatively strong emission lines. 
The y-axis shows the log of the ratio between the flux from the eBOSS-DAP and the dust-corrected SDSS-provided measurement for the same object for a particular line. The x-axis plots the S/N of that observation in the eBOSS-DAP. The 1-sigma spread of the data is plotted in pink points, which can be visually compared to the expected value at a specific S/N, shown in orange and green lines at the 1-sigma and 2-sigma levels, respectively. The average offset for lines with S/N $>3$ has been given for each line in the lower right of each panel. For access to summary tables for this figure see \hyperref[sec:data_availability]{Data Availability}
\label{fig:trumpet_sdss}}
\end{figure*}

To compare the two sets of emission line measurements, we first corrected the {\tt spZline} measurements for Milky Way dust attenuation using the same process as the eBOSS-DAP.
We then plot the flux of an emission line found by the eBOSS-DAP divided by the flux in the {\tt spZline} catalog and compare it to the S/N of that line, as shown in Figure \ref{fig:trumpet_sdss}. 
Random variations are expected to produce a shape in which the standard deviation at any S/N is given by $\sigma \simeq \frac{\sqrt{2}}{\mathrm{S/N}}$. In Figure \ref{fig:trumpet_sdss}, we compare the strong optical emission line fluxes of [O~II]~3726,3729, H$\beta$, [O~III]~5008, H$\alpha$ and [N~II]~6584, and [S~II]~6718,6733. For most lines, the standard deviation of the flux measurements (pink points) decreases with S/N as predicted (orange lines), demonstrating the expected level of agreement between the eBOSS-DAP and the {\tt spZline} catalog. 

 H$\beta$ shows the most prominent difference in the two line flux measurements, being 0.128 dex higher on average in the eBOSS-DAP. This can be attributed to the fact that H$\beta$ lies in a strong stellar absorption feature, which is modeled differently in the two codes. The SDSS code uses a Principal Component Analysis (PCA) approach and the ELODIE empirical stellar library \citep{ELODIE2001} to fit the stellar continuum \citep{Bolton:2012}. The ELODIE library is known to be incomplete, especially for very hot stars and at non-solar metallicities \citep{Maraston2011}. In addition, PCA spectral reconstruction can sometimes produce non-physical spectra since the eigenvectors summed to create the final spectrum can be added or subtracted. Thus, it is likely that the eBOSS-DAP, with its non-negative combination of modern stellar population synthesis models (see \S\ref{subsubsec:C3K}), provides an improved stellar continuum fit and a more accurate H$\beta$ flux. 

 Additionally, H$\alpha$ shows a slight negative preference in the low S/N regime, while [N~II] 6585 shows a slight positive preference. This is likely due to SDSS assigning flux from H$\alpha$ to [N~II] 6585 in low S/N data. [S~II]~6718,6733 also has a slight positive preference, meaning that the eBOSS-DAP assigned it more flux in low S/N data than SDSS did.

\subsubsection{Repeat Spectra Comparisons}\label{subsubsec:trumpets}

\begin{figure*}[!ht]
\centering
\includegraphics[width=0.8\textwidth]{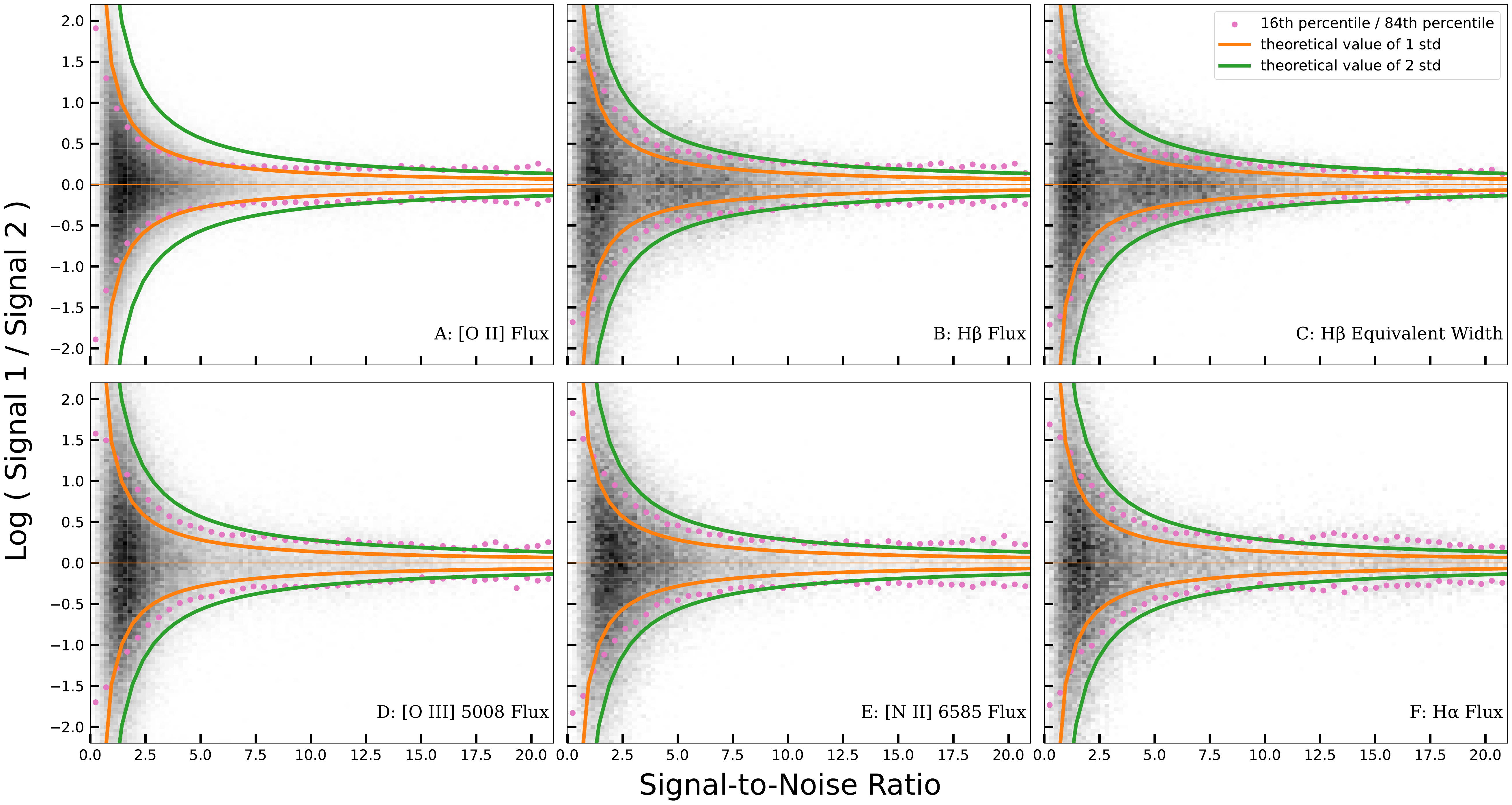}
\caption{ A comparison of line fluxes and EWs measured by the eBOSS-DAP in repeat galaxy spectra as a function of the emission line S/N. The data are shown in grayscale; pink dots show the 1-$\sigma$ width of the distribution. Orange and green lines show the expected 1-$\sigma$ and 2-$\sigma$ spread of the data given the line S/N. Good agreement between the pink points and orange lines indicates that the errors on the emission line measurements are correct. {\bf Panels A, B, D - F:} These panels compare fluxes of five relatively strong emission lines. At high S/N errors on the lines appear to be underestimated; this is a consequence of additional spectrophotometry errors; see \S\ref{subsubsec:trumpets}.
{\bf Panel C:} This panel compares the variation in the H$\beta$ equivalent width rather than flux. This variation is smaller than in the H$\beta$ flux, as we expect EWs to be unaffected by spectrophotometric errors. For access to summary tables for this figure see \hyperref[sec:data_availability]{Data Availability}
\label{fig:trumpet_strong}}
\end{figure*}

\begin{figure*}[!ht]
\centering
\includegraphics[width=0.8\textwidth]{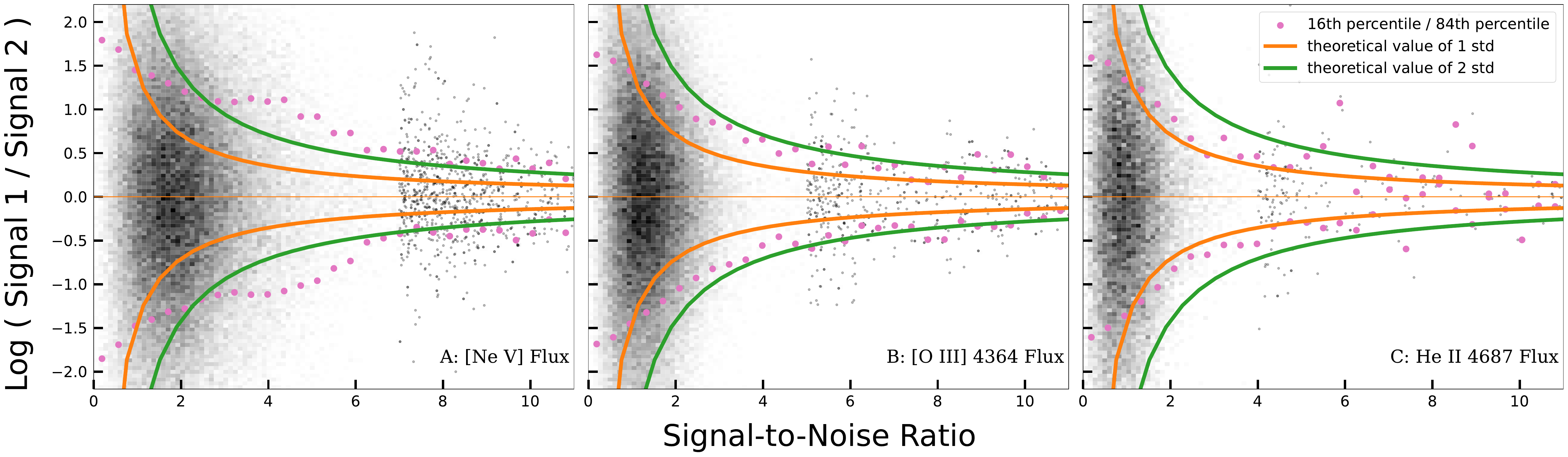}
\caption{
A comparison of line fluxes measured by the eBOSS-DAP for three weak emission lines: [Ne~V]~3426, [O~III]~4364, and He~II~4687. These lines are only measured at S/N$>5$ in $\sim1$\% of galaxies. The data is shown in grayscale and plotted as individual points above S/N $= 7, 5, 4$ in panels A, B, and C, respectively. The colored lines are the same as in Figure \ref{fig:trumpet_strong}. The discrepancy between the observed (pink points) and predicted (orange line) spread of the data at S/N = 2-6 in [Ne~V] is likely due to spurious line detections at very blue wavelengths; see \S\ref{subsubsec:trumpets}. For access to summary tables for this figure see \hyperref[sec:data_availability]{Data Availability}
\label{fig:trumpet_weak}}
\end{figure*}

Our final quality assessment was to make use of the \ndupes\ repeat spectra that eBOSS took to test consistency in measurements. In Figures \ref{fig:trumpet_strong} and \ref{fig:trumpet_weak} we plot the flux (or equivalent width) of an emission line divided by its value in a repeat observation and compare this ratio to the average S/N of the two observations.
In Figure \ref{fig:trumpet_strong}, we show the strong optical emission line fluxes of [O~II]~3726,3729, H$\beta$, [O~III]~5008, H$\alpha$ and [N~II]~6584. The standard deviation of the repeat data (pink points) decreases with S/N as predicted (orange lines), but shows some deviations at high S/N. 
This is due to variations in the spectrophotometric calibration between the repeat spectra (see \S\ref{subsec:Spectra_qual}.) 
Both statistical error and variations in spectrophotometric calibration drive differences in the fluxes measured in repeat spectra. 
In the low S/N regime, emission line differences are dominated by statistical error, which is well captured by the eBOSS-DAP (the pink 1-$\sigma$ values of the data line up well with the orange theoretical lines).
However, in the high S/N regime, the spectrophotometry errors become dominant, and then the difference between the repeat spectra becomes larger than our measured errors (the pink 1-$\sigma$ values of the data lie outside the orange theoretical lines).
Notably, equivalent widths are free from spectrophotometry errors. 
Consequently, the H$\beta$ EW differences (Panel C) show improved agreement of the pink points and orange lines at high S/N, although the errors are still slightly underestimated.

In Figure \ref{fig:trumpet_weak}, we show a comparison of the fluxes from repeat spectra of three intrinsically weak emission lines: [Ne V]~3426, [O~III]~4364, and He~II~4687. These lines are only detected at S/N$>5$ in $\sim1$\% of the galaxies with repeat spectra, making our comparison statistics more marginal.
Nonetheless, we observe some differences from Figure~\ref{fig:trumpet_strong}: for [Ne~V] (and to a lesser extent [O~III]~4364) at S/N$\simeq2-6$ the pink points deviate from the orange lines, suggesting that the measured error of the line fluxes substantially underestimates the true error. 
Errors in the continuum fit can impact low EW lines such as these; however, the eBOSS-DAP fits the continuum simultaneously with the emission lines, and therefore, the associated error is included in the flux errors. Therefore, this is unlikely to be the source of the discrepancy. 
Notably, the strong emission lines in Fig.~\ref{fig:trumpet_strong} show reasonable agreement between the predicted and actual errors at S/N = $2-6$. For these weak lines, we suspect that rare spurious detections may outnumber real line detections in this S/N regime. The wavelength of [Ne~V]~3426 also means that the line is commonly observed at the blue edge of the spectrograph, where the spectrophotometry errors are substantial (see \S\ref{subsec:Spectra_qual}). 
Users of the catalog are advised to exercise caution when conducting science with these and other weak lines and to utilize repeat spectra to better understand the appropriate detection threshold.

Unfortunately, it is not possible to accurately report the spectrophotometry error for an individual line flux measurement, as this depends on a complex interplay of plate drilling errors, telescope guiding errors, the airmass and seeing at the time of observation, errors in spectral typing the standard stars on the plate, etc. However, we used the repeat spectra to compute a scale factor that can be applied to the reported uncertainty of each line to approximately account for the additional uncertainty arising from spectrophotometric errors. These values are computed from the ratio of the measured errors to the 1-$\sigma$ dispersion of the repeat spectra (e.g., the ratio of the pink points to the orange lines) for emission lines with S/N $>3$. Specifically, we fit $\sigma_{\mathrm{measured}} = \sqrt{(\sqrt{2}/\mathrm{(S/N)})^{2} + (\sqrt{2}\,a)^{2}}$, where $a$ is the scale factor reported in Table \ref{table:error_scale}. In Table \ref{table:error_scale}, we list this recommended scale factor for some commonly used emission lines.  We also report the scaling that should be applied to the errors of various common line ratios. These correction factors have lower values than for the line fluxes because they are only sensitive to changes in the shape of the spectrum rather than its zero-point. Readers are referred to \S\ref{subsec:Spectra_qual} for additional details on the spectrophotometric calibration.

\begin{table}[ht]
\centering
\begin{threeparttable}
    \caption{Uncertainty Scaling}
    \label{table:error_scale}
    \begin{tabular}{|| c c ||}
        \hline
         & Recommended \\
        Name & Uncertainty Scaling\\ [0.5ex]
        \hline
        {[Ne V] 3427} & 0.114 \\
        {[O II]~3727, 3729} & 0.097 \\
        {[Ne III] 3869} & 0.129 \\
        {[O III] 4364} & 0.138 \\
        {He II 4687} & 0.188 \\
        H$\beta$ & 0.151 \\
        {[O III] 4960} & 0.143 \\
        {[O III] 5008} & 0.143 \\
        He I 5877 & 0.123 \\
        {[O I] 6302} & 0.115 \\
        H$\alpha$ & 0.187 \\
        {[N II] 6585} & 0.159 \\
        {[S II] 6718} & 0.163 \\
        {[S II] 6732} & 0.159 \\
        log (({[O II] 3727 + 3729}) / {[O III] 5008}) & 0.077 \\
        log ({[O III] 5008} / H$\beta$) & 0.061 \\
        log ({[N II] 6585} / H$\alpha$) & 0.065 \\
        H$\alpha$ / H$\beta$ & 0.08 \\
        $\sigma_\ast$ & 1.29 \\

        \hline
    \end{tabular}
    \begin{tablenotes}
        \small
        \item {Note: For line fluxes, add the scale factor $a$ times the flux in quadrature with the reported flux error, such that $\sigma_{\mathrm{corrected}} = \sqrt{\sigma_{\mathrm{eBOSS-DAP}}^{2} + (F_{\mathrm{eBOSS-DAP}}\times a)^{2}}$. The scaling on $\sigma_\ast$ is multiplicative rather than additive: $\sigma_{\mathrm{\ast,corrected}} = \sigma_{\mathrm{\ast,eBOSS-DAP}} \times a$. See \S\ref{subsubsec:trumpets} for how these values were derived.}
    \end{tablenotes}
\end{threeparttable}
\end{table}

\begin{figure}[!ht]
\plotone{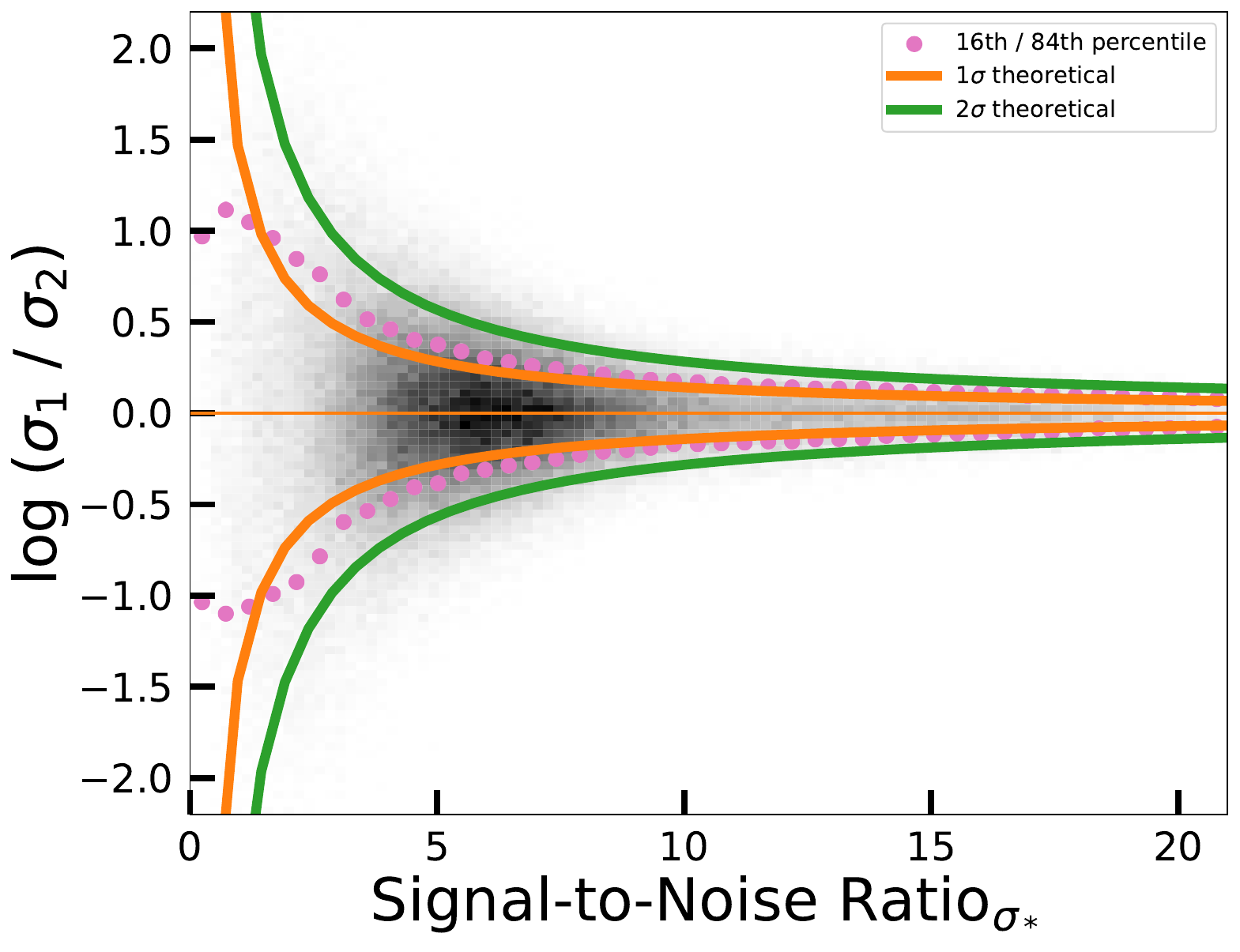}
\caption{
A comparison of the stellar velocity dispersions measured by the eBOSS-DAP in repeat spectra. The color coding is the same as in Figure \ref{fig:trumpet_strong}. The reported velocity dispersion errors require a scaling correction of 1.29 to match the observed variation in the pink dots (see \S\ref{subsubsec:trumpets}).
\label{fig:trumpet_disp}}
\end{figure}

Our final validation step was to assess the quality of our stellar continuum kinematics, in particular the stellar velocity dispersion $\sigma_{*}$, by comparing measurements made from repeat spectra. 
Figure~\ref{fig:trumpet_disp} shows that the reported kinematic errors explain the variation in the pairs, requiring only a scaling of 1.29 on the error. Unlike the line-flux scale factors, this $\sigma_*$ scaling is multiplicative: we fit $\sigma_{\mathrm{measured}} = \sqrt{2}/\mathrm{(S/N)} \times a$, where $a = 1.29$. The stellar velocity dispersion is measured with an error of less than 50 \kms\ for 82.5\% of the sample.

\subsection{Sample Characterization}\label{subsec:characterization}

As part of this investigation, we have made a slate of figures of the data to provide a clear picture of the make-up of the eBOSS-DAP catalog and illustrate the catalog's possible uses: a mass--stellar velocity dispersion diagram, a BPT diagram, a star formation rate (SFR) vs stellar mass diagram, a mass--metallicity diagram, and a H$\delta_{A}$ vs Dn4000 diagram.

The stellar masses are from a catalog of SED-derived parameters produced by D.~Miller et al., in prep, using the Code for Investigating GALaxy Emission \citep[CIGALE;][]{CIGALE}. The inputs to the SED fitting are the SDSS redshift, the Legacy Survey $gri$ photometry \citep{Dey:2019}, the WISE
\citep{Wright:2010} W1-W3 band photometry and the Dn4000 index measured by the eBOSS-DAP. 
The SEDs were computed by using a grid of parametric star-formation histories (a delayed $\tau$ model plus a burst), \citet{sps} SSP models with a Chabrier IMF \citep{chabrier_IMF}, nebular emission, a \citet{Calzetti} dust attenuation law, and a \citet{Draine} dust emission law. Typical errors on the stellar masses are $\pm$0.07 dex. 

\subsubsection{Mass--\texorpdfstring{$\sigma_\ast$}{sigma-*}}\label{subsubsec:mass-sig}
\begin{figure}[!ht]
\plotone{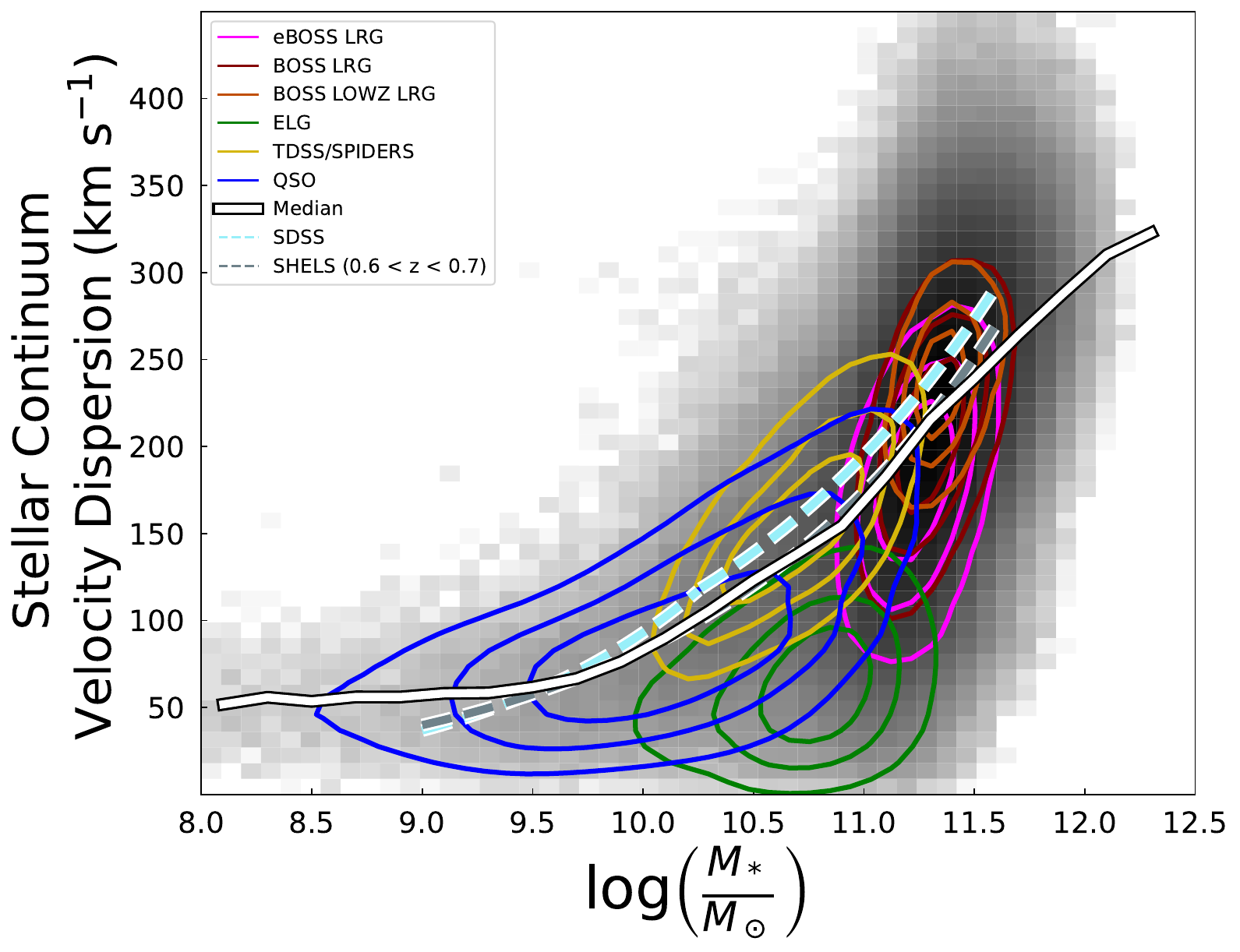}
\caption{Stellar velocity dispersion vs.\ stellar mass for eBOSS galaxies with a stellar velocity dispersion error $< 50$~\kms.
The grayscale is a 2D log-scaled histogram of the full sample.
The contours show the different target classes; the outermost contour for each class encloses 75\% of the total sample.
The median relation is shown in white.
The dashed lines show the median relation for the SDSS-I sample (light blue) and a high-$z$ SHELS sample (gray) found in \citet{zahid}.
The SDSS-I sample has a redshift range $0.02 < z < 0.2$, while the SHELS sample has a higher range ($0.6 < z < 0.7$).
The eBOSS sample has a redshift range of $0.0005 < z < 1.12$ with a peak at $z = 0.55$.
\label{fig:m-vdisp}}
\end{figure}

In Figure \ref{fig:m-vdisp} we show a plot of the stellar continuum velocity dispersion measured by the eBOSS-DAP versus stellar mass, M$_{*}$, measured from the photometry with CIGALE (D.\ Miller et al, in prep.).
The eBOSS-DAP yields 1,449,940 stellar velocity dispersion measurements with error $< 50$~\kms, or 76\% of the sample.
Included in Figure \ref{fig:m-vdisp} are best-fit lines to the $0.6 < z < 0.7$ Smithsonian Hectospec Lensing Survey \citep[SHELS;][]{shels1, shels2, shels3} sample and the $0.02 < z < 0.2$ SDSS main galaxy sample \citep{SDSS-MGS} found by \citet{zahid}.
The median relation for eBOSS is shown as a white line over the same mass range as these two samples ($9.0 < \mathrm{Log}(\frac{\mathrm{M}_{*}}{\mathrm{M}_{\odot}}) < 11.5$).
Our data, which peaks at $z \simeq 0.55$, broadly agree with both the low-$z$ SDSS sample and the higher-redshift SHELS sample. The eBOSS ELGs (shown in green) fall somewhat below the mean relation.  These $z\sim0.8$ star forming galaxies were selected based on their blue colors,  which may have favored galaxies with small bulges and face-on orientations, leading to lower velocity dispersions.  
As an additional check, we compare the stellar velocity dispersion to the H$\beta$ emission-line width in Appendix~\ref{sec:add_plots_two} (Figure~\ref{fig:hb-sc-disp}). The two measurements agree well in the mean across all target classes with the notable exception of the ELGs, which have stellar velocity dispersions $\sim$25 \kms\ smaller than their gas velocity dispersions, on average.  The ELGs have the lowest continuum S/N of any of the target classes and it is unclear whether this difference is a systematic error or real physical effect.  Notably, galaxies targeted as QSOs also have blue colors, but they tend to be compact and brighter than the ELGs; they do not show any discrepancies in either Figure~\ref{fig:m-vdisp} or Figure~\ref{fig:hb-sc-disp}.

\subsubsection{BPT}\label{subsubsec:BPT}
\begin{figure}[!ht]
\plotone{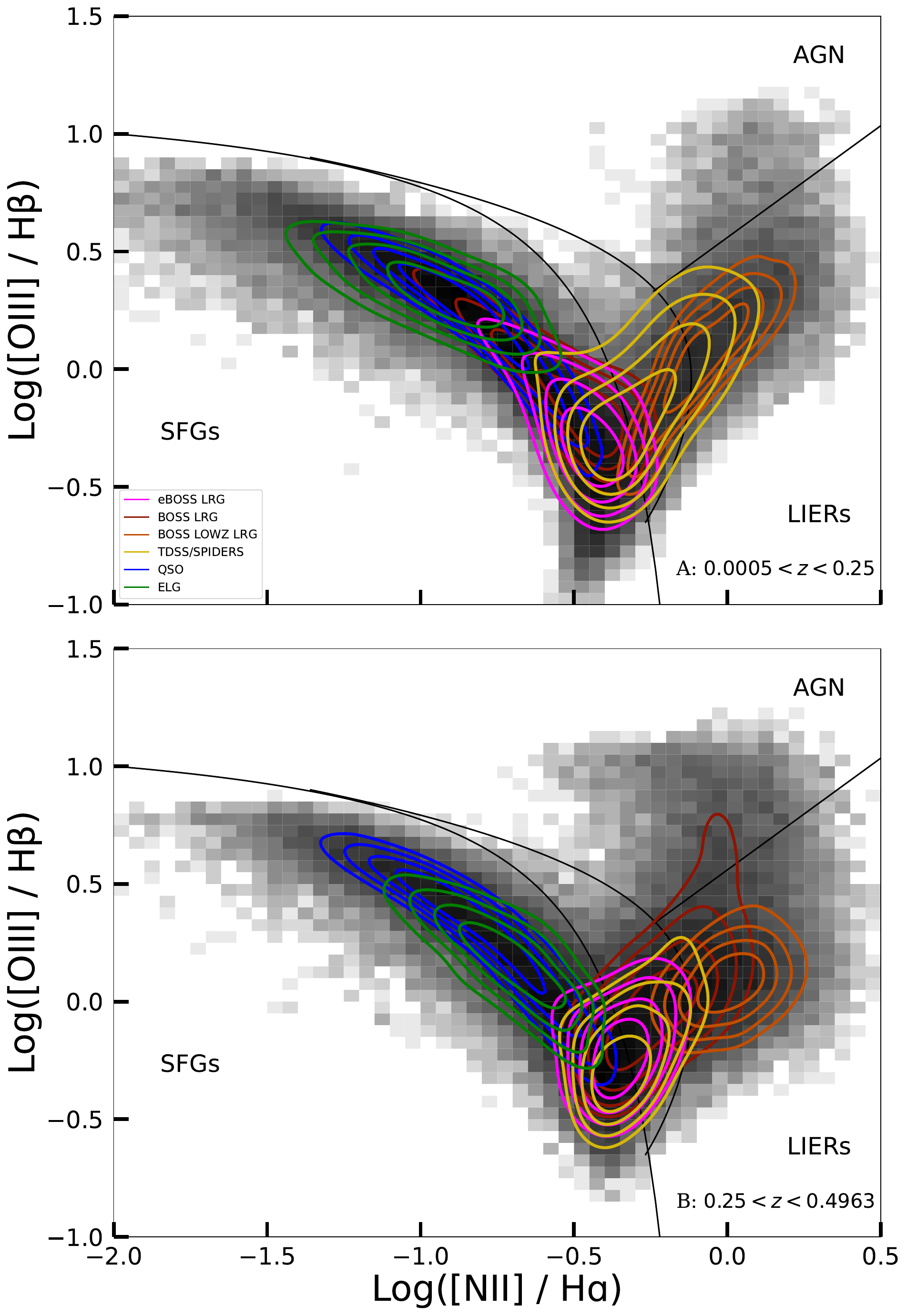}
\caption{BPT diagram in two redshift ranges ($0.0005 < z < 0.25$ and $0.25 < z < 0.4963$). The eBOSS data are shown as 2D log-scaled histograms. Contours show the location of different target classes; they begin at 50\% of the data. The diagram is divided into regions representing different ionization sources following \citet{LawBPT}. `AGN' are Active Galactic Nuclei, `SFGs' are Star-Forming Galaxies, `LIERs' are Low Ionization Emission Regions, and finally, the unlabeled region between SFGs and AGN is composite galaxies. 
\label{fig:BPT_kde}}
\end{figure}

Our next example figure is the Baldwin, Phillips, and Terlevich (BPT) diagram \citep{OG_BPT} using the updated classification lines from \citet{LawBPT}. 
The BPT uses a combination of $\log (\frac{\mathrm{[O~III]~5008}}{\mathrm{H\beta}})$ (R3) and $\log(\frac{\mathrm{[N~II]~6584}}{\mathrm{H\alpha}})$ (N2) to identify the origin of emission lines in galaxies. 
It is typically understood that N2 acts as a proxy for metallicity and R3 acts as a proxy for ionization level.
Star-Forming Galaxies (SFGs) span a broad range of metallicity and ionization parameter, but the two are generally anti-correlated; e.g., high metallicity galaxies have low ionization parameter \citep{SFG-anticorr}.

AGN are typically found in massive metal-rich galaxies, and they can be a powerful source of hard ionizing photons, leading to high ionization parameters.
Thus, the presence of an AGN will increase the R3 ratio as hard ionization from AGN will increase [OIII] \citep{kewleyN2}. 
Additionally, as AGN emission lines trace the Narrow Line Region (NLR) rather than interstellar medium (ISM) gas, [N~II] emission will be much higher in AGN than in SFGs.
This is because the NLR is far denser and hotter than the gas in the ISM, leading to much more collisional excitation, which [N~II] is particularly sensitive to.
As such, AGN will lie in the high metallicity (high N2) and high ionization (high R3) region, opposite from Star-Forming Galaxies (SFGs).
Additionally, there are two other classifications: composite galaxies, which are a mix of AGN-like and SFG-like emission, and Low-Ionization Emission Line Regions (LIERs).
LIERs are galaxies that see highly elevated N2 with low R3.
It is unknown what mechanism drives emission in LIER galaxies; they exhibit AGN-like N2 levels, but this emission is sometimes detected far from the nuclear region, leading to debate about the cause of this radiation \citep{BelfioreLIER}.

We first trimmed the sample down such that all BPT lines (H$\beta$, [O~III]~5008, H$\alpha$, and [N~II]~6584) were at S/N $>$ 3, and then trimmed to a redshift range of $0.0005 < z < 0.4963$. 
We chose this redshift cut-off to prevent any possible systematic errors from switching the line tying from H$\alpha$ to H$\beta$ (see \S\ref{subsubsec:linelist} for details).
This resulted in \BPTpass~ spectra or $\sim$ 12.4\% of the sample in that redshift range. In Figure \ref{fig:BPT_kde} we show the BPT diagram for this sample for redshifts below and above $z=0.25$. 
As shown in Fig. \ref{fig:BPT_kde}, the different target classes lie primarily in specific parts of the BPT diagram, with ELGs and compact blue galaxies mis-targeted as quasars (QSOs) lying primarily in the star-forming region, and the other target classes having some level of AGN contribution.
Of note is that, in higher redshift bins, LRG classes are more likely to host AGN.

The BPT is usable up to $z = 0.5459$, at which point H$\alpha$ is redshifted out of the observable window for eBOSS. Alternative AGN classification diagrams, such as MEx \citep{MEx} or OH-NO \citep{OHNO}, can be used at higher redshifts, although they suffer from reduced completeness and purity \citep{cleri2025}. 

\subsubsection{Star-Formation Rate vs Stellar Mass}\label{subsubsec:sfr-vs-M}
\begin{figure}[!ht]
\plotone{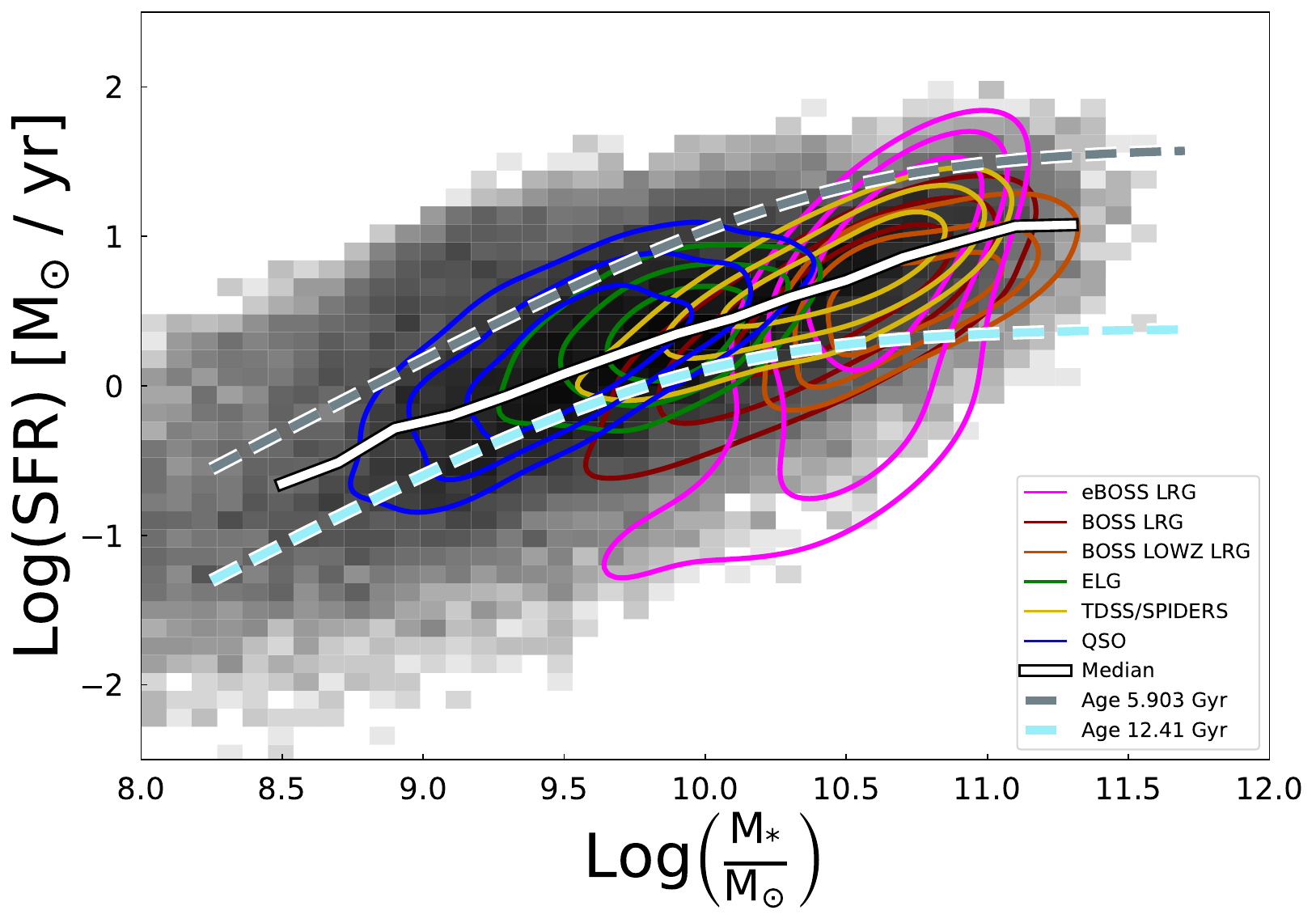}
\caption{The Star-Formation Rate (SFR; found through H$\alpha$ luminosity) vs stellar mass (M$_{*}$; found with CIGALE) diagram for star-forming galaxies at $0.0005 < z < 0.4963$. The eBOSS data for BPT-selected star-forming galaxies with S/N $> 3$ [O III]~5008 and [N~II]~6584 lines and S/N $> 5$ H$\beta$ and H$\alpha$ lines is shown as a 2D log-scaled histogram (grayscale). Colored contours show the location of different target classes; contours begin at 50\% of the data.
The median line has been plotted in white; orange and blue lines show the star-forming main sequence from \citet{m-sfr} at ages of 12.41 Gyr($z=0.1$) and 5.903 Gyr($z=1$), respectively. 
\label{fig:SFR}}
\end{figure}

Our third diagram is a Star-Formation Rate (SFR) versus stellar mass diagram, which is commonly used to identify star formation modes in galaxies: galaxies on the `main sequence' are forming stars steadily, those above it are in a starburst phase, while those below may be quenching or already passive.
Additionally, the star-forming main sequence evolves over redshift, moving towards higher average star-formation rates for galaxies at higher redshifts \citep[c.f.,][]{m-sfr}.
To make this diagram, we require [O III]~5008 and [N~II]~6584 S/N $> 3$ and H$\beta$ and H$\alpha$ S/N $> 5$. The Balmer lines have a higher S/N cut as they are used to compute the extinction correction of the spectra before determining the star formation rate.
We then trim the sample to include only galaxies that were classified as a BPT SFG using the \citet{LawBPT} lines.
This resulted in 44,406 galaxies or $\sim6\%$ of the sample in this redshift range ($0.0005 < z < 0.4963$), or $\sim47\%$ of galaxies that pass the BPT cuts.

We compute the extinction corrected H$\alpha$ luminosity by first computing E(B - V) from $\frac{\mathrm{H}\alpha}{\mathrm{H}\beta}$.
We then use the same {\tt dust\_extinction} package to find the correction and apply it to the H$\alpha$ flux. 
The H$\alpha$ flux was then converted to luminosity and corrected for the fiber filling fraction (see \S\ref{subsec:appature_bias} and Fig.\ref{fig:infiber}).
We then calculate star formation rates for this subsample using the extinction corrected H$\alpha$ luminosity and the \citet{sfr} calibration. 
We plot the 2D log-scaled histogram of the SFRs versus the masses of the subsample, as well as the star-forming main sequence from \citet{m-sfr} for ages of 5.903 Gyr ($z=1$) and 12.41 Gyr ($z=0.1$) in Figure \ref{fig:SFR}.

Our median lies comfortably between the two star-forming main sequences, chosen by finding the ages of the universe at the maximum and minimum redshifts of our sample. 
Additionally, we find that galaxies targeted as ELGs and QSOs have characteristically lower masses than the LRG targets, and they are more likely to be on the main sequence. 

Of note is that we only show BPT SFGs in this figure. This is because we use H$\alpha$ as our star formation rate indicator, and it is not a reliable indicator for BPT LIERs and AGN.
These two populations typically fall below the main sequence (c.f., D.\ Miller et al., in prep.).

\subsubsection{Mass--Metallicity Relation}\label{subsubsec:MZR}
\begin{figure}[!ht]
\plotone{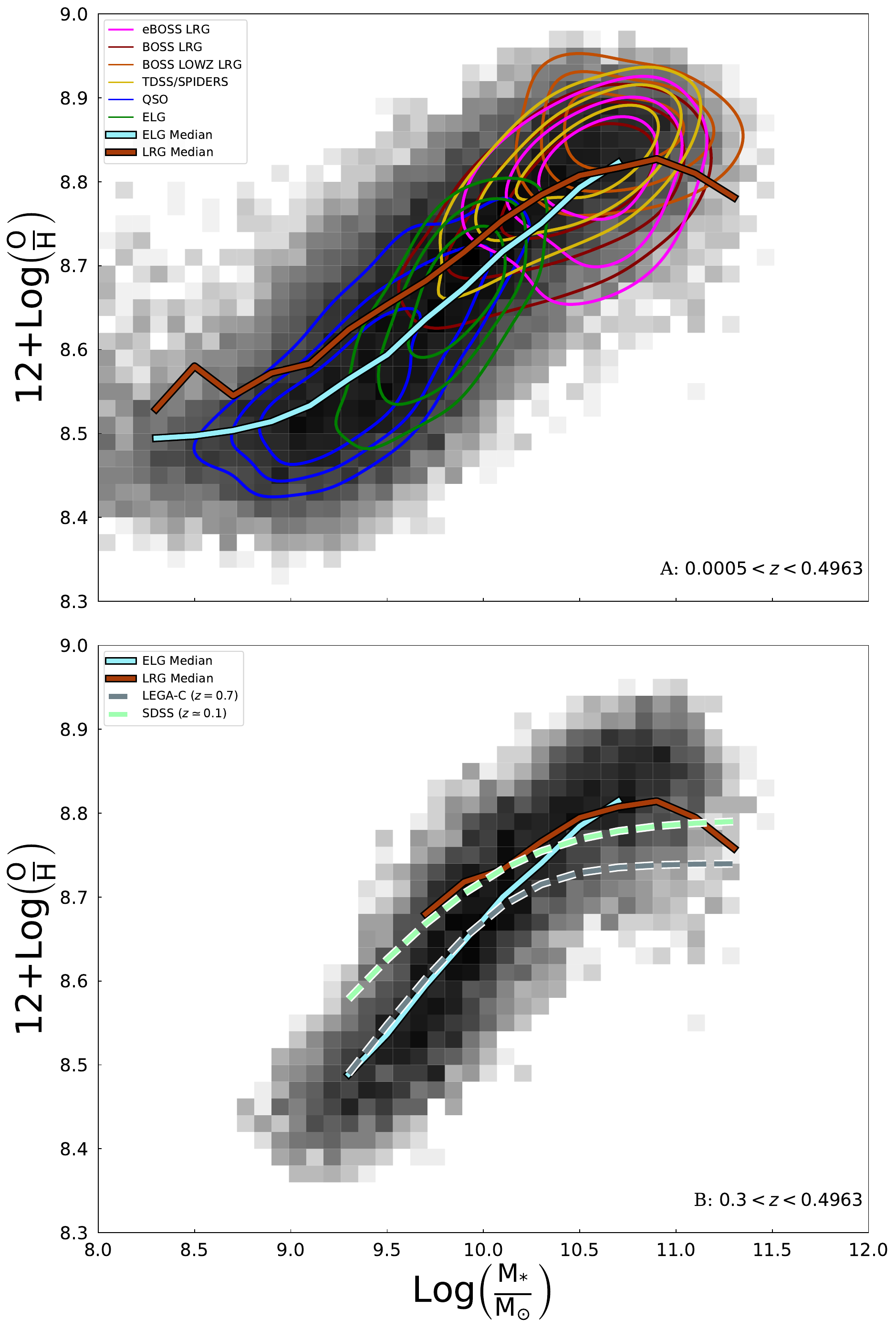}
\caption{The mass--metallicity relation for eBOSS galaxies. This figure follows the plotting scheme of Fig. \ref{fig:SFR}. Metallicities are determined using the empirical calibrations of \citet{nakajima2022}, which employ the R23 calibration and the O32 parameter to break the double-branching. \textbf{Panel A}: The data in the redshift range $0.0005 < z < 0.4963$ is shown in grayscale, with contours representing different target classes.  \textbf{Panel B}: The data is trimmed to the redshift range $0.3 < z < 0.4963$ to limit the redshift evolution and remove cases of H~II regions classified as galaxies (see \S\ref{subsec:known_errs}). In both panels, the rust-red solid lines denote the medians for LRG and TDSS/SPIDERS spectra, whereas the light-blue solid lines denote the medians for ELG and QSO spectra. For comparison, we plot the MZR found in LEGA-C for $z \sim 0.7$ by \citet{Zach} in a gray dashed line and the SDSS-I MZR at $z \simeq 0.1$ found by \citet{z=0MZR} in a green dashed line. The differences in trends between LRGs and ELGS highlight the importance of sample selection. \label{fig:MZR}}
\end{figure}

Our penultimate sample characterization is the stellar mass -- gas-phase metallicity relation (MZR).
The MZR is extremely tight ($\pm0.1$ dex) at $z\sim0.1$ \citep{mass-metal_1}, and it has been shown to evolve mildly with redshift \citep{Zahid2014, Zach}. 
Galaxies classified as AGN according to the \citet{LawBPT} BPT diagram are removed from the sample, as are galaxies with [O II] 3727, 3729, H$\beta$, [O~III] 5008, H$\alpha$, and [N~II]~6584 lines with S/N $< 3$. 
These cuts yield 46,098 spectra, corresponding to roughly 73\% of the 63,182 BPT SFGs.
Metallicities are calculated according to the log($\frac{\mathrm{[O~II]~3727,3729 +[O~III]~5008}}{\mathrm{H\beta}}$) (R23) diagnostic calibration of \citet{nakajima2022}, and the double-branching is broken using the excitation parameter log($\frac{\mathrm{[O~III]~5008}}{\mathrm{[O~II]~3727,3729}}$) (O32).

We plot our findings in Fig.~\ref{fig:MZR}.
Panel A shows the MZR for our sample in the redshift range $0.0005 < z < 0.4963$ along with target classification contours.
Here we see that LRGs have both higher masses and higher metallicities than ELGs and QSO targets.
We grouped LRGs and TDSS/SPIDERS spectra (metal-rich, redder galaxies), as well as ELGs and QSOs (metal-poor, bluer galaxies), to find mass--metallicity relations for both broad classes of objects.
These relations are observed to exhibit the expected correlation, including both the power-law rise at low stellar masses in ELGs and the metallicity saturation and plateau at higher stellar masses in LRGs with an average $\sigma_{\mathrm{ELG}} \simeq 0.101$ dex and $\sigma_{\mathrm{LRG}} \simeq 0.108$ dex. ELGs show lower metallicity at a given mass than LRGs for all but the highest masses. At $10^{9.5}~\textrm{M}_{\odot}$ , the metallicity difference is 0.06 dex, highlighting the importance of sample selection.

For Panel B we trimmed the sample further to the redshift range $0.3<z<0.4963$ resulting in 13,866 spectra. This was done to avoid cases of H~II regions classified as galaxies (see \S\ref{subsec:known_errs}), which cause the metal-rich but low-mass outliers seen in Panel A. (These galaxies have artificially low stellar masses due to deblending problems in the Legacy Survey photometry see \S\ref{subsec:known_errs}.)
Another benefit of the trim was the ability to restrict the redshift evolution of the MZR, which is a major source of the scatter in Panel A. 
As seen in Panel B, the correlation of the MZRs is tighter in this redshift range with an average $\sigma_{\mathrm{ELG}} \simeq 0.09$ dex and $\sigma_{\mathrm{LRG}} \simeq 0.101$ dex.

For comparison, in Figure~\ref{fig:MZR} panel B we plot the median MZR of the SDSS-I sample ($z \simeq 0.1$) found by \citet{z=0MZR} and the median MZR of the LEGA-C survey ($z = 0.7$) found by \citet{Zach}.  The LEGA-C sample is K-band selected and relatively mass-complete \citep{vanderWel2016}. Both samples use very similar metallcity measurements to ours, mitigating strong line abundance calibration systematics.   Given the redshift of the eBOSS sample used in this comparison ($z\sim0.4$), we expect the median MZR of the ELGs and LRGs to lie between those of the SDSS-I ($z\sim0.1$) and LEGA-C ($z\sim0.7$) samples. However, this is not what we find. The LRGs show a fairly good match to the SDSS-I MZR over most of the mass range with a slightly elevated metallicity at $\log(M_*) = 10.5 - 11$. The LRGs were selected for their large 4000-\AA\ breaks and red colors, a selection which favors older and dustier star-forming galaxies. Both stellar population age and dust attenuation may have residual correlations with metallicity, accounting for the fact that $z \sim0.4$ eBOSS LRGs have an MZR more similar to $z\sim0.1$ galaxies. In contrast, the eBOSS ELG MZR matches that of the $z\sim0.7$ LEGA-C sample until about $10^{10}~\textrm{M}_{\odot}$ after which the median metallicity becomes higher than that of LEGA-C and even exceeds the median metallicity of SDSS-I galaxies in the most massive bin. The ELGs were selected based on their blue colors, which may be linked with less dust and metals and higher rates of star formation due to recent gas inflows, causing them to better match the MZR of high-redshift galaxies. The higher metallicities seen above $\log(M_*)$ are more difficult to explain and may have to do with survey volume differences between LEGA-C and eBOSS.  In any case, the impact of sample selection effects on the MZR is evident.  A more thorough   
Analysis of these differences will be the subject of future work.

\subsubsection{H\texorpdfstring{$\delta_{A}$}{delta-a} vs Dn4000}\label{subsubsec:hdela-vs-dn4000}
\begin{figure}[!ht]
\plotone{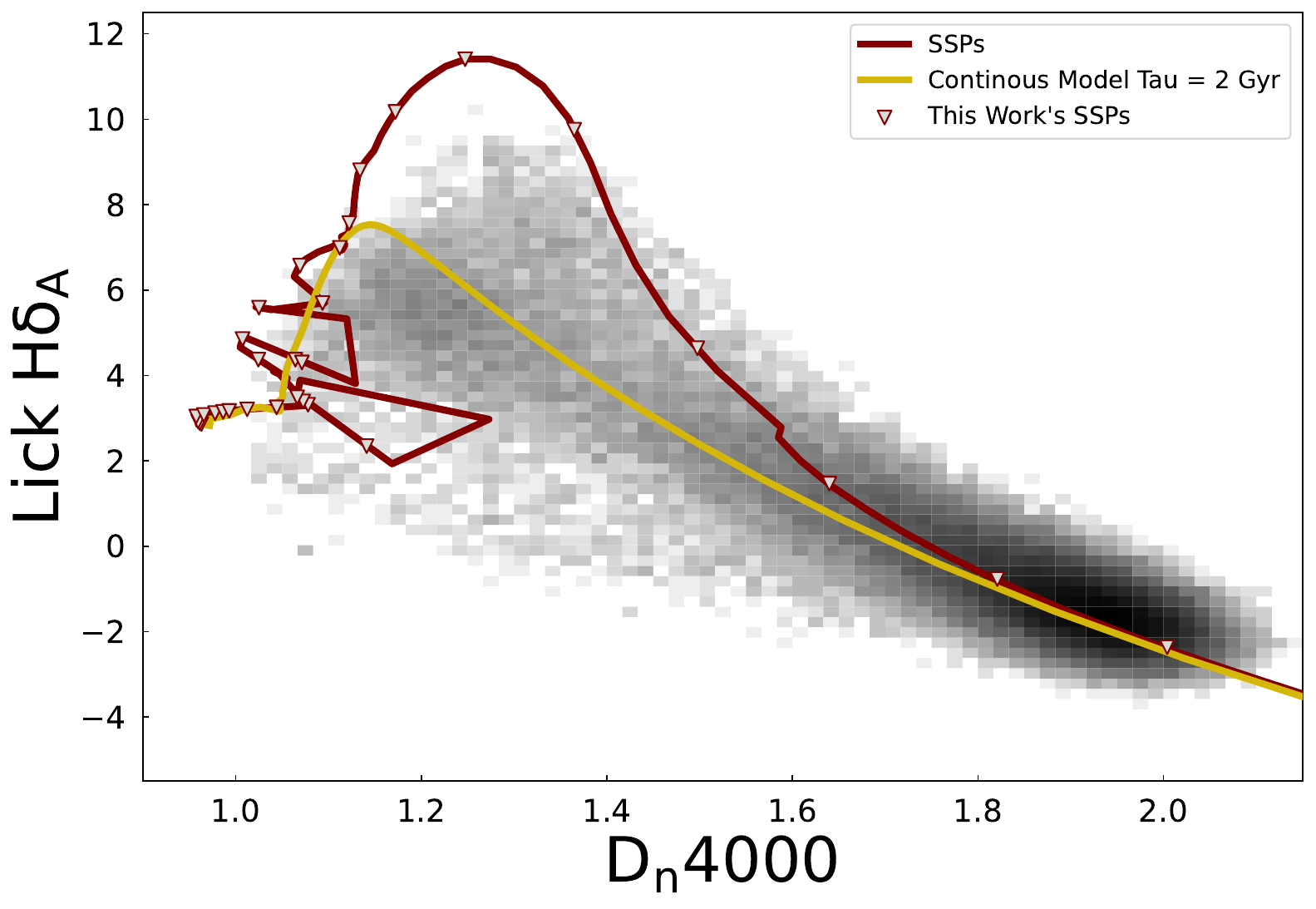}
\caption{A spectral index diagram of H$\delta_{A}$ vs Dn4000. Spectra that pass the H$\delta_{A}$ error $< 0.8$ and Dn4000 error $< 0.03$ cuts from \citet{spind_cut} (approximately 3.3\% of the sample) are shown as a 2D log-scaled histogram (grayscale). Overplotted is an evolutionary sequence for a continuous star formation sequence (yellow) as well as a simple stellar population (dark red). The red triangles denote the exact values of the solar-metallicity SSPs used in this work. 
\label{fig:spind}}
\end{figure}

Our final example is the H$\delta_{A}$ versus Dn4000 spectral indices plot (Fig. \ref{fig:spind}). 
To start, we applied the cuts found in \citet{spind_cut} of H$\delta_{a}$ error $< 0.8$ and Dn4000 error $< 0.03$.
We then removed spectra with unphysical index values and limited the redshift to $0.05 < z < 0.8$, resulting in 65,184 spectra or merely $\sim3.4$\% of our full sample.  The majority of eBOSS galaxies have stellar continua that are  too low S/N for the line indices to be measured accurately.

By comparing the Dn4000 and H$\delta_A$ indices, we gain insight into the average stellar populations of this relatively bright sub-sample of eBOSS galaxies. Dn4000 traces the 4000-\AA\ break and indicates the mean age of the stellar population; a low Dn4000 value suggests a young stellar population, while a high Dn4000 value indicates older stars. The H$\delta_{A}$ index traces the strength of stellar Balmer absorption, which peaks at ages of a few 100 Myr, when A-stars dominate the spectrum.

In Figure \ref{fig:spind}, we overplot the Dn4000 and H$\delta_{a}$ values of two C3K stellar population models: Solar metallicity SSPs and a continuous star formation model in red and yellow, respectively.
A continuous star formation history model is a good match to the bulk of our data, but some galaxies with extreme Balmer absorption lie closer to the SSP models; such galaxies are likely post-starburst galaxies \citep{French2021}. Many of the galaxies that lie well below the continuous models appear to be blazars: early-type galaxies with an additional power law continuum contribution arising from a relativistically beamed jet. 

\section{Data Use Recommendations} \label{sec:DUC} 
We include a \textit{Recommended Usage Guidelines} table (see Table \ref{table:recommended_usage}) that summarizes best practices for the scientific use of the eBOSS-DAP catalog. While specific science goals may demand more specialized criteria, these recommendations offer a general foundation for reliable use of the catalog.

{\renewcommand{\arraystretch}{1.5}
\begin{table*}[htbp]
\centering
\caption{Recommended Usage Guidelines for the eBOSS-DAP Catalog}
\label{table:recommended_usage}

\begin{tabular}{|| p{0.12\textwidth} | p{0.15\textwidth} | p{0.68\textwidth} ||}
\hline
\bf{Category} &
\bf{Recommended Regime} &
\bf{Notes} \\
\hline\hline

Wavelength Coverage & 3800--10200\,\AA\ (observed frame) &
Spectrophotometry better than 25\% is achieved in this range for most galaxies (see Fig. \ref{fig:Bowtie} and \S\ref{subsec:Spectra_qual} for details.)  \\

\hline

Emission Line Fluxes &

S/N $\geq 3$ for lines with $\lambda_{\mathrm{observed}} > 4500$\,\AA; S/N $\geq 5$ for lines with $\lambda_{\mathrm{observed}} \leq 4500$\,\AA. &

The higher threshold in the blue reflects an increased rate of spurious detections at these wavelengths. \newline

If using absolute flux values or ratios of lines widely separated in wavelength, it will be important to account for spectrophotometric error using the numbers given in Table~\ref{table:error_scale}. \newline

Emission lines have been corrected for foreground Milky Way reddening 
(see \S\ref{subsec:eBOSS-DAP}.)\\

\hline

Global Galaxy Quantities (e.g., a SFR based on $L_{H\alpha}$) &
Require the fraction of light in the fiber to be $\gtrsim 25$\% and apply aperture corrections &
The eBOSS spectra are obtained through 2\arcsec\ diameter fibers that do not typically encompass the entire galaxy (see \S\ref{subsec:appature_bias} and Fig.~\ref{fig:infiber}).  The fraction of light in the fiber has been computed from the ratio of the $i$-band flux synthesized from spectrum and the $i$-band photometry and stored under the heading {\tt infiber} in the \texttt{eboss\_dap\_auxdata} file. Note that for low S/N spectra, this quantity may 
be noisy. \\

\hline

High EW Emission Lines &  EW$ < 100$~\AA\ or $\sigma_{line} \gtrsim 100$~\kms &
Very high EW narrow emission lines (such as H$\alpha$ and [O~III] in starbursting dwarf galaxies) are sometimes flagged as bad and clipped off by the {\tt IDLSpec2D} pipeline (see \S\ref{subsec:known_errs} and Fig.~\ref{fig:cutoff_spec}.)  Some of these clipped lines will have erroneous line fluxes reported in the eBOSS-DAP catalog.  Techniques for identifying problem cases and a method to recover the clipped lines are briefly discussed in \S\ref{subsec:known_errs}. \\

\hline

Emission Line Velocity Dispersions & Dispersion error $< 50$~\kms; values of $-999.0$ indicate the dispersion was not fit successfully&
The reported emission line velocity dispersions have been corrected for instrumental resolution as follows: 
$\sigma = \sqrt{\sigma_{\mathrm{raw}}^2 - \sigma_{\mathrm{corr}}^2}$ where $\sigma_{\mathrm{raw}}$ is the measured Gaussian line width and  $\sigma_{\mathrm{corr}}$ is the instrumental resolution at line center. 
The instrumental resolution ranges from $\sim$40 - 80 \kms; line widths below this will have large measurement errors.  In cases where $\sigma_{\mathrm{raw}} < \sigma_{\mathrm{corr}}$, the reported value is set to $\sigma = 0$ \kms. 
\newline

Line widths of all the emission lines are tied together during the fit with a 75\% variation allowed (see \S\ref{subsubsec:linelist}). \\

\hline

Stellar Velocity Dispersion &
Dispersion error $< 50$~\kms; values of $-999.0$ indicate the dispersion was invalid or exceeded 450~\kms &
The eBOSS-DAP applies the same instrumental resolution correction described above (for the difference between the BOSS spectrograph and the CKC SSP template resolution) and reports the corrected value. \newline

Corrected values at or above 450 km s$^{-1}$ are considered non-physical and are set to $-999.0$.  Velocity dispersion values below the instrumental resolution ($\sigma_{*} \lesssim 60$~\kms) will have large errors. In cases where $\sigma_{*,\mathrm{raw}} < \sigma_{*,\mathrm{corr}}$, the reported value is set to $\sigma_* = 0$ \kms.\newline

Repeat spectra suggest that the velocity dispersion errors are slightly underestimated (see Fig.~\ref{fig:trumpet_disp}). Scale uncertainties as suggested in Table~\ref{table:error_scale}. \\

\hline

Absorption Line Indices & Apply an error cut appropriate to the index of interest (e.g., D$_n$4000 error $< 0.03$ and H$\delta_A$ error $< 0.8$) & Line indices have been measured from the data at its native resolution -- i.e., they have not been corrected to a fiducial resolution (such as the original Lick resolution of 8.4 \AA), nor are correction factors provided. \newline 

Only a small fraction of the eBOSS sample has sufficient continuum S/N for reliable index measurements; for example, the suggested D$_n$4000/H$\delta_A$ cuts retain only $\sim$3.4\% of the full sample (see \S\ref{subsubsec:hdela-vs-dn4000}). Beware of selection effects induced by S/N cuts. \\

\hline

Photometry &
$z > 0.1$ &
Below this redshift, galaxy deblending by both the SDSS and the Legacy Survey photometric pipelines can be unreliable, leading to errors in the photometric data (i.e., H~II regions can be classified as separate galaxies; see \S\ref{subsec:known_errs} and Fig.~\ref{fig:cutout}, for an example). This can also cause strong aperture bias. At lower redshifts, visual inspection of the photometry may be required.   
\\

\hline
Additional Quality Flags & {\tt FIBERMASKED = F} \newline 
[...]$\_{\mathrm{ERR}} > 0$ & 
As described in \S\ref{subsubsec:masking}, we have retained some mask bits set by the {\tt IDLSpec2d} pipeline in the \texttt{fibermasked} field in the \texttt{eboss\_dap\_auxdata} file. These mask bits signal potential problems with the flatfield, arc, or trace, although most of the spectra appear to be of good quality. The small faction of galaxies with \texttt{fibermasked = T} should be removed for a maximally conservative analysis. \newline 

Error values set to -999 on any measured quantity indicate a problem with the measurement. 
\\
\hline

Galaxy scaling relations and other population studies & Consider target selection effects \newline 

Remove duplicate spectra & 
SDSS-IV contains several different galaxy surveys with distinct target selection algorithms (see \S\ref{sec:Data} and Appendix \ref{apx:target_class}). Target selection can have significant systematic effects on galaxy scaling relations; for example, see the mass-metallicity relation in Fig.~\ref{fig:MZR}. Targeting information can be found in the 
\texttt{target\_class} and \texttt{target\_text} fields in the \texttt{eboss\_dap\_auxdata} file as well
as in the  SDSS {\tt spAll} file. \newline

Roughly 10\% of the eBOSS spectra are duplicates.  To retain only the highest quality spectrum of each galaxy, select on \texttt{SPECPRIMARY = 1} in the \texttt{eboss\_dap\_auxdata} file.
\\

\hline

\end{tabular}
\end{table*}
}

\section{Summary} \label{sec:summary}
This work presents the initial release and analysis of data and software products from the eBOSS Data Analysis Pipeline (eBOSS-DAP), designed to extend the utility of spectral data from the Extended Baryon Oscillation Spectroscopic Survey \citep[eBOSS;][]{Dawson:2016}. 
Previously, these data had primarily been utilized for measuring cosmological parameters. 
By using the eBOSS-DAP, we were able to extend the functionality of the survey and robustly characterize the eBOSS sample. 
Our main findings are as follows:
\begin{enumerate}
    \item We characterized the eBOSS galaxy sample which includes galaxies targeted using a variety of color-based target selection algorithms (See \S\ref{sec:Data}).   There are \nsample~ spectra with \ndupes\ duplicates (1,704,313 unique galaxies) with a range of $z = 0$ to $z = 1.12$ (Fig.~\ref{fig:redshift_hist}) and an average of 52\% of their light down the fiber (Fig.~\ref{fig:infiber}). These spectra have an average S/N that peaks at S/N = 1.88 with an extended tail towards higher values (Fig.~\ref{fig:snr_hist}). We grouped the galaxies into 6 classes based on their target selection. The properties of these target classes are elucidated in 
    Figures \ref{fig:HB_EW} and \ref{fig:m-vdisp} -- \ref{fig:MZR}.
    
    \item We made extensive use of the duplicate spectra to characterize the eBOSS survey's spectrophotometric calibration (\S\ref{subsec:Spectra_qual}). We show that spectrophotometry errors are strongly wavelength-dependent. For most eBOSS galaxies, relative flux calibration error is less than 10\% between 4488 – 8397 ~\AA~ with errors exceeding 50\% at wavelengths less than 3665 ~\AA~ or greater than 10283~\AA\ (Fig.~\ref{fig:Bowtie}) 

    \item We provide a pipeline overview in which we detail how the MaNGA-DAP was modified to become the eBOSS-DAP (\S\ref{sec:methods}). The eBOSS-DAP utilizes algorithms from the MaNGA-DAP to process one-dimensional galaxy spectra from the eBOSS dataset, extracting comprehensive emission line fluxes, kinematics, continuum spectral indices, and spectral template weights. Like the MaNGA-DAP, the eBOSS-DAP can be easily adapted to work on datasets beyond eBOSS. Some key features of the eBOSS-DAP include: flexibility in adjusting emission line lists, spectral indices, and stellar templates, and tools for modifying fits, such as tying emission line parameters for improved precision. 

    \item Our final data products include emission line flux measurements (Fig.~\ref{fig:EML_HIST}), spectral index measurements (Fig.~\ref{fig:SPIND_HIST}), kinematics, and best-fit continuum models.   We create three catalogs: a full sample (\npassed\ spectra) emission line catalog with a reduced emission line list (\nlineslow\ lines), a high equivalent width sample (\nhighew\ spectra) emission line catalog which makes use of the full line list (\nlineshigh~ lines), and a catalog of the spectral indices and the weights of the stellar continuum templates.  In addition, we store the stellar continuum and emission-line fits for each spectrum as FITS files. All of these data products, as well as a suite of tools including the eBOSS-DAP source code, are publicly available. (See \hyperref[sec:data_availability]{Data Availability} for details.)

    \item We provide catalog validation metrics, such as comparisons of emission-line fluxes in repeat spectra (Figures \ref{fig:trumpet_strong} and \ref{fig:trumpet_weak}) and comparisons with fluxes measured by the SDSS pipeline \citep{Bolton:2012} (Fig. \ref{fig:trumpet_sdss}). Generally, we find the expected level of agreement between the fluxes given the errors, except at high S/N, where the spectrophotometric error becomes dominant. We provide average correction terms for a few commonly used lines and line ratios to approximately account for the spectrophotometric error (Table \ref{table:error_scale}). 

    \item We use the eBOSS-DAP data to construct various galaxy scaling relations (Figures \ref{fig:m-vdisp} - \ref{fig:spind}), highlighting the 6 target classes.  We caution users of the data to account for sample selection effects when pursuing scientific applications. 

    \item We provide a table that contains Recommended Usage Guidelines for the eBOSS-DAP catalog (\S\ref{sec:DUC}). These guidelines summarize the key caveats discussed throughout this paper, providing a starting point for the appropriate use of the catalog. Individual science cases will have their own specific requirements beyond these general guidelines. 

\end{enumerate}

In summary, we have produced an extensive set of  well-characterized emission line measurements for the 1.7 million galaxies in the eBOSS sample. This data will be useful for a variety of statistical studies of galaxy evolution at $z<1$.  The eBOSS-DAP catalogs will be of particular value for AGN studies due to the sample makeup and the expanded line list.  In Matthews Acu\~{n}a et al, in prep, we focus on utilizing the high quantity of [Ne V] 3347, 3427 detections to study coronal line detections in AGN.

\section*{Acknowledgments}
The authors thank the anonymous referee for their constructive comments which have signficantly improved the paper.
OMA, CAT, and DM gratefully acknowledge support for this work from NSF grant AST-2107725.
OMA thanks I. McConachie, I. Laseter, and M. Maseda for helpful comments. 
AD-S gratefully acknowledges support for this work from NSF grant 2107726.
BL gratefully acknowledges support for this work from NSF grant 2107727.
ZJL acknowledges support from the National Science Foundation Graduate Research Fellowship under grant No. 2137424.

This research uses services or data provided by the Astro Data Lab at NSF’s NOIRLab. NOIRLab is operated by the Association of Universities for Research in Astronomy (AURA), Inc. under a cooperative agreement with the National Science Foundation.

Funding for the Sloan Digital Sky Survey IV has been provided by the Alfred P. Sloan Foundation, the U.S.Department of Energy Office of Science, and the Participating Institutions. SDSS-IV acknowledges support and resources from the Center for High-Performance Computing at the University of Utah. The SDSS website is available at \href{www.sdss.org}{www.sdss.org}. SDSS-IV is managed by the Astrophysical Research Consortium for the Participating Institutions of the SDSS Collaboration including the Brazilian Participation Group, the Carnegie Institution for Science, Carnegie Mellon University, the Chilean Participation Group, the French Participation Group, Harvard-Smithsonian Center for Astrophysics, Instituto de Astrofísica de Canarias, The Johns Hopkins University, Kavli Institute for the Physics and Mathematics of the Universe (IPMU)/University of Tokyo, Lawrence Berkeley National Laboratory, Leibniz Institut für Astrophysik Potsdam (AIP), Max-Planck-Institut für Astronomie (MPIA Heidelberg), MaxPlanck-Institut für Astrophysik (MPA Garching), Max-PlanckInstitut für Extraterrestrische Physik (MPE), National Astronomical Observatories of China, New Mexico State University,New York University, University of Notre Dame, Observatário Nacional/MCTI, The Ohio State University, Pennsylvania State University, Shanghai Astronomical Observatory, United Kingdom Participation Group, Universidad Nacional Autónoma de México, University of Arizona, University of Colorado Boulder, University of Oxford, University of Portsmouth, University of Utah, University of Virginia, University of Washington, University of Wisconsin, Vanderbilt University, and Yale University.

\section*{software}
This research made use of
{\tt Astropy}, a community-developed core Python package for Astronomy \citep{astropy:2013,astropy:2018,astropy:2022}; 
{\tt dust-extinction} \citep{dust_extinction};
{\tt dustmaps} \citep{dustmaps};
{\tt FSPS} \citep{fsps_1,fsps_2} and its Python bindings package {\tt Python-FSPS}\footnote{\href{https://github.com/dfm/python-fsps}{\tt https://github.com/dfm/python-fsps}};
{\tt IPython} \citep{PER-GRA:2007};
{\tt MaNGA-DAP} \citep{MaNGA-DAP, Belfiore_MaNGA};
{\tt matplotlib} \citep{Hunter:2007};
{\tt numpy} \citep{numpy}; 
{\tt pandas} \citep{reback2020pandas,mckinney-proc-scipy-2010};
{\tt SciPy} \citep{2020SciPy-NMeth};
and {\tt Seaborn} \citep{Waskom2021}

Additionally, computing resources from the University of Wisconsin--Madison Center for High Throughput Computing \citep{CHTC} were used to run the pipeline.

\hypertarget{sec:data_availability}{}
\section*{Data Availability}\forcehref{sec:data_availability}\label{sec:data_availability}
All catalogs in this paper are available at \url{https://datalab.noirlab.edu/data/sdss#sdss-iv-eboss-dap-value-added-catalog}. 

The \texttt{eBOSS-DAP} is available at \url{https://github.com/owenmatthewsa/ebossdap}. Summary tables of figures \ref{fig:Bowtie}, \ref{fig:trumpet_sdss},\ref{fig:trumpet_strong}, and \ref{fig:trumpet_weak} are available at this same location in addition to tables used in \ref{subsubsec:linelist}. The data used in additional plots within this article are available upon request. All fits reported in this paper or in the catalog can be reproduced for any galaxy by running the eBOSS-DAP code on the original spectra.

\bibliographystyle{aasjournal_notitle}
\bibliography{main}
\clearpage
\mbox{} 
\clearpage

\appendix

\section{Target Selection}\label{apx:target_class}
The BOSS and eBOSS surveys targeted galaxies and quasars for spectroscopy using a wide variety of algorithms \citep{Dawson:2013, Dawson:2016}.  Targeting information for each galaxy can be found in the the SDSS {\tt spAll} file as  hexidecimal target mask bits stored in {\tt BOSS\_TARGET1, ANCILLARY\_TARGET1, ANCILLARY\_TARGET2, EBOSS\_TARGET0, EBOSS\_TARGET1}, and {\tt EBOSS\_TARGET2}. We have translated the hexidecimal mask bits into text format and stored them in the {\tt eBOSS\_DAP\_auxdata} table under the heading {\tt `target\_text'}. In total there are 595 unique target types. For convenience, we have grouped them into 8 broad `target classes' which are defined in Table~\ref{table:target_classes} and stored under the  heading `{\tt target\_class}'. These target classes are shown as colored contours Figures \ref{fig:redshift_hist} - \ref{fig:infiber} and \ref{fig:m-vdisp} - \ref{fig:MZR} . Additional information on SDSS-IV target selection can be found at \url{https://www.sdss4.org/dr17/spectro/targets/}.

\begin{table}[H]
\centering
\scriptsize
\caption{Target Classes}
\label{table:target_classes}
\begin{tabular}{|| p{0.12\textwidth} | p{0.07\textwidth} | p{0.05\textwidth} | p{0.68\textwidth} ||}
\hline
\textbf{Target Class Name \newline (description)} &
\textbf{Number of Galaxies (\% of sample)} &
\textbf{Median Redshift} &
\textbf{SDSS-IV Target Types Included} \\
\hline\hline

QSO  & 61,300 (3.22\%)  & 0.30 & 
{\raggedleft BLAZGRQSO, BLAZGXQSO, GAL\_NEAR\_QSO, KQSO\_BOSS, QSO1\_BAD\_BOSS, QSO1\_EBOSS\_CORE, QSO1\_EBOSS\_FIRST,
QSO1\_PTF, QSO1\_REOBS, QSO1\_VAR\_S82, QSO\_BONUS, QSO\_BONUS\_MAIN, QSO\_BOSS\_TARGET, QSO\_CORE, QSO\_CORE\_ED,
QSO\_CORE\_LIKE, QSO\_CORE\_MAIN, QSO\_DEEP, QSO\_EBOSS\_CORE, QSO\_EBOSS\_FIRST, QSO\_EBOSS\_KDE,
QSO\_EBOSS\_W3\_ADM, QSO\_FIRST\_BOSS, QSO\_GRI, QSO\_HIZ, QSO\_KDE, QSO\_KDE\_COADD, QSO\_KNOWN, QSO\_KNOWN\_MIDZ,
QSO\_LIKE, QSO\_NN, QSO\_PTF, QSO\_RADIO, QSO\_RIZ, QSO\_SDSS\_TARGET, QSO\_SUPPZ, QSO\_UKIDSS, QSO\_VAR,
QSO\_VAR\_FPG, QSO\_VAR\_LF, QSO\_VAR\_SDSS, QSO\_WISE\_FULL\_SKY, QSO\_WISE\_SUPP, QSO\_XD\_KDE\_PAIR,
RADIO\_2LOBE\_QSO, TEMPLATE\_QSO\_SDSS1, WISE\_BOSS\_QSO} \\

\hline
 
LRG3 (SDSS-III/BOSS Luminous Red Galaxies) & 1,024,272 (53.86\%)  & 0.54 & 
{\raggedright GAL\_CMASS, GAL\_CMASS\_ALL, GAL\_CMASS\_COMM, GAL\_CMASS\_SPARSE} \\

\hline
 
LRG4 (SDSS-IV/eBOSS Luminous Red Galaxies) & 200,368 (10.54\%)  & 0.70 & 
{\raggedright FAINT\_HIZ\_LRG, HIZ\_LRG, LRG1\_IDROP, LRG1\_WISE, LRG\_IZW, LRG\_RIW, LRG\_ROUND3} \\

\hline
 
LRGL (BOSS low-$z$ Luminous Red Galaxies) & 369,507 (19.43\%)  & 0.33 & GAL\_LOZ \\

\hline
 
ELG (Emission Line Galaxies) & 180,025 (9.47\%)  & 0.82 & 
{\raggedright ELG, ELG1\_EBOSS, ELG1\_EXTENDED, ELG1\_NGC, ELG1\_SGC, ELG\_DECALS\_TEST1, ELG\_DESI\_TEST1,
ELG\_DES\_TEST1, ELG\_GRIW\_TEST1, ELG\_SCUSS\_TEST1, ELG\_SDSS\_TEST1, ELG\_TEST1, ELG\_UGRIZW\_TEST1,
ELG\_UGRIZWbright\_TEST1, FAINT\_ELG, SEQUELS\_ELG, SEQUELS\_ELG\_LOWP} \\

\hline
 
TDSP (TDSS and SPIDERS Targets) & 37,594 (1.98\%)  & 0.27 & 
{\raggedright TDSS\_A, TDSS\_B, TDSS\_CP, TDSS\_FES\_DE, TDSS\_FES\_HYPQSO, TDSS\_FES\_HYPSTAR, TDSS\_FES\_NQHISN,
TDSS\_FES\_VARBAL, TDSS\_PILOT, TDSS\_PILOT\_SNHOST, TDSS\_RQS1, TDSS\_RQS2, TDSS\_RQS3, TDSS\_RQS3v,
TDSS\_TARGET, SPIDERS\_CODEX\_CLUS\_CFHT, SPIDERS\_CODEX\_CLUS\_DECALS, SPIDERS\_CODEX\_CLUS\_PS1, SPIDERS\_PILOT,
SPIDERS\_RASS\_AGN, SPIDERS\_RASS\_CLUS, SPIDERS\_TARGET, SPIDERS\_XCLASS\_CLUS, SPIDERS\_XMMSL\_AGN,
EFEDS\_COSMOQSO, EFEDS\_CSC2, EFEDS\_EROSITA\_AGN, EFEDS\_EROSITA\_CLUS, EFEDS\_GAIAWISE\_QSO,
EFEDS\_HSC\_REDAGN, EFEDS\_HSC\_REDMAPPER, EFEDS\_HSC\_WERGS, EFEDS\_KIDS\_QSO, EFEDS\_SDSSWISE\_QSO,
EFEDS\_SDSS\_REDMAPPER, EFEDS\_WISE\_AGN, EFEDS\_WISE\_VARAGN, EFEDS\_XMMATLAS, S82X\_BRIGHT\_TARGET,
S82X\_GTRADMZ4\_TARGET, S82X\_LSSTZ4\_TARGET, S82X\_PETERS15\_COLORVAR\_TARGET,
S82X\_RICHARDS15\_PHOTOQSO\_TARGET, S82X\_SACLAY\_BDT\_TARGET, S82X\_SACLAY\_HIZ\_TARGET,
S82X\_SACLAY\_VAR\_TARGET, S82X\_TILE1, S82X\_TILE2, S82X\_TILE3, S82X\_UNWISE\_TARGET, S82X\_WISE\_TARGET,
S82X\_XMM\_TARGET} \\

\hline
 
ANCI (Ancillary Targets) & 28,310 (1.49\%)  & 0.31 & 
{\raggedright 25ORI\_WISE, 25ORI\_WISE\_W3, BLAZGRFLAT, BLAZGX, BLAZGXR, BLAZR, BLAZXR, BLUE\_RADIO, BRIGHTGAL, CHANDRAV1, CLUSTER\_MEMBER, CXOBRIGHT, CXOGRIZ, CXORED, ELAIS\_N1\_FIRST, ELAIS\_N1\_GMRT\_GARN,
ELAIS\_N1\_GMRT\_TAYLOR, ELAIS\_N1\_JVLA, ELAIS\_N1\_LOFAR, FAINTERM, IAMASERS, KOE2023BSTAR, KOE2023\_STAR,
KOE2068BSTAR, KOE2068\_STAR, KOEKAP\_STAR, LBG, ODDBAL, OTBAL, PTF\_GAL, RM\_TILE1, RM\_TILE2, RQSS\_SF,
RQSS\_SFC, SEQUELS\_PTF\_VARIABLE, SEQUELS\_TARGET, SN\_GAL1, SN\_GAL2, SN\_GAL3, SN\_LOC, SPEC\_SN, SPOKE,
STRIPE82BCG, TAU\_STAR, VARBAL, VARS, WHITEDWARF\_NEW, WHITEDWARF\_SDSS, WISE\_COMPLETE, XMMBRIGHT,
XMMGRIZ, XMMHR, XMMRED, XMMSDSS, XMM\_PRIME, XMM\_SECOND} \\

\hline
 
MISC (Other Targets) & 458 (0.02\%)  & 0.23 & STD\_FSTAR, TEMPLATE\_GAL\_PHOTO \\

\hline
\hline

\end{tabular}

\end{table}

\clearpage
\newpage

\section{Line List}\label{apx:line_list}
\begin{longtable}{||c c c c c c||} 
\caption[line list]{Emission Line List} \label{table:line_list} \\
\endfirsthead
\multicolumn{6}{c}
{{\bfseries \tablename\ \thetable{} -- continued from previous page}} \\
\hline 
\endhead
\hline \multicolumn{6}{|r|}{{Continued on next page}} \\ \hline
\endfoot
\hline 
\endlastfoot
\hline
\rule{0pt}{2.5ex} 
Name & Vacuum & Flux Tie & Integration & Dispersion & In Low EW  \\
(* Unavailable in MaNGA; & Wavelength &  & Range & Tie & Sample \\
\textsuperscript{\textdagger} Unavailable in SDSS) & & & & A $|$ B\footnote{A $|$ B denotes where the line is tied. Case A is redshifts less than z = 0.4963, Case B is for redshifts greater than z = 0.4963, [n-m] signifies the range multiplied by (A$|$B), in all cases the velocity is tied to be equal to the line that the dispersion is tied to} & (H$\beta$ EW $<$ 10)
\\ [0.5ex] 
\hline\hline
{[}C III{]} 1906\textsuperscript{*}\textsuperscript{\textdagger} & 1906.680 & - &  1902.2 -- 1907.6 & [0.57 - 1.75] (H$\alpha$ $|$ H$\beta$) & No \\
{[}C III{]} 1909\textsuperscript{*} & 1908.734 & - &  1907.6 -- 1914.7 & {[}C III{]} 1906 & No \\
{[}C II{]} 2325C\textsuperscript{*}\textsuperscript{\textdagger} & 2326.113 & - &  2320.6 -- 2326.4 & {[}C II{]} 2325D & No \\
{[}C II{]} 2325D\textsuperscript{*}\textsuperscript{\textdagger} & 2327.645 & - &  2326.4 -- 2328.3 & [0.57 - 1.75] (H$\alpha$ $|$ H$\beta$) & No \\
{[}C II{]} 2328\textsuperscript{*}\textsuperscript{\textdagger} & 2328.838 & - &  2328.4 -- 2331.8 & {[}C II{]} 2325D & No \\
{[}Fe II*{]} 2365\textsuperscript{*}\textsuperscript{\textdagger} & 2365.552 & - &  2361.9 -- 2368.6 & {[}Fe II*{]} 2396 & No \\
{[}Fe II*{]} 2396\textsuperscript{*}\textsuperscript{\textdagger} & 2396.350 & - &  2392.1 -- 2400.1 & [0.57 - 1.75] (H$\alpha$ $|$ H$\beta$) & No \\
{[}Ne IV{]} 2423\textsuperscript{*}\textsuperscript{\textdagger} & 2422.560 & - &  2417.1 -- 2424.2 & {[}Ne IV{]} 2425 & No \\
{[}Ne IV{]} 2425\textsuperscript{*}\textsuperscript{\textdagger} & 2425.140 & - &  2424.2 -- 2430.2 & [0.57 - 1.75] (H$\alpha$ $|$ H$\beta$) & No \\
{[}O II{]} 2470\textsuperscript{*}\textsuperscript{\textdagger} & 2471.027 & - &  2467.5 -- 2474.8 & [0.57 - 1.75] (H$\alpha$ $|$ H$\beta$) & No \\
{[}Fe II*{]} 2613\textsuperscript{*}\textsuperscript{\textdagger} & 2612.654 & - &  2608.7 -- 2617.2 & [0.57 - 1.75] (H$\alpha$ $|$ H$\beta$) & No \\
{[}Fe II*{]} 2626\textsuperscript{*}\textsuperscript{\textdagger} & 2626.451 & - &  2622.2 -- 2629.5 & {[}Fe II*{]} 2613 & No \\
{[}Fe II*{]} 2632\textsuperscript{*}\textsuperscript{\textdagger} & 2632.110 & - &  2629.9 -- 2635.4 & {[}Fe II*{]} 2613 & No \\
He II 2734\textsuperscript{*}\textsuperscript{\textdagger} & 2733.700 & - &  2729.7 -- 2737.7 & [0.57 - 1.75] (H$\alpha$ $|$ H$\beta$) & No \\
{[}Mg V{]} 2784\textsuperscript{*}\textsuperscript{\textdagger} & 2783.700 & - &  2780.2 -- 2787.2 & [0.57 - 1.75] (H$\alpha$ $|$ H$\beta$) & No \\
{[}Mg II{]} 2796\textsuperscript{*} & 2796.350 & - &  2787.9 -- 2797.7 & {[}S III{]} 9533 & Yes \\
{[}Mg II{]} 2803\textsuperscript{*}\textsuperscript{\textdagger} & 2803.531 & - &  2797.7 -- 2809.8 & {[}Mg II{]} 2796 & Yes \\
{[}Fe IV{]} 2830\textsuperscript{*}\textsuperscript{\textdagger} & 2830.300 & - &  2827.8 -- 2832.8 & {[}Fe IV{]} 2836 & No \\
{[}Fe IV{]} 2836\textsuperscript{*}\textsuperscript{\textdagger} & 2836.000 & - &  2832.8 -- 2839.2 & [0.57 - 1.75] (H$\alpha$ $|$ H$\beta$) & No \\
He I 2945\textsuperscript{*}\textsuperscript{\textdagger} & 2945.103 & - &  2943.1 -- 2949.1 & [0.57 - 1.75] (H$\alpha$ $|$ H$\beta$) & No \\
{[}O III{]} 3134\textsuperscript{*}\textsuperscript{\textdagger} & 3134.200 & - &  3126.0 -- 3141.0 & [0.57 - 1.75] (H$\alpha$ $|$ H$\beta$) & No \\
He I 3189\textsuperscript{*}\textsuperscript{\textdagger} & 3188.670 & - &  3178.7 -- 3196.0 & [0.57 - 1.75] (H$\alpha$ $|$ H$\beta$) & No \\
He II 3204\textsuperscript{*}\textsuperscript{\textdagger} & 3204.019 & - &  3195.1 -- 3211.7 & [0.57 - 1.75] (H$\alpha$ $|$ H$\beta$) & No \\
{[}Ne V{]} 3347\textsuperscript{*}\textsuperscript{\textdagger} & 3346.783 &  0.366 * NeV\_3427 &  3336.7 -- 3355.0 & {[}Ne V{]} 3427 & Yes \\
{[}Ne V{]} 3427\textsuperscript{*}\textsuperscript{\textdagger} & 3426.864 & - &  3415.7 -- 3437.0 & [0.57 - 1.75] (H$\alpha$ $|$ H$\beta$) & Yes \\
{[}Fe VII{]} 3586\textsuperscript{*}\textsuperscript{\textdagger} & 3586.000 & - &  3576.0 -- 3596.0 & [0.57 - 1.75] (H$\alpha$ $|$ H$\beta$) & No \\
{[}O II{]} 3727 & 3727.092 & - &  3716.3 -- 3738.3 & [0.57 - 1.75] (H$\alpha$ $|$ H$\beta$) & Yes \\
{[}O II{]} 3729 & 3729.875 & - &  -1.0 -- -1.0 & {[}O II{]} 3727 & Yes \\
H12\textsuperscript{\textdagger} & 3751.217 & - &  3746.2 -- 3756.2 & [0.57 - 1.75] (H$\alpha$ $|$ H$\beta$) & No \\
{[}Fe VII{]} 3760\textsuperscript{*}\textsuperscript{\textdagger} & 3760.000 & - &  3756.6 -- 3766.6 & [0.57 - 1.75] (H$\alpha$ $|$ H$\beta$) & No \\
H11\textsuperscript{\textdagger} & 3771.701 & - &  3761.7 -- 3781.7 & [0.57 - 1.75] (H$\alpha$ $|$ H$\beta$) & No \\
H$\theta$\textsuperscript{\textdagger} & 3798.976 & - &  3789.0 -- 3809.0 & [0.57 - 1.75] (H$\alpha$ $|$ H$\beta$) & No \\
H$\eta$\textsuperscript{\textdagger} & 3836.472 & - &  3826.5 -- 3846.5 & [0.57 - 1.75] (H$\alpha$ $|$ H$\beta$) & Yes \\
{[}Ne III{]} 3870 & 3869.860 & - &  3859.9 -- 3879.9 & [0.57 - 1.75] (H$\alpha$ $|$ H$\beta$) & Yes \\
He I 3890\textsuperscript{\textdagger} & 3889.749 & - &  -1.0 -- -1.0 & H$\zeta$ & No \\
H$\zeta$\textsuperscript{\textdagger} & 3890.151 & - &  3880.2 -- 3900.2 & [0.57 - 1.75] (H$\alpha$ $|$ H$\beta$) & No \\
{[}Ne III{]} 3969 & 3968.590 &  0.30 * NeIII\_3870 &  -1.0 -- -1.0 & {[}Ne III{]} 3870 & Yes \\
H$\epsilon$ & 3971.195 & - &  3961.2 -- 3981.2 & [0.57 - 1.75] (H$\alpha$ $|$ H$\beta$) & Yes \\
He I 4027\textsuperscript{*}\textsuperscript{\textdagger} & 4027.175 & - &  4023.3 -- 4031.1 & [0.57 - 1.75] (H$\alpha$ $|$ H$\beta$) & No \\
{[}S II{]} 4070\textsuperscript{*}\textsuperscript{\textdagger} & 4069.750 & - &  4063.0 -- 4076.0 & [0.57 - 1.75] (H$\alpha$ $|$ H$\beta$) & No \\
H$\delta$ & 4102.892 & - &  4092.9 -- 4112.9 & [0.57 - 1.75] (H$\alpha$ $|$ H$\beta$) & Yes \\
H$\gamma$ & 4341.684 & - &  4331.7 -- 4351.7 & [0.57 - 1.75] (H$\alpha$ $|$ H$\beta$) & Yes \\
{[}O III{]} 4364\textsuperscript{*} & 4364.440 & - &  4356.4 -- 4373.4 & [0.57 - 1.75] (H$\alpha$ $|$ H$\beta$) & Yes \\
He I 4473\textsuperscript{*}\textsuperscript{\textdagger} & 4472.760 & - &  4464.2 -- 4481.2 & [0.57 - 1.75] (H$\alpha$ $|$ H$\beta$) & No \\
{[}Fe III{]} 4659\textsuperscript{*}\textsuperscript{\textdagger} & 4659.400 & - &  4652.7 -- 4664.4 & [0.57 - 1.75] (H$\alpha$ $|$ H$\beta$) & No \\
He II 4687 & 4687.015 & - &  4677.0 -- 4697.0 & [0.57 - 1.75] (H$\alpha$ $|$ H$\beta$) & Yes \\
{[}Ar IV{]} + He I 4713\textsuperscript{*}\textsuperscript{\textdagger} & 4713.300 & - &  4852.7 -- 4872.7 & [0.57 - 1.75] (H$\alpha$ $|$ H$\beta$) & No \\
{[}Fe III{]} 4742\textsuperscript{*}\textsuperscript{\textdagger} & 4741.500 & - &  4709.3 -- 4717.3 & [0.57 - 1.75] (H$\alpha$ $|$ H$\beta$) & No \\
H$\beta$ & 4862.683 & - &  4738.5 -- 4744.5 & {[}S III{]} 9533 & Yes \\
{[}O III{]} 4960 & 4960.295 &  0.331 * OIII\_5008 &  4919.0 -- 4928.0 & {[}O III{]} 5008 & Yes \\
He I 4924\textsuperscript{*}\textsuperscript{\textdagger} & 4923.500 & - &  4950.3 -- 4970.3 & [0.57 - 1.75] (H$\alpha$ $|$ H$\beta$) & No \\
{[}Fe III{]} 4988\textsuperscript{*}\textsuperscript{\textdagger} & 4987.700 & - &  4984.0 -- 4991.4 & [0.57 - 1.75] (H$\alpha$ $|$ H$\beta$) & No \\
{[}O III{]} 5008 & 5008.240 & - &  4998.2 -- 5018.2 & [0.57 - 1.75] (H$\alpha$ $|$ H$\beta$) & Yes \\
{[}Fe II{]} 5018\textsuperscript{*}\textsuperscript{\textdagger} & 5016.250 & - &  5012.6 -- 5019.9 & [0.57 - 1.75] (H$\alpha$ $|$ H$\beta$) & No \\
{[}N I{]} 5199\textsuperscript{\textdagger} & 5199.349 & - &  5189.3 -- 5209.3 & [0.57 - 1.75] (H$\alpha$ $|$ H$\beta$) & No \\
{[}N I{]} 5202\textsuperscript{\textdagger} & 5201.705 & - &  -1.0 -- -1.0 & {[}N I{]} 5199 & No \\
{[}Fe VII{]} 5723\textsuperscript{*}\textsuperscript{\textdagger} & 5723.200 & - &  5867.2 -- 5887.2 & [0.57 - 1.75] (H$\alpha$ $|$ H$\beta$) & No \\
{[}N II{]} 5757\textsuperscript{*}\textsuperscript{\textdagger} & 5757.000 & - &  5715.7 -- 5730.7 & [0.57 - 1.75] (H$\alpha$ $|$ H$\beta$) & No \\
He I 5877 & 5877.252 & - &  5753.5 -- 5760.5 & [0.57 - 1.75] (H$\alpha$ $|$ H$\beta$) & Yes \\
{[}Fe VII{]} 6086\textsuperscript{*}\textsuperscript{\textdagger} & 6086.000 & - &  6076.0 -- 6096.0 & [0.57 - 1.75] (H$\alpha$ $|$ H$\beta$) & No \\
{[}O I{]} 6302 & 6302.046 & - &  6296.0 -- 6308.0 & [0.57 - 1.75] (H$\alpha$ $|$ H$\beta$) & Yes \\
{[}S III{]} 6314\textsuperscript{*} & 6313.500 & - &  6309.5 -- 6317.5 & [0.57 - 1.75] (H$\alpha$ $|$ H$\beta$) & No \\
{[}O I{]} 6366 & 6365.536 &  0.32 * OI\_6302 &  6355.5 -- 6370.5 & {[}O I{]} 6302 & Yes \\
{[}Fe X{]} 6374\textsuperscript{*}\textsuperscript{\textdagger} & 6374.000 & - &  6370.0 -- 6380.0 & [0.57 - 1.75] (H$\alpha$ $|$ H$\beta$) & No \\
{[}N II{]} 6550 & 6549.860 &  0.326 * NII\_6585 &  6542.9 -- 6556.9 & {[}N II{]} 6585 & Yes \\
H$\alpha$ & 6564.608 & - &  6557.6 -- 6571.6 & [0.57 - 1.75] (H$\alpha$ $|$ H$\beta$) & Yes \\
{[}N II{]} 6585 & 6585.270 & - &  6575.3 -- 6595.3 & [0.57 - 1.75] (H$\alpha$ $|$ H$\beta$) & Yes \\
He I 6680\textsuperscript{*}\textsuperscript{\textdagger} & 6679.750 & - &  6674.7 -- 6684.7 & [0.57 - 1.75] (H$\alpha$ $|$ H$\beta$) & No \\
{[}S II{]} 6718 & 6718.295 & - &  6711.3 -- 6725.3 & [0.57 - 1.75] (H$\alpha$ $|$ H$\beta$) & Yes \\
{[}S II{]} 6733 & 6732.674 & - &  6725.7 -- 6739.7 & [0.57 - 1.75] (H$\alpha$ $|$ H$\beta$) & Yes \\
He I 7068\textsuperscript{\textdagger} & 7067.657 & - &  7057.1 -- 7077.1 & [0.57 - 1.75] (H$\alpha$ $|$ H$\beta$) & No \\
{[}Ar III{]} 7138 & 7137.760 & - &  7127.8 -- 7147.8 & [0.57 - 1.75] (H$\alpha$ $|$ H$\beta$) & No \\
{[}O II{]} 7321\textsuperscript{*}\textsuperscript{\textdagger} & 7321.200 & - &  7316.6 -- 7325.8 & {[}Ar III{]} 7138 & No \\
{[}O II{]} 7332\textsuperscript{*}\textsuperscript{\textdagger} & 7332.200 & - &  7327.4 -- 7337.0 & [0.57 - 1.75] (H$\alpha$ $|$ H$\beta$) & No \\
{[}Ar III{]} 7753\textsuperscript{\textdagger} & 7753.240 & - &  7743.2 -- 7763.2 & [0.57 - 1.75] (H$\alpha$ $|$ H$\beta$) & No \\
P$\eta$\textsuperscript{\textdagger} & 9017.384 & - &  9007.4 -- 9027.4 & [0.57 - 1.75] (H$\alpha$ $|$ H$\beta$) & No \\
{[}S III{]} 9071\textsuperscript{\textdagger} & 9071.100 &  0.41 * SIII\_9533 &  9061.1 -- 9081.1 & {[}S III{]} 9533 & Yes \\
P$\zeta$\textsuperscript{\textdagger} & 9231.546 & - &  9221.5 -- 9241.5 & [0.57 - 1.75] (H$\alpha$ $|$ H$\beta$) & No \\
{[}S III{]} 9533\textsuperscript{\textdagger} & 9533.200 & - &  9525.5 -- 9540.9 & [0.57 - 1.75] (H$\alpha$ $|$ H$\beta$) & Yes \\
P$\epsilon$\textsuperscript{\textdagger} & 9548.588 & - &  9540.9 -- 9556.3 & [0.57 - 1.75] (H$\alpha$ $|$ H$\beta$) & No 

\end{longtable}
\newpage

\section{Catalog Tables} \label{sec:cat_tables}
\begin{table}[htbp]
\centering
\caption{Emission Line Catalog Descriptions\label{table:eml_table}}
\begin{tabular}{||c c c p{0.45\textwidth}||}
\hline
Column Name & Unit & Format & Column Description \\[0.5ex]
\hline\hline
PMF\_STRING & - & STRING & The combined Plate-MJD-Fiber of the spectrum \\
PLATE & - & STRING & The Plate \\
MJD & - & STRING & The Modified Julian Date of when the spectrum was taken \\
FIBER & - & STRING & The Fiber \\
Z & - & DOUBLE & The redshift \\
SC\_VELOCITY & $\mathrm{km~s^{-1}}$ & DOUBLE & The stellar continuum velocity as measured by the DAP \\
SC\_VELOCITY\_ERR & $\mathrm{km~s^{-1}}$ & DOUBLE & The error in the stellar continuum velocity as measured by the DAP \\
SC\_DISPERSION & $\mathrm{km~s^{-1}}$ & DOUBLE & The intrinsic stellar continuum dispersion corrected for instrumental dispersion as measured by the DAP \\
SC\_DISPERSION\_ERR & $\mathrm{km~s^{-1}}$ & DOUBLE & The error in the stellar continuum dispersion as measured by the DAP \\
SC\_CORRECTION & $\mathrm{km~s^{-1}}$ & DOUBLE & The stellar continuum correction applied to the SC\_DISPERSION \\
SNR & - & DOUBLE & The median S/N ratio of spectrum \\
LINEID\_FLUX & $10^{-17}\,\mathrm{erg\,s^{-1}\,cm^{-2}}$ & DOUBLE & The flux of a line given by LINEID as seen in Table~\ref{table:line_list} \\
LINEID\_FLUX\_ERR & $10^{-17}\,\mathrm{erg\,s^{-1}\,cm^{-2}}$ & DOUBLE & The error in flux of a line given by LINEID as seen in Table~\ref{table:line_list} \\
LINEID\_EW & \AA & DOUBLE & The restframe Gaussian equivalent width of a line given by LINEID as seen in Table~\ref{table:line_list} \\
LINEID\_EW\_ERR & \AA & DOUBLE & The error in the restframe Gaussian equivalent width of a line given by LINEID as seen in Table~\ref{table:line_list} \\
LINEID\_INT\_EW & \AA & DOUBLE & The restframe integrated equivalent width of a line given by LINEID as seen in Table~\ref{table:line_list} \\
LINEID\_INT\_EW\_ERR & \AA & DOUBLE & The error in the restframe integrated equivalent width of a line given by LINEID as seen in Table~\ref{table:line_list} \\
LINEID\_SIGMA & $\mathrm{km~s^{-1}}$ & DOUBLE & The intrinsic $\sigma$ corrected for instrumental dispersion of a line given by LINEID as seen in Table~\ref{table:line_list} \\
LINEID\_SIGMA\_ERR & $\mathrm{km~s^{-1}}$ & DOUBLE & The error in $\sigma$ of a line given by LINEID as seen in Table~\ref{table:line_list} \\
LINEID\_SIGMA\_CORR & $\mathrm{km~s^{-1}}$ & DOUBLE & The correction applied to the $\sigma$ of a line given by LINEID as seen in Table~\ref{table:line_list} \\
\hline
\end{tabular}
\end{table}

\begin{table*}
\centering
\caption{Spectral Index Catalog Descriptions\label{table:spind_table}}
\begin{tabular}{||c c p{0.5\textwidth}||}
\hline
Column Name & Format & Column Description \\[0.5ex]
\hline\hline
PMF\_STRING & STRING & The combined Plate-MJD-Fiber \\
PLATE       & STRING & The Plate \\
MJD         & STRING & The Modified Julian Date \\
FIBER       & STRING & The Fiber \\
Z           & DOUBLE & The redshift \\
SNR         & DOUBLE & The median S/N of spectrum \\
SPIND       & DOUBLE & The value of the Spectral Index given by the Index column in Table 4 of \citet{MaNGA-DAP} \\
SPIND\_ERR  & DOUBLE & The error in the Spectral Index given by the Index column in Table 4 of \citet{MaNGA-DAP} \\
\hline
\end{tabular}
\end{table*}

\begin{table*}
\centering
\caption{AUXDATA Catalog Descriptions\label{table:auxdata_table}}
\begin{tabular}{||c c p{0.5\textwidth}||}
\hline
Column Name & Format & Column Description \\[0.5ex]
\hline\hline
\multicolumn{3}{||c||}{\textit{Template Weight Columns}} \\
\hline
PMF\_STRING               & STRING  & The combined Plate-MJD-Fiber \\
PLATE                     & INT     & The Plate \\
MJD                       & INT     & The Modified Julian Date \\
FIBER                     & INT     & The Fiber \\
E(B-V)                    & DOUBLE  & The E(B-V) of the Milky Way at the position of the spectrum used as input to the eBOSS-DAP, see Step \textbf{(2)} in \S\ref{subsec:workflow} \\
RCHI2                     & DOUBLE  & The reduced $\chi^2$ of the best-fit stellar continuum model \\
FIBERMASKED               & BOOLEAN & Flag for spectra masked with a full fiber mask (T for masked, F for unmasked), see \S\ref{subsubsec:masking} \\
AgeXX\_XXXXMetY\_Y        & DOUBLE  & Weight given to the SSP template with age XX.XXXX Gyr and metallicity Y.Y $Z_\odot$ \\
AgeC\_XXXXMetY\_Y+nebcont & DOUBLE  & Weight given to the continuous template with age 0.XXXX Gyr and metallicity Y.Y $Z_\odot$, including nebular continuum emission \\
multicoeff\_N             & DOUBLE  & Polynomial coefficient of order $N$, in units identical to the spectrum \\
\hline
\multicolumn{3}{||c||}{\textit{Photometric \& Spectrophotometric Quantities (from spAll\footnote{\url{https://data.sdss.org/datamodel/files/BOSS\_SPECTRO\_REDUX/RUN2D/spAll.html}})}} \\
\hline
SPECPRIMARY               & BYTE    & Flag set to 1 for the primary observation among repeat/duplicate spectra of the same object; use to remove duplicates \\
SN\_MEDIAN\_G             & FLOAT   & Median S/N per pixel in the $g$ band \\
SN\_MEDIAN\_R             & FLOAT   & Median S/N per pixel in the $r$ band \\
SN\_MEDIAN\_I             & FLOAT   & Median S/N per pixel in the $i$ band \\
SN\_MEDIAN\_ALL           & FLOAT   & Median S/N per pixel across all bands \\
MODELFLUX\_G              & FLOAT   & Best-fit model flux in the $g$ band \\
MODELFLUX\_R              & FLOAT   & Best-fit model flux in the $r$ band \\
MODELFLUX\_I              & FLOAT   & Best-fit model flux in the $i$ band \\
MODELFLUX\_IVAR\_G        & FLOAT   & Inverse variance of MODELFLUX\_G \\
MODELFLUX\_IVAR\_R        & FLOAT   & Inverse variance of MODELFLUX\_R \\
MODELFLUX\_IVAR\_I        & FLOAT   & Inverse variance of MODELFLUX\_I \\
SPECTROFLUX\_G            & FLOAT   & Spectroscopic flux in the $g$ band \\
SPECTROFLUX\_R            & FLOAT   & Spectroscopic flux in the $r$ band \\
SPECTROFLUX\_I            & FLOAT   & Spectroscopic flux in the $i$ band \\
PETROTH90\_I              & FLOAT   & Petrosian 90\% light radius in the $i$ band (arcsec) \\
PETROTH90ERR\_I           & FLOAT   & Uncertainty on PETROTH90\_I (arcsec) \\
LAMBDA\_EFF               & FLOAT   & Effective wavelength of the fiber offset correction (\AA) \\
RA                        & DOUBLE  & Right Ascension (J2000, degrees) \\
DEC                       & DOUBLE  & Declination (J2000, degrees) \\
\hline
\multicolumn{3}{||c||}{\textit{spZline Emission-Line Fluxes\footnote{\url{https://data.sdss.org/datamodel/files/SPECTRO\_REDUX/RUN2D/PLATE4/spZline.html}}}} \\
\hline
SPZ\_OII\_3727\_FLUX      & FLOAT   & [O\,\textsc{ii}] 3727 flux ($10^{-17}$ erg s$^{-1}$ cm$^{-2}$) \\
SPZ\_H\_beta\_FLUX        & FLOAT   & H$\beta$ flux ($10^{-17}$ erg s$^{-1}$ cm$^{-2}$) \\
SPZ\_OIII\_5007\_FLUX     & FLOAT   & [O\,\textsc{iii}] 5007 flux ($10^{-17}$ erg s$^{-1}$ cm$^{-2}$) \\
SPZ\_H\_alpha\_FLUX       & FLOAT   & H$\alpha$ flux ($10^{-17}$ erg s$^{-1}$ cm$^{-2}$) \\
SPZ\_NII\_6584\_FLUX      & FLOAT   & [N\,\textsc{ii}] 6584 flux ($10^{-17}$ erg s$^{-1}$ cm$^{-2}$) \\
SPZ\_SII\_6717\_FLUX      & FLOAT   & [S\,\textsc{ii}] 6717 flux ($10^{-17}$ erg s$^{-1}$ cm$^{-2}$) \\
SPZ\_SII\_6731\_FLUX      & FLOAT   & [S\,\textsc{ii}] 6731 flux ($10^{-17}$ erg s$^{-1}$ cm$^{-2}$) \\
\hline
\multicolumn{3}{||c||}{\textit{Target Classification \& Fiber Aperture}} \\
\hline
INFIBER                   & FLOAT   & Fraction of galaxy light within the 2\arcsec\ fiber aperture ($i$ band) \\
INFIBER\_SN               & FLOAT   & Signal to noise of INFIBER \\
TARGET\_TEXT              & STRING  & Full SDSS target type string \\
TARGET\_CLASS             & STRING  & Target class (QSO, LRG3, LRG4, LRGL, ELG, TDSP, ANCI, MISC); see Table~\ref{table:target_classes} \\
\hline
\end{tabular}
\end{table*}

\clearpage

\section{Additional Spectrum Fit Plots} \label{sec:add_plots}
\begin{figure*}[!ht]
\centering
\includegraphics[width=0.9\textwidth]{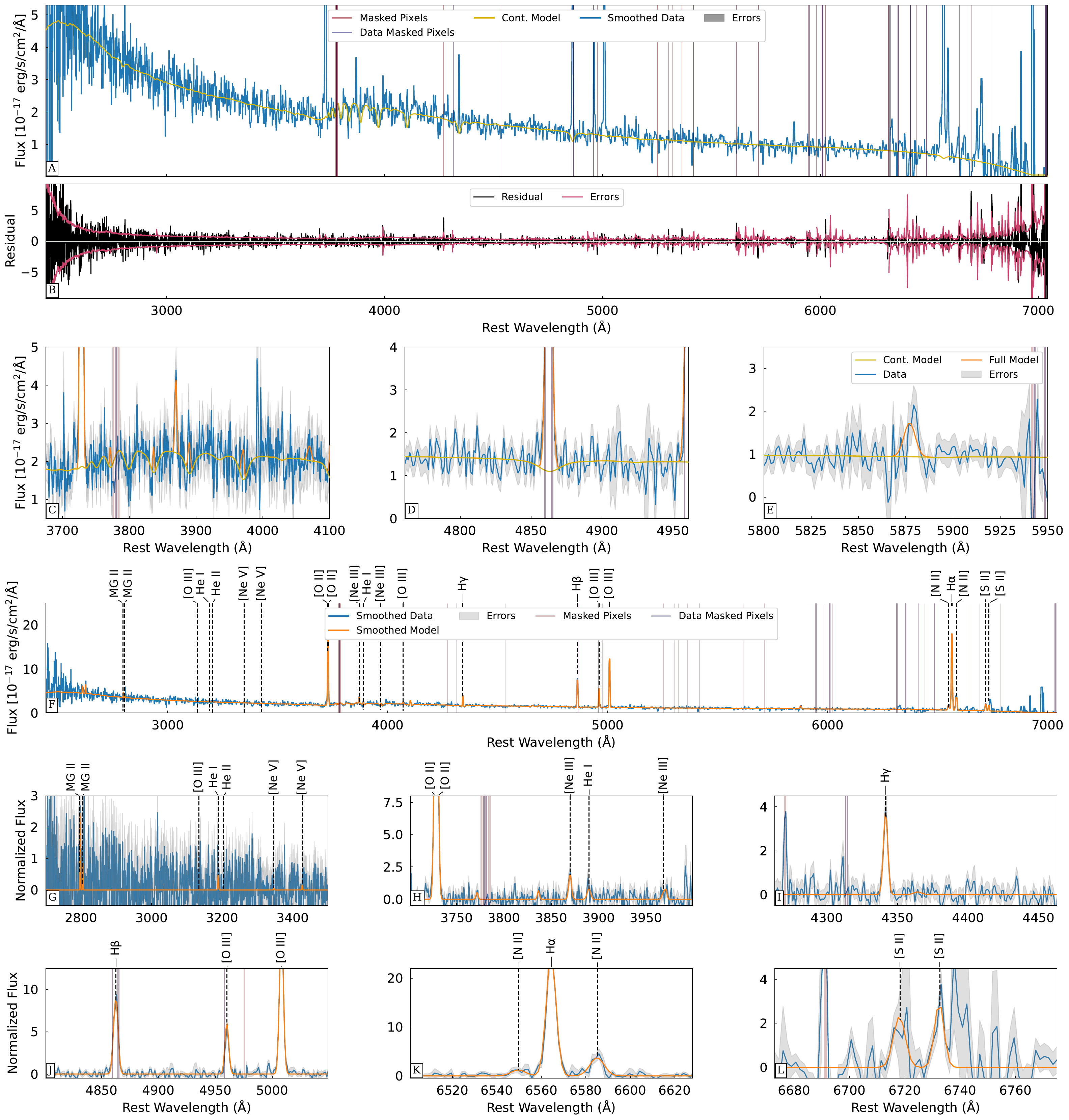}
\caption{Sample star-forming galaxy fit for the spectra spec-7386-56769-0560.fits, which was categorized as an SFG using the BPT AGN Diagnostic. All plotting information is identical to figure \ref{fig:AGN_fit}.
\label{fig:SFG_fit}}
\end{figure*}

\clearpage

\begin{figure*}[!ht]
\centering
\includegraphics[width=0.9\textwidth]{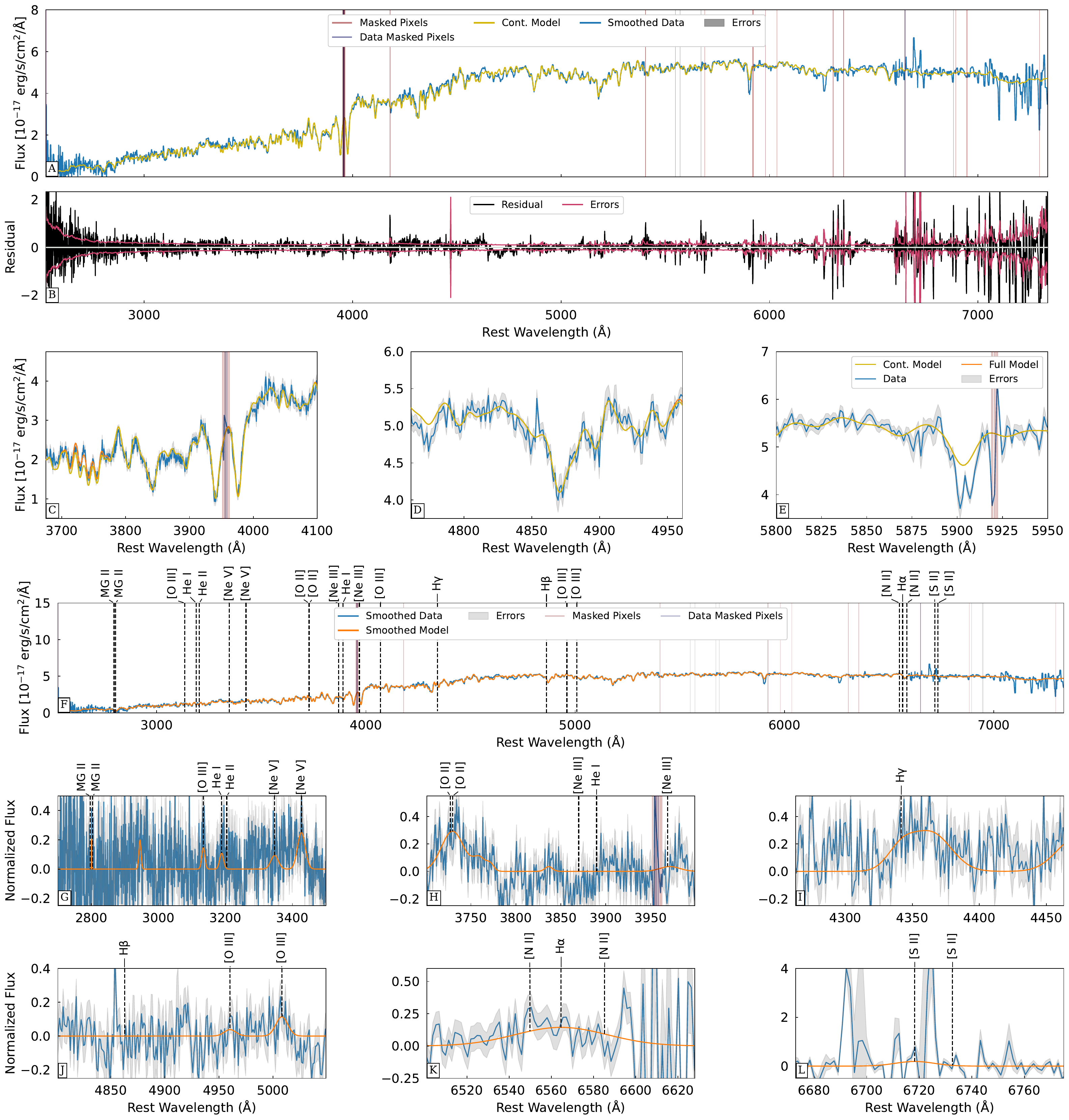}
\caption{Sample galaxy fit plots for the spectra spec-6138-56598-0432.fits, which we classified as a red dead galaxy. All plotting information is identical to figure \ref{fig:AGN_fit}.
\label{fig:RD_fit}}
\end{figure*}
\clearpage

\section{Survey Choice Error Analysis Plots} \label{sec:survey_analysis}

To evaluate the impact of the EW-based line-set division and the redshift-based anchor-line change on the consistency of our emission--line measurements, we selected a random subsample of 4171 spectra satisfying $9 < \mathrm{EW}_{\mathrm{H}\beta} < 11$~\AA\ and $0.1 < z < 0.4$, ensuring that all objects lie near the equivalent--width boundary while retaining all relevant lines within a regime of low spectrophotometric error.
In all comparisons below, we compute symmetric percent differences and normalize by the propagated measurement error using only those emission lines detected at $\mathrm{S/N} > 3$.

\begin{figure*}[!ht]
\centering
\includegraphics[width=0.8\textwidth]{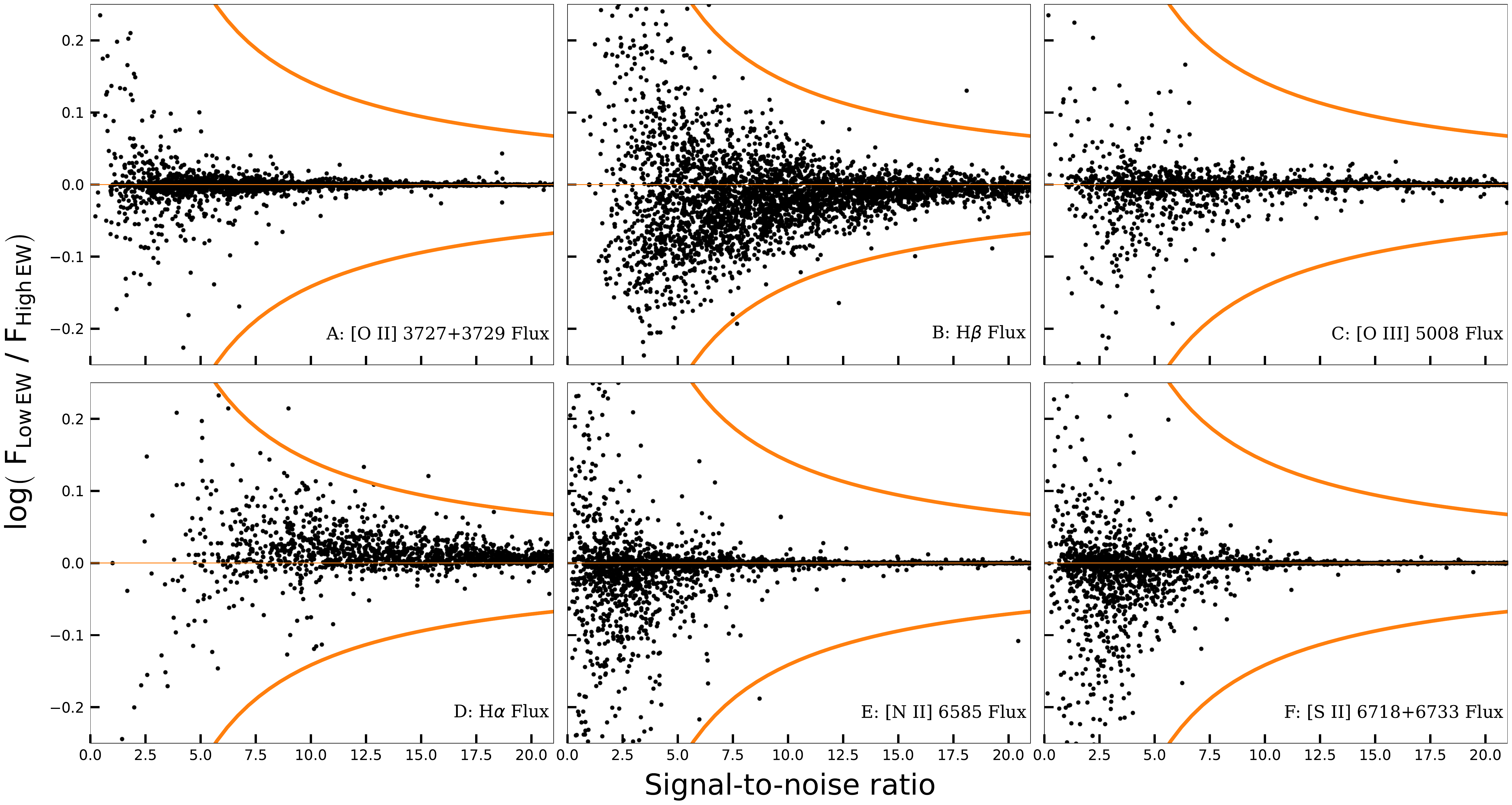}
\caption{
Comparison of line flux measurements obtained by running the eBOSS--DAP on a random subsample of 4171 spectra satisfying $9 < \mathrm{EW}_{\mathrm{H}\beta} < 11$~\text{\AA} and $0.1 < z < 0.4$, fit using H$\beta$ as the tying line. 
For each of the six strong emission lines, the y-axis shows 
$\log\!\left(\mathrm{F}_{\mathrm{Low\,EW}} / \mathrm{F}_{\mathrm{High\,EW}}\right)$, 
where $\mathrm{F}_{\mathrm{Low\,EW}}$ is the flux measured using the reduced \nlineslow--line set and $\mathrm{F}_{\mathrm{High\,EW}}$ is the flux measured using the full \nlineshigh--line set for the identical object and line. 
The x-axis shows the signal--to--noise ratio of that line measured in the full \nlineshigh--line set. 
The orange lines show the expected 1$\sigma$ flux uncertainties as a function of S/N.  The data points do not show a significant systematic offset and they lie well inside the orange lines, indicating that the difference induced by our choice of line list is significantly smaller than the random error.
\label{fig:trumpet_ew}}
\end{figure*}

We first assess the effect of the EW division by refitting the spectra in our comparison sample under two different modeling choices: one using the full set of \nlineshigh\ emission lines and one using the reduced set of \nlineslow\ lines.
For this analysis, we use H$\beta$ as the tying line since this configuration is used for 61\% of the eBOSS--DAP catalog.
Because the underlying galaxy sample is identical in both cases, any differences in the recovered fluxes directly reflect the impact of the emission line set used.
Across the six prominent lines considered (\([\mathrm{O\,II}]\,3727{+}3729\), H$\beta$, \([\mathrm{O\,III}]\,5008\), H$\alpha$, \([\mathrm{N\,II}]\,6585\), and \([\mathrm{S\,II}]\,6718{+}6733\)), the mean percent difference between the full and reduced line-set fits is consistent with zero for all six lines, with the largest offset being 0.7\% for [N~II]~6585.
The mean difference normalized by the propagated measurement error is also small for all lines, reaching at most $\sim$8\% of the measurement error for H$\beta$.
We therefore find no systematic offset introduced by the choice of line list; any differences between the full and reduced line sets are consistent with random measurement noise.
Exact values for each line are given in Table~\ref{table:survey_analysis_ew}.
As illustrated in Figure~\ref{fig:trumpet_ew}, the flux ratios show negligible systematic trend with S/N above $\mathrm{S/N} > 3$, confirming that the choice between the full and reduced line sets has no significant effect on the recovered fluxes.

\begin{figure*}[!ht]
\centering
\includegraphics[width=0.8\textwidth]{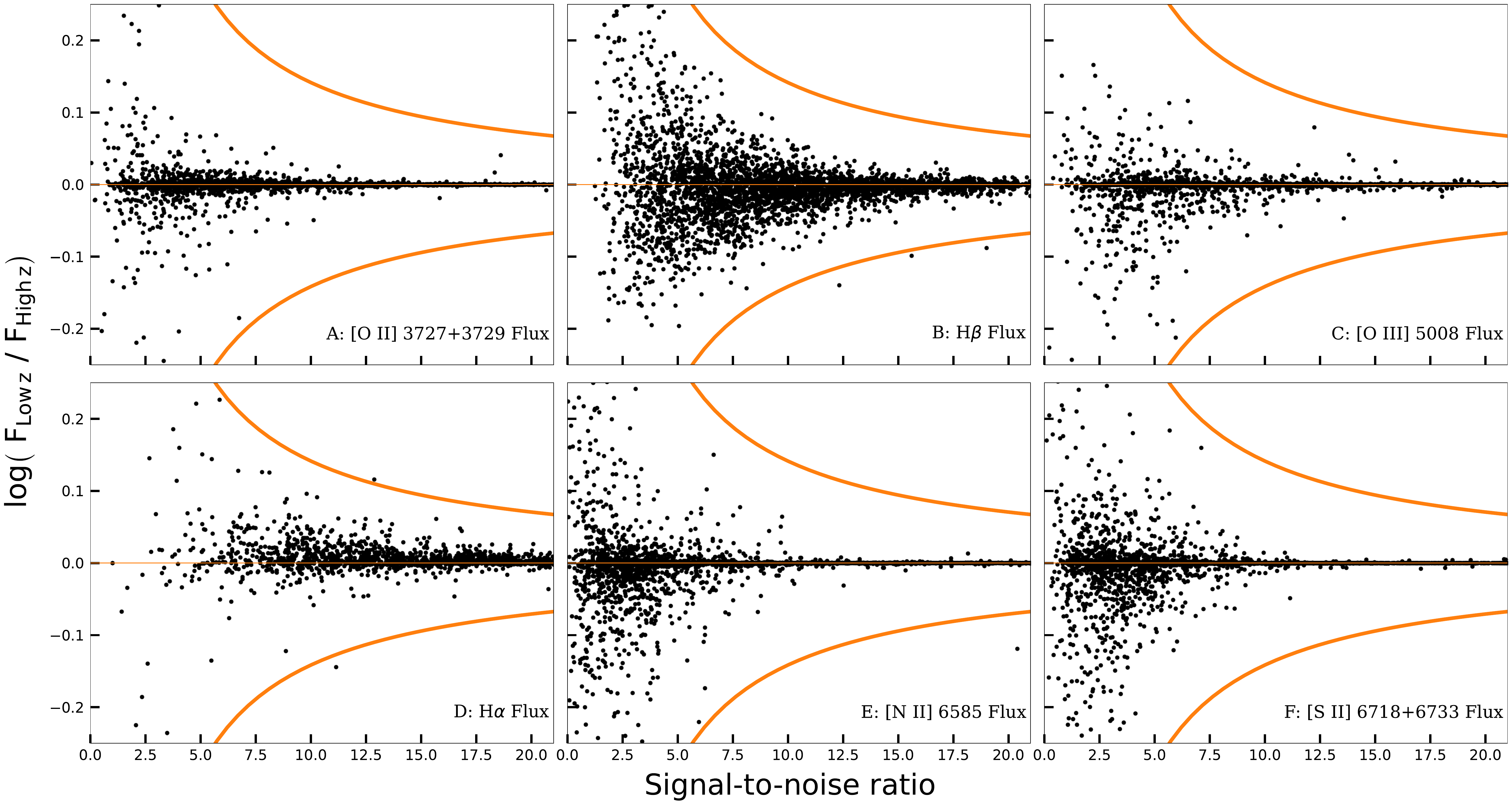}
\caption{
Comparison of line--flux measurements obtained by running the eBOSS--DAP on the same random subsample of 4171 spectra satisfying $9 < \mathrm{EW}_{\mathrm{H}\beta} < 11$~\text{\AA} and $0.1 < z < 0.4$, using the two line--tying prescriptions employed in the eBOSS--DAP: tying all kinematics to H$\alpha$ for $z < 0.4963$ and tying all kinematics to H$\beta$ for $z > 0.4963$. 
For each of the six strong emission lines, the y-axis shows 
$\log\!\left(\mathrm{F}_{\mathrm{Low\,z}} / \mathrm{F}_{\mathrm{High\,z}}\right)$, 
where $\mathrm{F}_{\mathrm{Low\,z}}$ is the flux measured using the H$\alpha$ tying configuration (low--$z$ prescription) and $\mathrm{F}_{\mathrm{High\,z}}$ is the flux measured using the H$\beta$ tying configuration (high--$z$ prescription) for the identical object and line. 
The x-axis shows the signal--to--noise ratio of that line measured using the H$\beta$ tying configuration. 
The orange curves show the expected 1$\sigma$ flux uncertainties as a function of S/N.  The data points do not show a significant systematic offset and they lie well inside the orange lines, indicating that the difference induced by our choice of anchor line is significantly smaller than the random error.
\label{fig:trumpet_z}}
\end{figure*}

We next assess the effect of the reference line change by taking the same set of spectra and refitting them with the reduced set of emission lines, tying all kinematics to H$\alpha$ instead of H$\beta$.
We then compare the emission line fluxes measured with the \nlineslow-line list using different reference lines (H$\alpha$ or H$\beta$) to isolate the impact of the reference line choice.
Across the same six emission lines, the mean percent difference between the H$\alpha$-tied and H$\beta$-tied fits is consistent with zero for all six lines, with the largest offset being 0.4\% for [N~II]~6585.
The mean difference normalized by the propagated measurement error is also small for all lines, reaching at most $\sim$3\% of the measurement error for H$\beta$.
We therefore find no systematic offset introduced by the choice of reference line; any differences between the H$\alpha$-tied and H$\beta$-tied fits are consistent with random measurement noise.
Exact values for each line are given in Table~\ref{table:survey_analysis_z}.
As shown in Figure~\ref{fig:trumpet_z}, the flux ratios remain tightly clustered with no significant trend with S/N, indicating that the choice of tying line has no meaningful effect on the recovered fluxes.

\begin{table*}[!ht]
\centering
\caption{Comparison of Emission-Line Fluxes: High-EW vs. Low-EW Line List (High-$z$, H$\beta$-tied)}
\label{table:survey_analysis_ew}
\begin{tabular}{|| l c c c c ||}
\hline
Line & $N_{\rm gal}$ & Mean \% Diff & Std \% Diff & Mean $\sigma$-ratio \\
\hline\hline
{[O~II]} 3727+3729 Flux & 4144 & 0.01  & 6.69  & 0.0016  \\
H$\beta$ Flux           & 4067 & 0.99  & 6.69  & 0.0849  \\
{[O~III]} 5008 Flux     & 3974 & 0.23  & 3.97  & 0.0067  \\
H$\alpha$ Flux          & 4166 & -0.55 & 3.44  & -0.0682 \\
{[N~II]} 6585 Flux      & 4080 & 0.74  & 14.08 & 0.0153  \\
{[S~II]} 6718+6733 Flux & 4111 & 0.71  & 7.39  & 0.0136  \\
\hline
\end{tabular}
\tablecomments{Mean \% Diff is the per-galaxy percent difference $(F_1-F_2)/((F_1+F_2)/2)\times100$, averaged over the sample; values far from 0 indicate a systematic offset between the two configurations. Std \% Diff is the scatter in that per-galaxy percent difference (random variation). Mean $\sigma$-ratio is the mean percent difference normalized by the propagated measurement error; values near 0 indicate the offset is consistent with measurement noise alone.}

\end{table*}
\begin{table*}[!ht]
\centering
\caption{Comparison of Emission-Line Fluxes: H$\alpha$-Tied vs. H$\beta$-Tied (Low-EW, Low-$z$)}
\label{table:survey_analysis_z}
\begin{tabular}{|| l c c c c ||}
\hline
Line & $N_{\rm gal}$ & Mean \% Diff & Std \% Diff & Mean $\sigma$-ratio \\
\hline\hline
{[O~II]} 3727+3729 Flux & 3998 & -0.19 & 7.00  & 0.0008  \\
H$\beta$ Flux           & 4024 & 0.12  & 5.13  & 0.0251  \\
{[O~III]} 5008 Flux     & 3996 & 0.33  & 4.10  & 0.0085  \\
H$\alpha$ Flux          & 4021 & -0.22 & 3.10  & -0.0288 \\
{[N~II]} 6585 Flux      & 3929 & 0.35  & 15.40 & 0.0106  \\
{[S~II]} 6718+6733 Flux & 3966 & 0.39  & 8.34  & 0.0082  \\
\hline
\end{tabular}
\tablecomments{See Table~\ref{table:survey_analysis_ew} for column definitions.}
\end{table*}

\clearpage

\section{H\texorpdfstring{$\beta$}{beta} vs.\ Stellar Continuum Velocity Dispersion} \label{sec:add_plots_two}

\begin{figure}[!ht]
\plotone{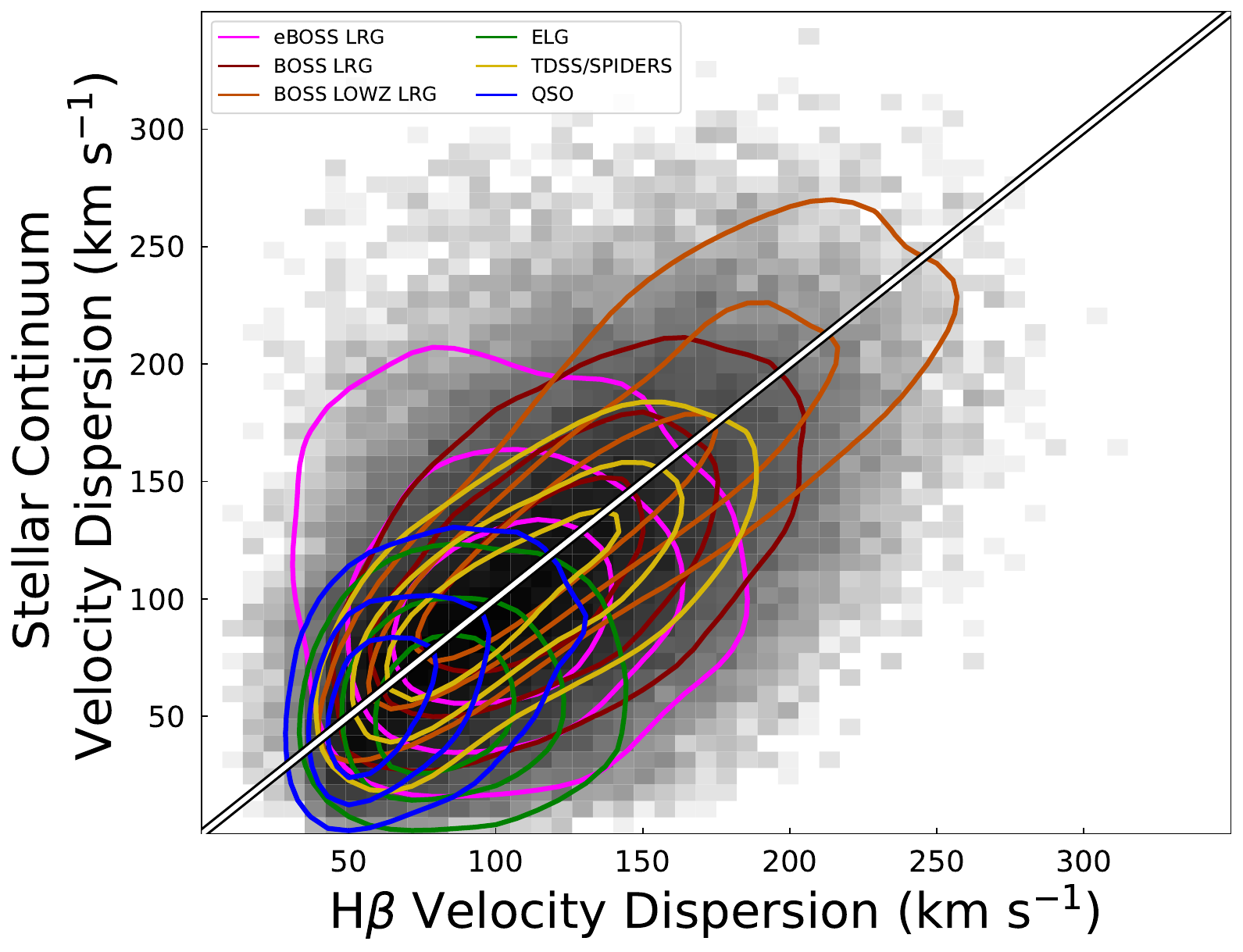}
\caption{H$\beta$ emission-line velocity dispersion vs.\ stellar continuum velocity dispersion for eBOSS galaxies with stellar continuum dispersion error $< 50$~\kms, H$\beta$ dispersion error $< 50$~\kms, H$\beta$ S/N $> 3$, and H$\beta$ EW $> 3$~\AA.
The grayscale is a 2D log-scaled histogram of the full sample.
The contours show the different target classes; the outermost contour for each class encloses 75\% of the total sample.
The one-to-one line is shown in white.
The strong agreement between the two independent dispersion measurements across most target classes confirms that the stellar continuum velocity dispersions measured by the eBOSS-DAP are reliable.  The one exception is the ELGs, which have stellar velocity dispersions that are 25~\kms\ lower than their gas dispersion, on average. This offset may be due to physical effects (e.g., the gas tracing more of the galaxy's rotation curve than the starlight). 
\label{fig:hb-sc-disp}}
\end{figure}

\end{document}